\documentclass[12pt]{article}
\usepackage{mathtools}
\usepackage{diagbox}
\usepackage{savesym}
\usepackage[nosort]{cite}
\usepackage{wasysym}
\usepackage{amscd}
\usepackage{graphicx}
\usepackage{pifont}
\usepackage{float}
\usepackage{subcaption}
\usepackage[all]{xy}
\usepackage{multicol}
\usepackage{multirow}
\usepackage{amsfonts}
\usepackage{picture}
\usepackage{amssymb}
\savesymbol{iint}
\savesymbol{iiint}
\usepackage{amsmath}
\restoresymbol{TXF}{iint}
\restoresymbol{TXF}{iiint}
\usepackage{color}
\usepackage{style}
\usepackage{hyperref}
\hypersetup{colorlinks=true}
\hypersetup{linkcolor=black}
\hypersetup{citecolor=black}
\hypersetup{urlcolor=black}
\usepackage{setspace}
\usepackage{wrapfig}
\usepackage{verbatim}
\usepackage{dsfont}
\usepackage[vcentermath]{youngtab}

\numberwithin{equation}{section}
\usepackage{float}
\restylefloat{table}
\allowdisplaybreaks
\usepackage{accents}

\usepackage{upgreek}

\newcommand{\R}{{\mathbb R}}

\def\tilde{\widetilde}

\def\hat{\widehat}

\def\bar{\overline}

\def\1{{\mathds 1}}

\def\CS{{\mathcal S}}
\def\CT{{\mathcal T}}

\newcommand{\Z}{\mathbb{Z}}

\newcommand{\dd}{\hspace{1pt}\mathrm{d}}

\usepackage{datetime}
\usepackage{booktabs}

\usepackage{enumitem} 
\usepackage{moreenum}

\newcommand{\ii}{\hspace{1pt}\mathrm{i}\hspace{1pt}}
\newcommand{\bea}{\begin{eqnarray}}
\newcommand{\eea}{\end{eqnarray}}
\newcommand{\Rf}[1]{Ref.~\cite{#1}}
\newcommand{\nn}{\nonumber}

\newcommand{\e}{\hspace{1pt}\mathrm{e}}

\def\SU{{\mathrm{SU}}}
\def\U{{\mathrm{U}}}
\def\O{{\mathrm{O}}}

\def\SO{{\mathrm{SO}}}
\def\Spin{{\mathrm{Spin}}}
\def\PO{{\mathrm{PO}}}

\def\eg{{\it e.g.}}

\newcommand{\Sec}[1]{Sec.~\ref{#1}} 
\newcommand{\Fig}[1]{Fig.~\ref{#1}} 
\newcommand{\eq}[1]{eqn.~(\ref{#1})}


\begin{document}

\institution{Caltech}{\centerline{${}^{1}$Walter Burke Institute for Theoretical Physics,
California Institute of Technology, Pasadena, CA 91125, USA}}
\institution{CMSA}{\centerline{${}^{2}$Center of Mathematical Sciences and Applications, Harvard University,  Cambridge, MA 02138, USA}}
\institution{IAS}{\centerline{${}^{3}$School of Natural Sciences, Institute for Advanced Study, Princeton, NJ 08540, USA}}

\title{An Effective Field Theory for\\ Fractional Quantum Hall Systems %effects 
near $\nu=5/2$}

\authors{Po-Shen Hsin\worksat{\Caltech}\footnote{e-mail: {\tt phsin@caltech.edu}},
Ying-Hsuan Lin\worksat{\Caltech}\footnote{e-mail: {\tt yhlin@caltech.edu}},
Natalie M. Paquette\worksat{\Caltech}\footnote{e-mail: {\tt nataliep@caltech.edu}},
Juven Wang\worksat{\CMSA,\IAS}{\footnote{e-mail: {\tt jw@cmsa.fas.harvard.edu}} }

}

\abstract{\noindent 
We propose an effective field theory (EFT) of fractional quantum Hall systems near the filling fraction $\nu=5/2$ that flows to pertinent IR candidate phases, including non-abelian Pfaffian, anti-Pfaffian,  and particle-hole Pfaffian states (Pf, APf, and PHPf). Our EFT has a 2+1$d$ O(2)$_{2,L}$ Chern-Simons gauge theory coupled to four Majorana fermions by a discrete charge conjugation gauge field, with Gross-Neveu-Yukawa-Higgs terms.  Including deformations via a Higgs condensate and fermion mass terms, we can map out a phase diagram with tunable parameters, reproducing the prediction of the recently-proposed percolation picture and its gapless topological quantum phase transitions.
Our EFT captures known features of both gapless and gapped sectors of time-reversal-breaking domain walls between Pf and APf phases.  Moreover, we find that Pf$\mid$APf domain walls have higher tension than domain walls in the PHPf phase. Then the former, if formed, may transition to the energetically-favored PHPf domain walls; this could, in turn, help further induce a bulk transition to PHPf. 
}

\preprint{CALT-TH-2020-021}

\maketitle

\pagenumbering{arabic}
    \setcounter{page}{2}

\setcounter{tocdepth}{3}
\tableofcontents

\section{Introduction}

One of the first non-abelian topologically ordered candidate states was observed experimentally in 1987 \cite{willett1987}. It is the filling fraction $\nu=5/2$ fractional quantum Hall (fQH) state
of an interacting electron gas in 2+1 spacetime dimensions (denoted as 2+1$d$). It has a
fractional quantized Hall conductance $\sigma_{xy}=5/2$ in units of $e^2/h$ where $e$ is the electron charge and $h$ is the Planck constant. There have been many proposed candidate states to describe the underlying topological orders of this system:
the major non-abelian candidates include
Moore-Read's Pfaffian state \cite{moore1991} (see also  \cite{1991Wen}), 
its particle-hole conjugate known as the 
anti-Pfaffian state \cite{levin2007,lee2007}, 
and a particle-hole symmetric state
known as the particle-hole Pfaffian state \cite{son2015}.
The particle-hole Pfaffian state \cite{son2015}
was originally 
proposed to be a particle-hole symmetric version of a composite fermion theory for the half-filled Landau level system \cite{halperin1993}. {Ref.~\cite{Jolicoeur200707051619,ZuckerFeldman2016} made earlier attempts to propose candidate wavefunctions
for the particle-hole Pfaffian state.}

In 2017, a remarkable experimental measurement by Banerjee {\it et al} \cite{Banerjee:2018qtz} suggested that the 
thermal Hall conductance of the $\nu=5/2$ fQH state is $\kappa_{xy}=5/2$
in units of $\pi^2k_B^2T/3h$, where $k_B$ is the Boltzmann constant and $T$ is the temperature.\footnote{The edge modes of the quantum Hall system can be understood via the bulk-boundary correspondence
of $2+1d$ Chern-Simons theory. 
In fact, the 
thermal Hall conductance $\kappa_{xy}=(c_L-c_R)\frac{\pi^2 k_B^2}{3 h}T$
is proportional to the chiral central charge 
$c_- \equiv c_L-c_R$, which is the difference between the left/right central charges 
$c_L$ and $c_R$.
It counts the  degrees of freedom of chiral modes 
 of the $(1+1)d$ edge conformal field theory (CFT) living on the boundary of a bulk-gapped $2+1d$ topological state \cite{kane1997}.
  For non-abelian fQH states, 
  the half-integer $\kappa_{xy}$ is attributable to an odd number of (1+1)$d$ chiral real Majorana-Weyl fermions on the boundary \cite{WenPhysRevLett.70.355}, 
in addition to $(1+1)d$ chiral bosons or chiral complex fermions.
}

{In this work,
we propose a unified bulk effective field theory (EFT) that give rise to various topological quantum 
field theories (TQFTs)
and their edge modes pertinent to the $\nu=5/2$ fQH system. We map the EFT parameters to
experimental quantities to produce a phase diagram in terms of the filling fraction (or the magnetic field) vs.~the disorder strength.
The phase diagram produced from our EFT turns out to be 
qualitatively similar to the previous theoretically proposed 
phase diagrams via the percolating phase transitions
from the $2+1d$ disordered systems with random puddles and domain walls of Pfaffian and anti-Pfaffian states 
\cite{Mross:2018MOSMH,Wang:2018WVH,Lian:2018xep}.
In the following, we first recall pertinent proposals from the literature.}

\subsection{Overview of theoretical proposals and questions}

 While both the theoretical proposals of Pfaffian state \cite{moore1991}
 and anti-Pfaffian state \cite{levin2007,lee2007}
 have a consistent fractional quantized Hall conductance $\sigma_{xy}=5/2$, their thermal Hall conductances, $\kappa_{xy}=7/2$ and $\kappa_{xy}=3/2$ respectively, seem to contradict with the result of \cite{Banerjee:2018qtz}. By contrast,
 the particle-hole Pfaffian state proposed by Son in 2015 \cite{son2015}\footnote{The particle-hole Pfaffian is analogous to the $T$-Pfaffian
 or $CT$-Pfaffian that occur on the surface of $3+1d$ topological superconductors, see \cite{chen2014Fidkowski, Metlitski2014xqa1406.3032}.} predicts both $\sigma_{xy} = 5/2$ and $\kappa_{xy}=5/2$, consistent with this recent experiment.
On the other hand,
vast numerical studies \cite{morf1998, rezayi2000, peterson2008, feiguin2009, wangh2009Haldane, storni2010DasSarma, rezayi2011Simon, papic2012HaldaneRezayi, zaletel2015MongPollmannRezayi, pakrouski2015PetersonNayak} on the 
$\nu=5/2$ fQH system
at low energy favor either
the Pfaffian state or the anti-Pfaffian state. The dilemma
between the experiment (favoring $\kappa_{xy}=5/2$)
and the numerical data
(favoring $\kappa_{xy}=7/2$ or
$3/2$) 
raises an important issue: can the seemingly contradictory experimental and numerical results be reconciled?

\Rf{ZuckerFeldman2016} argued that the numerical simulations are simplified systems lacking both disorder (say, induced by impurities of experimental samples) and Landau-level mixing (LLM), which occur in real laboratory experiments.
\Rf{ZuckerFeldman2016} further suggested that the particle-hole Pfaffian may be stabilized
by disorder, {\it i.e.} LLM and impurities that break particle-hole symmetry. However, \Rf{ZuckerFeldman2016} did not provide analytic details on how disorder can help realize this possibility in practice. 

Building on this suggestion, \Rf{Mross:2018MOSMH,Wang:2018WVH,Lian:2018xep} investigated the possibility
of particle-hole Pfaffian (PHPf) topological order emerging from disordered puddle systems 
of Pfaffian (Pf) and 
anti-Pfaffian (APf) states\footnote{For the sake of brevity, below we abbreviate
Pfaffian state as Pf, anti-Pfaffian state as APf, and
particle-hole Pfaffian as PHPf. See Appendix A of \Rf{Lian:2018xep} for the systematic list of data of the pertinent $\nu=5/2$-quantum Hall liquids
in terms of $2+1d$ bulk topological quantum field theories (TQFTs) and $(1+1)d$ edge theories.} 
with percolating random domain walls.

We recall that: 
\begin{enumerate}
\item
Neither the Pf nor the APf state has particle-hole (PH) symmetry \cite{Greiter:1991raWenWilczekPRL}.
Both Pf and APf
have their lower Landau levels fully occupied with spin-polarized electrons (which contribute $\sigma_{xy}=2$).
However, in the absence of LLM, if we assume that spin-polarized electrons in the highest, half-filled Landau level (so there is another contribution of $\sigma_{xy}=1/2$ and $\nu = 5/2$ in total) interact only through two-body interactions, then exact PH symmetry \emph{is} present in the \emph{idealized} Hamiltonian.\footnote{In the literature, there are two conventions for naming the Landau levels. One convention is to call the lowest level the zeroth Landau level (which here is fully occupied, with  spin-up and spin-down polarized electrons contributing $\sigma_{xy}=2$), and call the next the first Landau level (which here is half-filled with polarized spin, contributing $\sigma_{xy}=1/2$)
  \cite{Mross:2018MOSMH,Lian:2018xep}. 
  Another convention instead calls the
  lowest Landau level the first Landau level, and the
  half-filled Landau level the
 second Landau level \cite{Wang:2018WVH}.
We use the first convention for this $\nu=5/2$ system.}
With the PH symmetry at $\nu=5/2$, the two PH symmetry-breaking states, Pf and APf, are related by
a PH transformation. Thus, they have the same energy and 
become two degenerate states at $\nu=5/2$. 
PH symmetry is broken away from $\nu=5/2$,
so either Pf or APf is favored on each side of
$\nu> 5/2$ and $\nu< 5/2$.  At $\nu=5/2$, if PH symmetry is spontaneously broken, one of Pf and APf 
is realized.

\item With LLM, PH symmetry is only approximate, so the critical $\nu$ may be shifted to $\nu_c=5/2 +\delta \nu$. Second-order perturbation theory from
LLM modifies the Hamiltonian and
induces PH-symmetry-breaking three-body interaction terms, so both Pf or APf can be candidate ground states near $\nu_c$.
Whether Pf or APf is the candidate ground state for $\nu$ near $\nu_c$ 
partly depends on the sign of the three-body terms. For a small deviation away from $\nu_c$, we gain 
quasiparticles for $\nu>\nu_c$,
and quasiholes for $\nu<\nu_c$.
If the quasiparticles of APf have a lower energy than those of Pf for $\nu>\nu_c$, then in turn 
quasiholes of Pf have a lower energy than those of APf for $\nu<\nu_c$,
due to their PH conjugate properties at $\nu_c$ (and vice versa).
As long as $\nu$ is  within the $\nu_c \simeq 5/2$  fractional quantized Hall plateau,
we assume Pf is favored for $\nu<\nu_c$ (and hence 
APf is favored for $\nu>\nu_c$) for simplicity \cite{levin2007}.\footnote{There are two cases:
(1) The quasiparticles of APf have a lower energy than those of Pf for $\nu>\nu_c$. Then 
quasiholes of Pf have a lower energy than quasiholes of APf for $\nu<\nu_c$. In this case, Pf is favored for $\nu<\nu_c$ and 
APf is favored for $\nu>\nu_c$.
(2) The quasiparticles of Pf have a lower energy than quasiparticles of APf for $\nu>\nu_c$. Then, quasiholes of APf have a lower energy than quashioles of Pf for $\nu<\nu_c$. In this case, APf is favored for $\nu<\nu_c$ and Pf is favored for $\nu>\nu_c$.
Numerical simulations have favored both possibilities (see the discussions in \cite{Wang:2018WVH} and the references therein), so we cannot  exclude (1) or (2). We will assume (1) without losing generality. 
}

\item Under the presence of spatial disorder (e.g., quenched disorder arising from the presence of impurities, or spatial variations in the chemical potential) 
and spatial density fluctuations
on the sample,
many puddles of Pf or APf of radii $\ell_0$ would form, with puddle sizes bounded by $\ell_B <\ell_0 < L$ where $\ell_B=\sqrt{\hbar c/eB}$ is the magnetic length under a magnetic field $B$, and $L$ is the sample size. The disorder-induced
puddles \cite{imry1975Ma} separate Pf and APf
into patterns analogous to that of islands and seas in an archipelago (see the picture illustration in Fig.~1
and Fig.~3 in \cite{Lian:2018xep}). 
 The boundaries of puddles
then form $(1+1)d$ domain walls (between Pf and APf regions)
hosting four gapless chiral real Majorana-Weyl fermions
(with chiral central charge $c_-= 4 \times \frac{1}{2}=2$) and two copies of the so-called gappable non-chiral double-semion theory of compact complex bosons (with $c_-=0$, and $c_L=c_R=1$).
It is proposed that the domain walls percolating in the bulk drive the bulk phase into the so-called 
\emph{percolating phase}.\footnote{Let us briefly define what we mean by dis/order, percolation, and de/localized.
\begin{itemize}
\item Order vs.~disorder: We use order to mean Landau-Ginzburg
symmetry-breaking order, as well as Wen's long-range entangled topological order (beyond Landau).
Disorder here is mainly used to mean quenched disorder 
caused by impurities or a spatially non-uniform chemical potential, inducing puddles of Pf or APf near $\nu_c$.
\item Percolation: When we say that a phase percolates, we mean that the
phase can extend through the whole bulk-boundary system (e.g.,
see Fig.~3 (a) and (c) of \cite{Lian:2018xep}).
When we instead say the domain walls percolate, we mean that  Pf$\mid$APf domain walls can extend through the whole bulk-boundary system (e.g.,
see Fig.~3 (b) of \cite{Lian:2018xep}).
\item Localized vs.~delocalized: When we say the neutral Majorana modes are delocalized, we mean that the Majorana modes can diffuse freely on the network of domain walls. The delocalization happens at the percolation transition (approximately near a percolation critical point). When the neutral Majorana modes are delocalized, the thermal Hall $\kappa_{xy}$ is unquantized, thus either causing a \emph{percolation transition} or a \emph{thermal metal phase}. When 
neutral Majorana modes are localized (on the domain walls), we have a quantized $\kappa_{xy}$.
\end{itemize} 
``Percolation'' is used to indicate when a \emph{spatial subregion} (e.g. Pf, APf, or domain walls) spreads in the spatial sample, whereas 
``(de)localization'' is used to indicate when \emph{zero energy modes or energetic modes} in the energy spectrum are de/localized in the spatial sample.
}
The question about the nature of the percolating phase becomes the question of understanding whether the domain wall degrees of freedom are \emph{localized} in the bulk or \emph{delocalized} through the whole bulk-boundary system (see the picture illustration in Fig.~3 in \cite{Lian:2018xep}).

\Rf{Mross:2018MOSMH, Wang:2018WVH} 
modeled the
$\nu=5/2$ system in terms of a checkerboard network (of alternating Pf and APf in each chequered pattern) known as a
Chalker-Coddington network model \cite{chalker1988} (previously used in modeling the integer quantum Hall plateau transition). \Rf{Lian:2018xep}
 performed perturbative and non-perturbative analyses of the $(1+1)d$ edge theory on the domain wall between Pf and APf states at different disorder energy scales, with particular focus on the emergent symmetries

\subsection{Comparison of three related proposals on disordered percolating systems}

We compare the results of \Rf{Mross:2018MOSMH, Wang:2018WVH, Lian:2018xep}, which we also summarize pictorially in 
\Fig{fig:Hall-kappa}
and \Fig{fig:Hall-kappa-1711} below:\footnote{There is an alternative
interpretation from 
\cite{Simon2018mhp180109687, Ma2018xvjFeldman1809.05488, simon2019partial, Simon1909.12844Zaletel, Mulligan2004.04161}
favoring the
anti-Pfaffian state (see also the criticism
\cite{Feldman2018ssm180503204} %Feldman
of Ref.~\cite{Simon2018mhp180109687}'s interpretation). 
Some of these works propose 
that partial- or non-thermal equilibrium 
of anti-Pfaffian edge modes
can explain the $\kappa_{xy}=5/2$ measurement \cite{Banerjee:2018qtz}, 
even though the anti-Pfaffian bulk state has  
$\kappa_{xy}=3/2$ at equilibrium.
We shall not discuss this scenario \cite{Simon2018mhp180109687, simon2019partial, Simon1909.12844Zaletel, Mulligan2004.04161},
since we wish to obtain
an effective bulk field theory motivated by the scenario of \cite{Lian:2018xep}.
}

\begin{figure}[!h]
  \centering
  \includegraphics[width=4.9in]{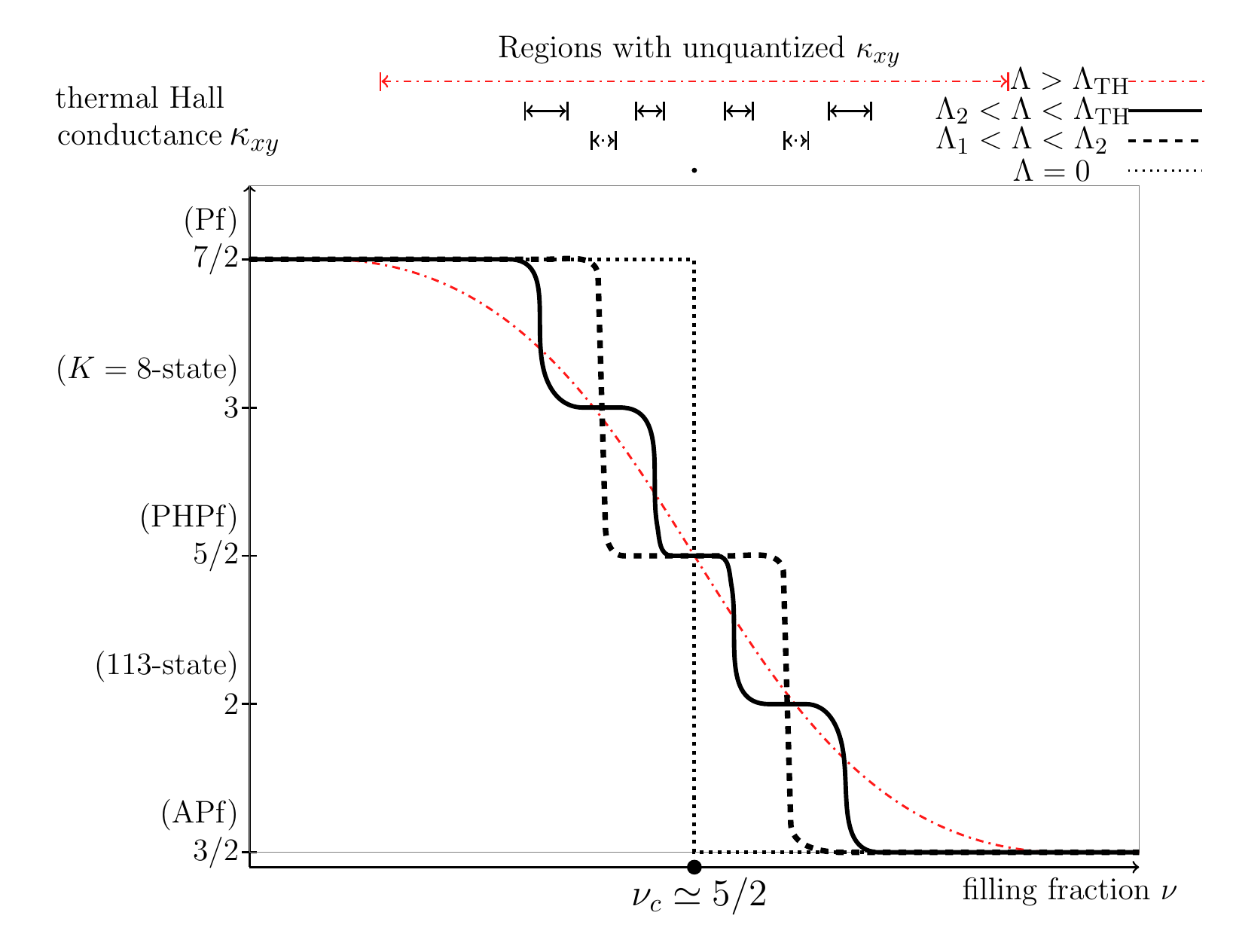}
  \caption{Thermal Hall conductance $\kappa_{xy}$ vs filling fraction $\nu$ for the scenario proposed in \Rf{Lian:2018xep}; see also \Fig{fig:Phase-Pf-APf-phase} for a phase diagram. 
  At different disorder energy scales $\Lambda$, we plot several curves of ($\nu$, $\kappa_{xy}$). 
  At $\Lambda=0$, the $\kappa_{xy}$ (drawn as a dotted line)
  jumps  at $\nu_c$ under a first order phase transition.
  From $0< \Lambda < \Lambda_1$, the jump can become smoother
  due to disorder.
   In the regime $\Lambda_1 < \Lambda < \Lambda_2$, 
 drawn as a dashed line, an intermediate $\kappa_{xy}= 5/2$ plateau phase appears.
  Finally, when $\Lambda_2 < \Lambda < \Lambda_{\rm{TH}}$, there are multiple plateau phases at $\kappa_{xy}= 3, 5/2$, and $2$.
  Notice that when $\Lambda >0$, all transitions between different quantized $\kappa_{xy}$ can have \emph{broadening}, where the jumps at transitions become smoother slopes. 
 On the top panel, we show different line intervals 
 which represent the extent of broadening over ranges of $\nu$ demarcated on the horizontal axis, for the values stated of $\Lambda$  on the top right corner.
 When $\Lambda > \Lambda_{\rm{TH}}$, the slope is smooth enough to become a thermal metal so there is no quantized $\kappa_{xy}$ between 7/2 and 3/2.
See Remark \ref{1801.10149}. \label{fig:Hall-kappa}}
\end{figure}

\begin{figure}[!h]
  \centering
  \includegraphics[width=3.15in]{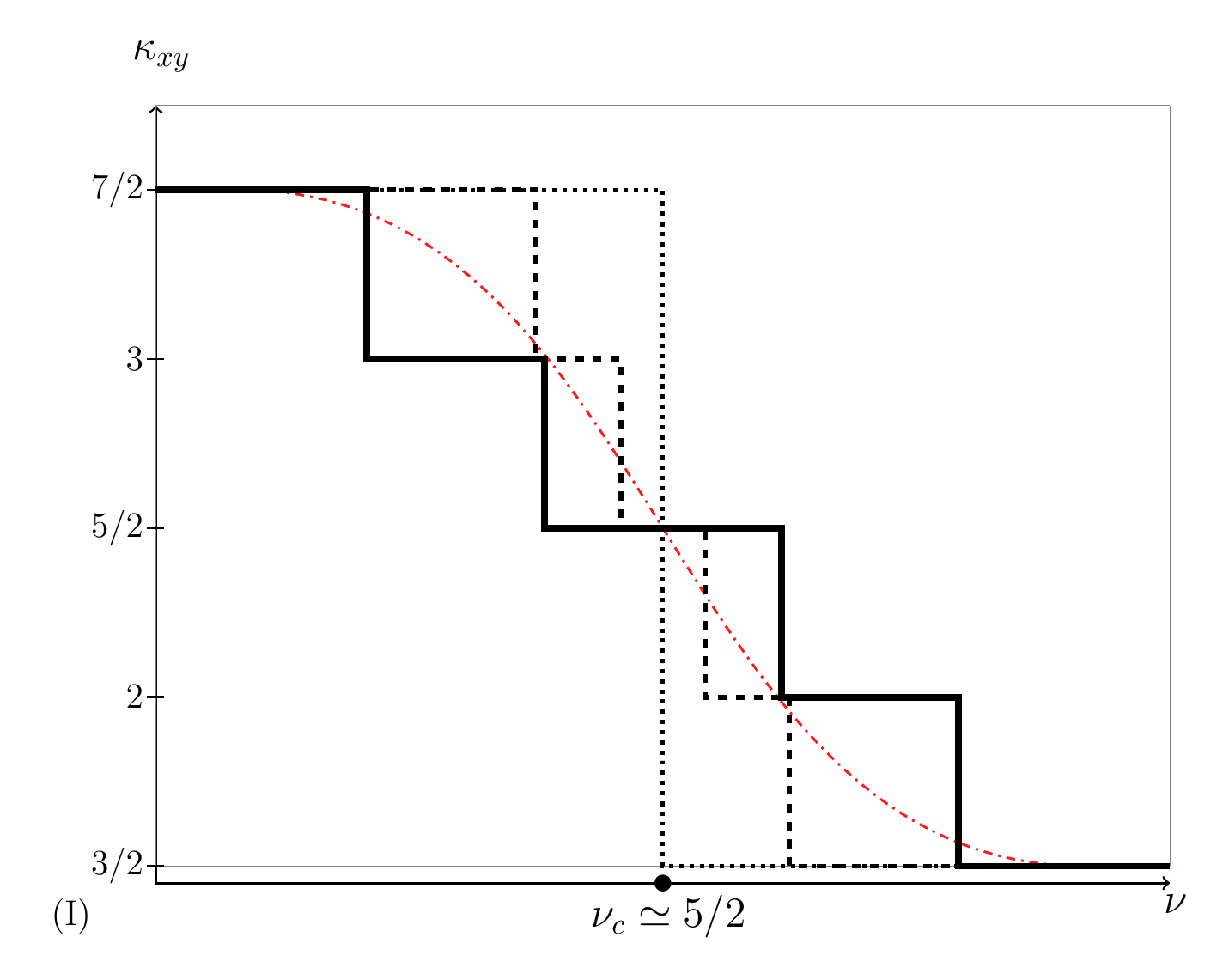}
    \includegraphics[width=3.15in]{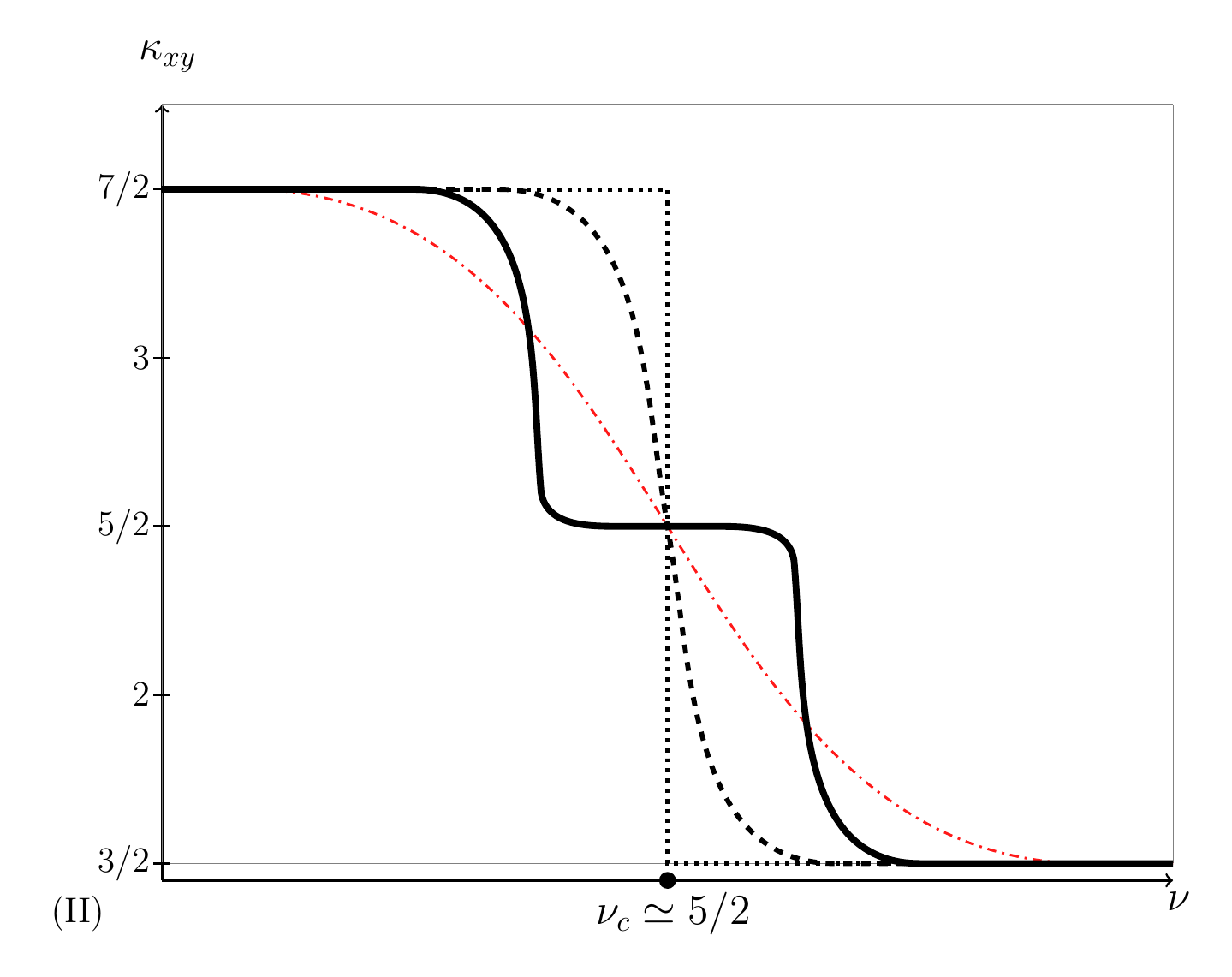}
  \caption{Thermal Hall conductances: Scenario (I) from \Rf{Mross:2018MOSMH} (left) and scenario (II) from  \Rf{Wang:2018WVH} (right). We use the same legend for drawing curves at different scales $\Lambda$ as in Figure \ref{fig:Hall-kappa}.
  At $\Lambda=0$, $\kappa_{xy}$ (drawn as a dotted line)
  jumps  at $\nu_c$ under a first order phase transition. For
  $\Lambda>0$, scenarios (I) and (II) differ.
  Scenario (I)'s $\kappa_{xy}$ has four jumps at the plateau
  for any $0<\Lambda<\Lambda_{\rm{TH}}$, and $\kappa_{xy}$ becomes smooth with non-quantized values for $\Lambda>\Lambda_{\rm{TH}}$.
  }\label{fig:Hall-kappa-1711}
\end{figure}

\begin{enumerate} [leftmargin=0mm, label=\textcolor{blue}{(\arabic*)}:, ref=\textcolor{blue}{(\arabic*)}]

\item \label{1711.06278} 
\Rf{Mross:2018MOSMH} proposed that a single first-order-like transition between Pf and APf occurs at $\nu_c$ and at zero disorder, due to an O(4) symmetry rotating four gapless chiral Majorana modes. The presence of these Majoranas induce a jump $\Delta \kappa_{xy} = \Delta c_- = 2$. In the presence of any nonzero disorder, which weakly perturbs  the first-order critical point, \Rf{Mross:2018MOSMH} proposed four consecutive continuous phase transitions
(e.g., second-order transitions). Each transition
causes $\kappa_{xy}$ to jump by 1/2, due to a single neutral chiral Majorana mode:
from Pf ($\kappa_{xy}=7/2$) $\to$ $\kappa_{xy}=3$
$\to$ $\kappa_{xy}=5/2$ $\to$ $\kappa_{xy}=2$ $\to$ APf ($\kappa_{xy}=3/2$).
\Rf{Mross:2018MOSMH} also expected the same universality class for disorder anisotropic models
and uniform models.
See the Fig.~1 phase diagram of \cite{Mross:2018MOSMH}.
We illustrate \Rf{Mross:2018MOSMH}'s thermal Hall prediction in \Fig{fig:Hall-kappa-1711}'s (I).

\item \label{1711.11557}
\Rf{Wang:2018WVH} suggested that for a finite range of $\nu \simeq \nu_c$, the Pf$\mid$APf domain walls percolate.\footnote{In \Rf{Wang:2018WVH}'s language, 
neither Pf nor APf percolates, but the Pf$\mid$APf domain walls percolate. However, in \Rf{Lian:2018xep}'s language,
not only the Pf$\mid$APf domain walls percolate,
but also both Pf and APf percolate --- because some regions
of Pf or APf extend through the whole bulk-boundary.
} 
If the charge neutral Majorana edge modes
can diffuse freely in the network of domain walls in the bulk-boundary system,
\Rf{Wang:2018WVH} proposed a 
 \emph{thermal metal} phase with an unquantized thermal Hall $\kappa_{xy}$ but 
 a divergent $\kappa_{xx}$ (and, as usual, a quantized Hall conductance $\sigma_{xy}=5/2$, and $\sigma_{xx}=0$ at zero temperature).
If the neutral Majorana modes
are localized,
\Rf{Wang:2018WVH} proposed a quantized $\sigma_{xy}=5/2$ phase with a quantized thermal Hall conductance $\kappa_{xy}=5/2$. \Rf{Wang:2018WVH} suggested that between the Pf and APf phases, there is a possible wide range of thermal metal behavior, even at low disorder.
By tuning $\nu$, in the absence of disorder, there is a first-order-like transition between Pf $\to$ APf.
At low disorder, there is a sequence of transitions from Pf $\to$ thermal metal $\to$ APf.
At larger disorder, there is a sequence of transitions from Pf ($\kappa_{xy}=7/2$) $\to$ thermal metal $\to$ $\kappa_{xy}=5/2$ $\to$ thermal metal $\to$ APf ($\kappa_{xy}=3/2$).
The intermediate thermal metal phase is a distinct key feature of 
\cite{Wang:2018WVH}'s proposal.
See the phase diagrams in Fig.~1 and Fig.~8  of \cite{Wang:2018WVH}.  We illustrate \Rf{Wang:2018WVH}'s thermal Hall prediction in \Fig{fig:Hall-kappa-1711}'s (II).

\item \label{1801.10149}
\Rf{Lian:2018xep}
 performed perturbative and non-perturbative analyses
on the $(1+1)d$ edge theory, and studied emergent symmetries on the domain wall between Pf and APf states at different disorder energy scales $\Lambda
= \overline{v}/\ell_0$ (which is related to
the inverse of the puddle size $\ell_0$ but proportional to
the mean value of the edge state velocity $\overline{v}$).  
Then \Rf{Lian:2018xep} proposed a more specific phase diagram of the $\nu=5/2$ disordered system, schematically shown in \Fig{fig:Phase-Pf-APf-phase}.
An example of \Rf{Lian:2018xep}'s thermal Hall prediction is illustrated in \Fig{fig:Hall-kappa}.

\begin{figure}[!t]
  \centering
    \includegraphics[width=5.9in]{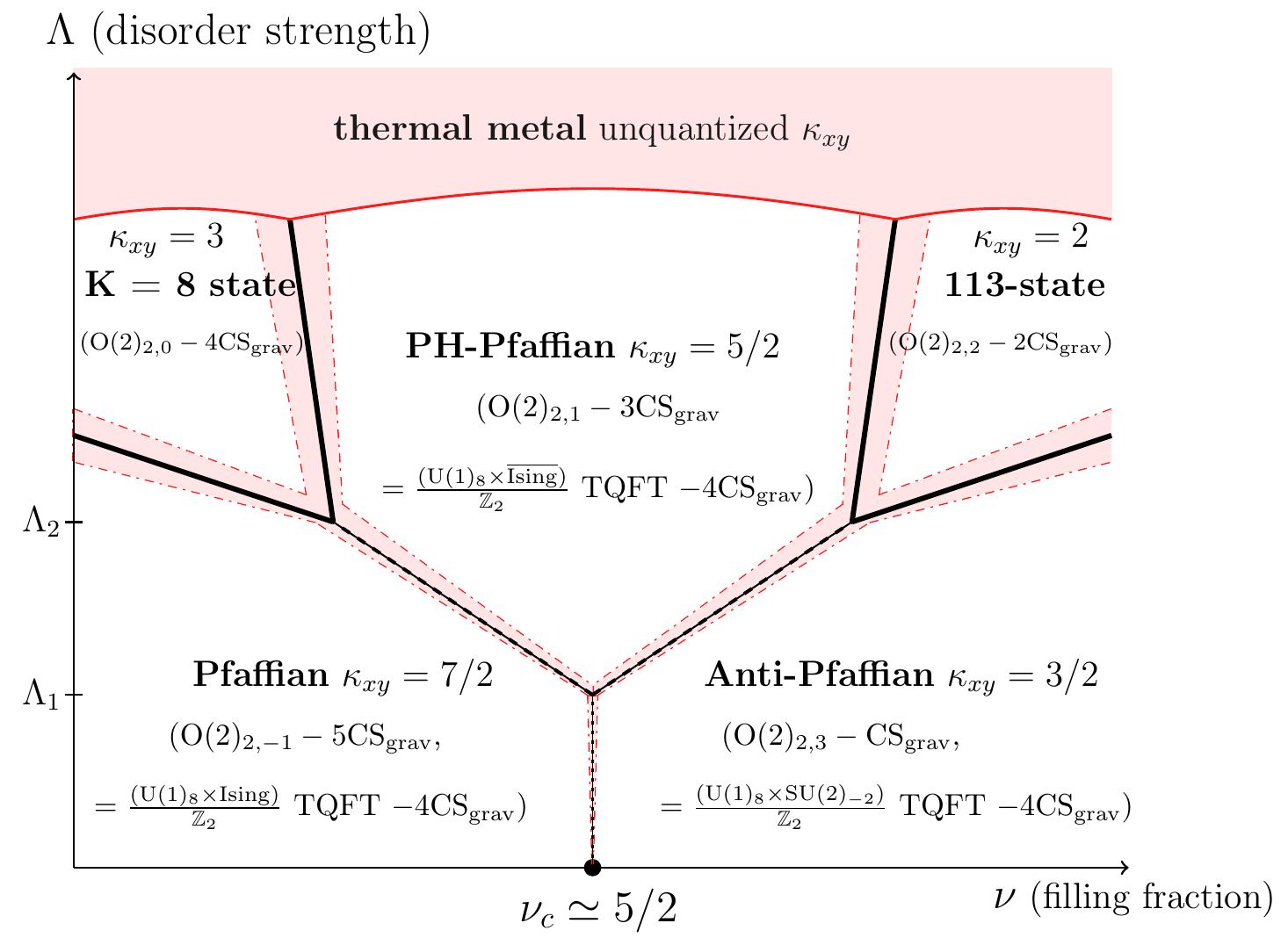}
  \caption{A schematic phase diagram similar to \Rf{Lian:2018xep}'s proposal.
  To see all these phases with varying $\kappa_{xy}$ requires that the  $\nu_c\simeq 5/2$ plateau spans a sufficient range around $\nu_c$. Previous work 
  \cite{Mross:2018MOSMH, Wang:2018WVH, Lian:2018xep} can obtain various quantized values of $\kappa_{xy}$ but cannot directly derive the bulk topological orders via the percolation transition argument. In this work, we propose a bulk 
  effective field theory (EFT) not only consistent with \cite{Mross:2018MOSMH, Wang:2018WVH, Lian:2018xep} but can reproduce all the implicated bulk topological orders.
  At zero disorder, $\Lambda=0$, the transition at $\nu_c$ is \emph{first order}.
  For $\Lambda>0$, there are different possibilities for transitions, depending on the microscopic details of samples.
  One scenario in \cite{Lian:2018xep} suggests that there are \emph{second order} phase transitions (drawn in solid black lines) between topological orders for $\Lambda>0$.
  Another scenario in \cite{Lian:2018xep} suggests that there can be \emph{first order} phase transitions between topological orders for $\Lambda>0$, but that disorder broadens these 
  first order transitions to regions
  (light red shaded regions) with unquantized $\kappa_{xy}$. 
   These broadened regions cannot merely be crossovers,
  because the topological orders and global symmetries are
  distinct on the two sides. The boundaries of these broadened regions
  (drawn as dash-dotted red curves) 
  could also be \emph{second order} phase transitions.
  At larger $\Lambda \gg \Lambda_2$, a 
  percolation transition from topological order to a thermal metal, also with unquantized $\kappa_{xy}$,
  is known to be a \emph{second order} phase transition
  (drawn as solid red curves).
  We propose a unified EFT in \eq{eqn:GNY} in \Sec{sec:EFT} and an upgraded version in \Sec{sec:moreK=8and113} to describe all phases in the phase diagram.
  }\label{fig:Phase-Pf-APf-phase}
\end{figure}
By the perturbative renormalization group (RG) analysis on disorder and scattering,
\Rf{Lian:2018xep} finds different emergent symmetries at 
different disorder energy scales. By a Berezinskii-Kosterlitz-Thouless (BKT)-type RG analysis, 
\Rf{Lian:2018xep} finds
for weak disorder
$$\Lambda<\Lambda_1 \simeq  \bar{v}(\bar{v}^2/W_v^*)^{-1/d_v},$$
there is an emergent O(4) symmetry among the four gapless chiral Majorana modes.\footnote{Here 
$\bar{v}$ is the average edge state velocity along the puddle, and
$W_v^*$ has the dimension of
$[\text{length}]^{-d_v}$ where the length scale is 
the correlation length of the BKT-like transition. 
The energy 
scale $\Lambda_1$  is around
$\ell_{\varphi}^{-1}$ set by the correlation length $\ell_{\varphi}$ 
of superconducting pairing fluctuation 
in the composite fermion picture of Pf and APf.
We thank Bert Haleprin pointing out that this $\Lambda_1$ is also related to the energy 
scale w$^{-1}$ of the domain wall width w, which can be solved from a setup with Haldane pseudopotential.} 
This describes a transition
\begin{equation}
\text{
Pf ($\kappa_{xy}=7/2$) $\to$ APf ($\kappa_{xy}=3/2$).
}
\end{equation}
(For $\Lambda=0$, this is a first order transition.
For $0<\Lambda<\Lambda_1$, this can be a second order transition or a first order transition with weak disorder broadening the transition.)
For 
$$\Lambda_1< \Lambda< \Lambda_2 \simeq \e^2/\epsilon\ell_B,$$
where $\Lambda_2$ is set by the electron's Coulomb interaction
and $\epsilon$ is a dielectric constant, we have two 
transitions
\begin{equation}
\text{Pf ($\kappa_{xy}=7/2$) $\to$ $\kappa_{xy}=5/2$ 
$\to$ APf ($\kappa_{xy}=3/2$).
}
\end{equation}
(Again, the two intermediate steps can be first order transitions but with disorder broadening, or second order transitions.) For the disorder scale 
$$\Lambda_2  < \Lambda < \Lambda_{\text{TH}},$$
we have four transitions
\begin{equation}
\text{
Pf ($\kappa_{xy}=7/2$) $\to$ $\kappa_{xy}=3$
$\to$ $\kappa_{xy}=5/2$ $\to$ $\kappa_{xy}=2$ $\to$ APf ($\kappa_{xy}=3/2$)
},
\end{equation}
all of which can be (broadened) first order transitions, or second order.

Finally, for $\Lambda > \Lambda_{\text{TH}}$, 
when the disorder is very strong,
the $\kappa_{xy}$ becomes unquantized and
we enter into the thermal metal (TH) regions (the light red area on the top of \Fig{fig:Phase-Pf-APf-phase}).
The percolation transition to the thermal metal phase guarantees the divergence of the correlation length, which therefore guarantees that the transitions from all topological orders to
the thermal metal (drawn as the red solid curves in \Fig{fig:Phase-Pf-APf-phase}) are second order phase transitions.

Note that the aforementioned disorder-broadening regions have unquantized $\kappa_{xy}$ and hence can behave similarly to a thermal metal as an intermediate phase. However, to be 
a precise thermal metal, one needs to check that $\kappa_{xx}$  diverges at zero temperature.

We expect the first-order disorder-broadening spreads to a region of size that is exponentially suppressed by 
$e^{-f(\Lambda_1,\Lambda_2)/\Lambda^2}$
with some functional form $f$ of $\Lambda_1$ 
and $\Lambda_2$ \cite{Wang:2018WVH},
which grows wider as the disorder increases (i.e. the light red area becomes wider in \Fig{fig:Phase-Pf-APf-phase} along the phase boundaries) \cite{imry1975Ma}.
What might be the outcomes of this phase boundary broadening?
\begin{itemize}
\item
One possibility is that the broadening region becomes a new intermediate phase, such as a thermal metal, with unquantized $\kappa_{xy}$, while the split phase boundaries 
(the dotted red lines in
\Fig{fig:Phase-Pf-APf-phase} along the phase boundaries)
become \emph{two} new second order phase transitions.
\item Another possibility is that 
 the percolation transition of the domain walls can be induced within the broadening region. Since at the percolation transition critical point, the domain wall size and correlation length diverge (at least for an infinite-sized system), this induces a new
\emph{single} second order transition within
the broadening.\footnote{In fact, 
our EFT can provide a second order phase transition 
 at the disorder scale $0< \Lambda <\Lambda_1$. In this case, the second order phase transition 
within the range $0< \Lambda <\Lambda_1$ 
can be understood as broadening of the 
first order phase transition at $\Lambda=0$ due to finite disorder. Within the broadening region, 
a new \emph{single} second order transition is induced; a similar statement holds for other second order transitions of our EFT
when $\Lambda >0$; see \Sec{sec:EFT}.}
\end{itemize}
Broadening regions cannot become crossovers between neighboring phases, because the bulk phases have different topological orders and/or global symmetries.
\end{enumerate}
{In fact, Remark \ref{1801.10149},
following the scenario from \Rf{Lian:2018xep}, 
can be regarded as a general scenario that recovers both of the two scenarios from Remarks \ref{1711.06278} and
\ref{1711.11557} in certain limits.
}

\end{enumerate}

The key point for us is that \Rf{Mross:2018MOSMH, Wang:2018WVH, Lian:2018xep} suggested that a $\kappa_{xy}=5/2$ plateau may be induced when 
Pf$\mid$APf domain walls percolate. However, 
\Rf{Mross:2018MOSMH, Wang:2018WVH, Lian:2018xep} 
 have \emph{not}  directly demonstrated that the resulting bulk order is indeed PHPf. Although PHPf has $\kappa_{xy}=5/2$, it remains an open question to show the bulk PHPf induces this $\kappa_{xy}=5/2$.
In this work, we propose a unified effective field theory that can be viewed as a \emph{parent} or \emph{mother} quantum field theory at some higher energy scale\footnote{The energy scale of our EFT is at an intermediate energy scale ($\sim \xi^{-1}$), somewhere 
above the IR low energy topological field theory ($\sim L^{-1}$) 
but below the
inverse magnetic length scale ${\ell_B}^{-1}$ of electrons
or the high-energy lattice cutoff scale $a_{\text{lattice}}^{-1}$ in the far UV. The length scales run from small to large as follows:
$$\hspace{-10mm}
\text{
lattice cutoff $a_{\text{lattice}}$ $<$ magnetic length $\ell_B$ $<$ phase-coherence length $\ell_{\varphi}$ 
 $\lesssim$ $\xi$ 
$<$
sample size $L$.
}
$$ 

 The corresponding energy scales, the inverse of the length scales, 
 run from large to small accordingly. 
The fluctuation length $\xi$ is the length scale of the chemical potential fluctuation due to the  impurity/doping in the system and it is roughly the length scale of disorder $\Lambda^{-1}$.
 
The $\ell_0$ is the puddle linear size which is the link size for the Chalker-Coddington network model \cite{chalker1988}.
The disorder energy $\Lambda = \overline{v}/\ell_0$ is tunable and set by the inverse of the tunable puddle size $\ell_0$ \cite{Lian:2018xep}.
When the length $\ell_0$ is large compared to 
the \emph{domain wall thickness} $\rm{w}$, the domain walls tend to expand and the energetics of the system warrant a more careful analysis \cite{Wang:2018WVH} to determine if the system prefers Pf or APf percolation, instead of domain wall percolation.
See more on energy and length scales in \Sec{Sec:energy-length-scale}. 
We discuss the tension of the domain walls in \Sec{sec:Domain-wall}.
\label{ft:length-scale}}, 
which at low energies can give rise to all the relevant IR TQFT phases listed in \Fig{fig:Phase-Pf-APf-phase}, including
$K=8$, PHPf and 113-state, etc. 

\subsection{Outline}

In the previous subsections, we have summarized several proposed phase diagrams in the literature for the $\nu=5/2$ fractional quantum Hall state. We will focus on reproducing the phase diagram of \cite{Lian:2018xep}, illustrated in \Fig{fig:Phase-Pf-APf-phase}. Our $2+1d$ EFT will also be able, in special limits, to reproduce phase diagrams arising from the other proposals \cite{Mross:2018MOSMH, Wang:2018WVH}, as will become clear in the subsequent sections. The EFT description also reproduces the $1+1d$ domain wall worldvolume theory predicted by \cite{Lian:2018xep}, and it additionally fixes the type of phase transitions at the various phase
boundaries (i.e. first order vs. second order). We also begin a preliminary study of the energetics of our EFT by performing  computation of the domain wall tension, valid in a semiclassical limit, in the relevant phases. The tension of the walls differs in the Pf$\mid$APf and PHPf phases of the theory due to the presence of the chiral Majorana fermions in the former regime. 

We conclude this introduction by summarizing the plan for the rest of this article.

In \Sec{sec:EFT}, we introduce our
effective field theory,  discuss its various IR phases,
and describe in detail how it maps to the phase diagram in \Fig{fig:Phase-Pf-APf-phase}.

In \Sec{sec:quantum-numbers},
we describe the anyon spectra in the various IR phases of our EFT in terms of TQFTs and their quantum numbers, which will be matched to the many-body wavefunctions later (in Appendix \ref{app:many-body-wavefunctions}).

In \Sec{sec:Domain-wall},
we analyze the {domain wall theory and excitations} in some detail. In particular, we study the gapless sectors and evaluate the tension of the walls.
 
In \Sec{sec:conclusion}, we conclude, make final remarks, and 
point out several future directions.

Several appendices contain additional background and some technical details used in the body of the paper. In Appendix \ref{app:gravCS}, we review the relation between the {gravitational Chern-Simons term and the thermal Hall response.}
In  Appendix \ref{app:Z2-gauge}, we describe abelian and 
non-abelian versions of {$\mathbb{Z}_2$ gauge theory in $2+1d$}. In  Appendix \ref{app:FermionPathIntegral}, we clarify some details about the {fermion path integral and counterterms}. In  Appendix \ref{app:gaugingoneform}, we discuss the procedure for {gauging a one-form symmetry in a $2+1d$ TQFT}. In  Appendix \ref{app:O2},
we systematically introduce {O(2)$_{2,L}$ Chern-Simons theories}, their Hall conductance, and other relevant physical properties. In Appendix \ref{app:many-body-wavefunctions}, we review the wavefunction descriptions of the IR TQFTs relevant for our study. In  Appendix \ref{App:One-loop}, we provide some additional details regarding our {one-loop computation} of the domain wall tension.

\section{Effective field theory near the critical filling fraction in $2+1d$}
\label{sec:EFT}

We now present our effective field theory (EFT).

\subsection{Gauge sector, global symmetry, and 't Hooft anomaly}
The 2+1$d$ EFT consists of three sectors:
\begin{itemize}
    \item $[\O(2)]$ gauge field with Chern-Simons (CS) term $\O(2)_{2,1}$ (in the notation of \cite{Cordova:2017vab}).\footnote{%\cred{}
    {Since the discussions here involving several orthogonal groups O($N$) for global or gauge groups, to avoid confusion,
    we may sometimes use the bracket $[\O(2)]$ to specify the gauge group or gauge sectors arising from 
    O(2)$_{2,L}$ CS gauge theory, 
    in contrast with the global symmetries groups (\eg O(2) and O(4)) have no brackets. In general, for a group $G_g$ which is dynamically gauged, we may denote it as $[G_g]$.}}
    \item Two Dirac fermions $\Psi^j$ with flavor index $j \in \{1,2\}$ in the determinant sign representation of the $[\O(2)]$ gauge group. Namely, fermions are odd ($-1$) under the $\det([\O(2)])= \pm 1.$ %\fixme{PH+JW}
    \item A real non-compact scalar $\phi \in \mathbb{R}$ field coupled to the Dirac fermions by a Yukawa term. The $\phi$ also has a Higgs potential.
\end{itemize}
   
To make the connection with fQH, the Chern-Simons gauge field is coupled to a background $\U(1)_\text{EM}$ gauge field.  
We begin by considering a particular mass term for the Dirac fermions and an {\it even} exponent scalar potential so that the EFT preserves particle-hole symmetry and captures the phase transition at the critical filling fraction $\nu_c$.  More general extra deformations including
\begin{itemize}
\item Particle-hole symmetry-breaking potential
(an {\it odd} exponent scalar potential) for $\phi$,
\item Majorana mass terms for four Majorana fermions, 
where each complex Dirac $\Psi^j=\eta_{j1} +\ii \eta_{j2}$ is written as two real Majorana $\eta_{j1}$ and $\eta_{j2}$, with $j \in \{1,2\}$,
\end{itemize}
will be considered in subsequent sections to produce the entire phase diagram of \Fig{fig:Phase-Pf-APf-phase}.

Explicitly, the theory is a 2+1$d$ gauged Gross-Neveu-Yukawa-Higgs theory coupled to a non-abelian Chern-Simons theory
%with an $\frac{\left(\rm{O}(2)\times \rm{O}(2)\right)}{\mathbb{Z}_2}$ global symmetry 
\begin{equation}\label{eqn:GNY}
\begin{aligned}
& \O(2)_{2,1} \;\text{CS}+\sum_{j=1,2} \bar\Psi^j(\ii\slashed{D}_{\cal C}-g\phi)\Psi^j-m(\bar\Psi^1\Psi^1-\bar\Psi^2\Psi^2)
\\
& \hspace{2.5in}
+\frac12 (\partial\phi)^2
+{\mu^2 \over 2}\phi^2-{\lambda \over 4}\phi^4-3\text{CS}_\text{grav}~
,
\end{aligned}
\end{equation}
{where ${\cal C}$ indicates that the fermions are odd 
under the charge conjugation ${\cal C}: \Psi_j \to -\Psi_j$ where
$[\mathbb{Z}_2^{\cal C}] \subset [\O(2)] = [\SO(2) \rtimes \mathbb{Z}_2^{\cal C}]$ 
becomes part of the 
gauge group.}\footnote{%\cred
{Originally there was a fermion parity symmetry 
$\mathbb{Z}_2^F$ where $(-1)^F: \Psi_j \to -\Psi_j$ in the Dirac fermion theory.}
%\cred
{But this $\mathbb{Z}_2^F$ is identified with the charge conjugation
$[\mathbb{Z}_2^{\cal C}]$ in the gauge group $[\O(2)] = [\SO(2) \rtimes \mathbb{Z}_2^{\cal C}]$, and thus
$[\mathbb{Z}_2^F]$ is gauged since $[\mathbb{Z}_2^{\cal C}]$ is gauged in [O(2)]. Beware that
here the gauged $[\mathbb{Z}_2^{\cal C}]$ and $[\mathbb{Z}_2^F]$ are different from the
familiar charge conjugation $C$-symmetry $\mathbb{Z}_2^C$ of the Dirac fermion: 
 where $C: \Psi_j \to \Psi_j^\dagger$, which remains ungauged.
 Readers should be careful to distinguish the ${\cal C}$ and $C$ transformations. 
}}
Since the entire [O(2)] is gauged, the fermions couple to the O(2)$_{2,1}$ Chern-Simons gauge theory by this $\mathbb{Z}_2^C$ gauging (see Appendix \ref{app:O2}).
Each of the complex Dirac fermions ($\Psi^1$ or $\Psi^2$), regarded as 
two real Majorana fermions respectively ($\Psi^j=\eta_{j1} +\ii \eta_{j2}$), enjoys an O(2) global symmetry.
{There is a faithful $\frac{\O(2)\times \O(2)}{\Z_2}$ global symmetry rotating the two Dirac fermions independently that we will explain later below.}

The theory can be obtained by starting from the decoupled semion theory $\U(1)_2=\SO(2)_2$ CS and the Gross-Neveu-Yukawa theory and then gauging the diagonal $\mathbb{Z}_2$ symmetry that acts on $\SO(2)_2$ as the charge conjugation symmetry $[\Z_2^{{\cal C}}]$ and acts on the Gross-Neveu-Yukawa sector as the fermion parity $[\Z_2^F]$.
This changes the gauge group from $[\SO(2)]$ to $[\O(2)]$.
We add a fermionic counterterm for the $\mathbb{Z}_2$ gauge field, which gives the $\mathbb{Z}_2$ level in $\O(2)_{2,1}$, and makes the resulting theory still depend on the spin structure. The spin structure dependence \emph{only} comes from this discrete gauged $\mathbb{Z}_2$ Chern-Simons level.%

Despite the appearance of the Chern-Simons term, the theory in fact has a \emph{particle-hole PH symmetry},
also called the \emph{time-reversal $CT$ 
symmetry}\footnote{The time-reversal symmetry 
in our field theory
language is an anti-unitary symmetry sometimes known as the $CT$ symmetry with its square %\cred
{$(CT)^2=(-1)^f$} giving the fermion parity $\Z_2^f$ 
of the whole theory.  The time-reversal $CT$
 indeed corresponds to the anti-unitary particle-hole (PH) conjugation transformation in the condensed matter literature of $\nu=5/2$ quantum Hall systems. We will see in Appendix \ref{app:O2} that among the $\O(2)_{2,L}$ gauge theories with
 $L \in \Z_8$ classes, only $L=1$ and $L=5$ produce time-reversal invariant theories.
 The  $\O(2)_{2,1}$ gauge theory will be later used for the particle-hole Pfaffian (PH-Pfaffian).
 } (possibly with a 't Hooft anomaly discussed later): $\Psi^i\rightarrow \epsilon^{ij}\gamma^0\Psi^j$ and $\phi\rightarrow-\phi$, so that $\phi$ transforms as a real-valued pseudoscalar.
To see this, we express the theory in (\ref{eqn:GNY}) as
\begin{equation}
{\O(2)_{2,1} \;\text{CS} { +} \left((\mathbb{Z}_2)_0\text{ coupled to Gross-Neveu-Yukawa}\right)\over \mathbb{Z}_2}~,
\end{equation}
where the quotient denotes gauging the $\mathbb{Z}_2$ one-form symmetry generated by the composite line given by the product of the $\O(2)$ Wilson line in the non-trivial one-dimensional representation and the $\mathbb{Z}_2$ electric line.\footnote{
Gauging this one-form symmetry identifies the $\mathbb{Z}_2$ gauge field in $(\mathbb{Z}_2)_0$ with the first Stiefel-Whitney class $w_1$ of the $\O(2)$ gauge field. 
Namely, the $\mathbb{Z}_2$ gauge field in $(\mathbb{Z}_2)_0$ is $w_1(E)$, with $E$ the $\O(2)$ gauge bundle.
The one-form symmetry involved is different from the center one-form symmetry of $\O(2)$.
Note the $\mathbb{Z}_2$ electric line in the $\mathbb{Z}_2$ gauge theory with matter is topological, while the magnetic line is not topological.
}
We briefly review the notion of gauging one-form symmetries in Appendix \ref{app:gaugingoneform}. 
The O(2)$_{2,1}$ Chern-Simons theory is time-reversal invariant by level/rank duality \cite{Hsin2016blu1607.07457, Cordova:2017vab}.
Each of the two theories in the numerator has time-reversal zero-form symmetry and $\mathbb{Z}_2$ one-form symmetry, where the zero-form and one-form symmetries do not have a mixed anomaly (since gauging the one-form symmetry reduces the two theories to SO(2)$_2$ and the Gross-Neveu-Yukawa theory, respectively, both of which are time-reversal invariant).
Therefore, the quotient theory is also time-reversal invariant.

As discussed in Appendix \ref{app:hall-conductivity}, the theory can couple to a background U(1)$_{\rm{EM}}$ electromagnetic gauge field $A$ to have a fractional quantum Hall conductivity $\sigma_{xy}=5/2$
under the U(1)$_{\rm{EM}}$ electromagnetic charge's transverse conductivity measurement. The U(1)$_{\rm{EM}}$ electromagnetic gauge field only couples to the Chern-Simons gauge field, and hence all the phases we discuss have the same Hall conductivity $\sigma_{xy}$.\footnote{
As discussed in Appendix \ref{app:hall-conductivity}, the Hall conductivity only depends on the first Chern-Simons level in $\O(2)_{2,L}$, while integrating out massive fermions in the sign representation only changes the second level $L$ \cite{Cordova:2017vab}.
}

We will consider phases with $\mu^2>0$, which implies that the real (pseudo-)scalar field $\phi$ condenses with a vacuum expectation value (vev):
\begin{equation}
\langle\phi\rangle=\pm v,\quad v\sim \mu/\sqrt{\lambda}>0~. 
\end{equation}
This spontaneously breaks the time-reversal $CT$ symmetry, and there can be two symmetry-breaking vacua exchanged by the (broken) symmetry 
transformation in the 2+1$d$ bulk.
The spontaneously broken time-reversal symmetry  $CT$ leads to a 1+1$d$ domain wall that interpolates between the two vacua.
We will investigate the domain walls in Section \ref{sec:Domain-wall}.

Let us elaborate more on the global symmetries and gauge group in \eq{eqn:GNY}:
\begin{itemize}
\item Continuous global symmetries ---

If we turn off the mass deformation $m=0$, then the theory has an enlarged $\O(4)/\Z_2 \equiv {\rm PO}(4)$ symmetry, where the four Majorana components from the two Dirac fermions transform in a vector representation of $\O(4)$. 
Let us explain why we mod out by a $\Z_2$ subgroup of the n\"aive rotational O(4) global symmetry. 
The $\Z_2$ center of $\O(4)$ (in fact the same as
$\Z_2^F \subset \SO(4) \subset \O(4)$, where
$\Z_2^F$ sends $\Psi_j \to -\Psi_j$)
is identified with the charge conjugation element $[\Z_2^{\cal{C}}]$ of the $[\O(2)]$ gauge group. 

If we turn on $m \neq 0$, there is a faithful $\frac{\left(\rm{O}(2)\times \rm{O}(2)\right)}{\mathbb{Z}_2}$ global symmetry in \eq{eqn:GNY}.
We mod out by a $\mathbb{Z}_2$ subgroup of the n\"aive $\rm{O}(2)\times \rm{O}(2)$ symmetry
because the diagonal $\mathbb{Z}_2$ center of $\rm{O}(2)\times \rm{O}(2)$
(in fact the same as
$\Z_2^F \subset \rm{SO}(2)\times \rm{SO}(2) \subset \rm{O}(2)\times \rm{O}(2)$, where
$\Z_2^F$ sends $\Psi_j \to -\Psi_j$)
is identified with the charge conjugation element of $[\Z_2^{\cal{C}}]$ of $[\O(2)]$ gauge group.

%\footnote{More precisely the faithful symmetry on local operators is $\frac{\left(\rm{O}(2)\times \rm{O}(2)\right)}{\mathbb{Z}_2}\subset {\rm PO}(4)$. In the following discussion we will ignore such distinctions.}

By contrast, if we allow the four Majorana fermions to all have different masses, then there is no continuous global symmetry (because the
fermion parity $\Z_2^F$ that flips the sign of the fermions is identified with a gauge rotation
$[\Z_2^{\cal{C}}]$). The different mass deformations considered in the subsequent sections can be organized by the breaking pattern of $\rm{PO}(4)$. 
%\cred
{The continuous global symmetry $\text{PO}(4)$ that transforms the fermions has the standard mixed anomaly with the time-reversal $CT$ symmetry given by the 3+1d $\theta$ term for a $\text{PO}(4)$ background gauge field with $\theta=(\pi,\pi)$.\footnote{
The gauge Lie algebra of PO(4) is $\mathfrak{su}(2)\times \mathfrak{su}(2)$, hence the two $\theta$ angles.
}}

\item Discrete global symmetries  ---
\begin{enumerate}
\item $\Z_2^f$ fermion parity symmetry: This $\Z_2^f$ should not to be confused with the already gauged $[\Z_2^F]$ (acting by $\Psi_j \to  -\Psi_j$ only in the Dirac fermion sectors). 
The $[\Z_2^F]$ and $[\Z_2^{\cal C}]$ charge conjugation
are identified and both dynamically gauged due to $\slashed{D}_{\cal C}$.
(Neither $\Psi_1$ nor $\Psi_2$ are gauge-invariant local fermionic operators).
In fact, the $\Z_2^f$ acts \emph{not}
on Gross-Neveu-Yukawa sector, but \emph{only} on the
O$(2)_{2,1}$ CS and $-3\text{CS}_\text{grav}$.
Note that \eq{eqn:GNY} is an intrinsically fermionic theory (defined on spin manifolds) because both O$(2)_{2,1}$ CS and $-3\text{CS}_\text{grav}$ are spin Chern-Simons actions  whose UV completion, say on a lattice, requires some gauge-invariant local fermionic operators. 

\item $\Z_4^{CT}$-symmetry: This is the particle-hole (PH) symmetry, also known as the $CT$-symmetry. This is an anti-unitary symmetry. 
 Its normal subgroup is the $\Z_2^f$ fermion parity, since $(CT)^2=(-1)^f$. 
As mentioned, $\Psi^i(t,x)\rightarrow \epsilon^{ij}\gamma^0\Psi^j(-t,x)$ and $\phi(t,x)\rightarrow-\phi(-t,x)$.\footnote{There are also other discrete charge $C$ and parity $P$ symmetries for our EFT as a Lorentz invariant QFT, which should be 
familiar to the readers. There are also other time-reversal symmetries given by composing 
this anti-unitary symmetry with additional $\mathbb{Z}_2$ subgroup unitary symmetries.} We will further explain how the $CT$ acts on the CS theories in TQFT sectors in \Sec{sec:quantum-numbers}. 
The 't Hooft anomaly of the $CT$ symmetry can be derived by adding a large time-reversal preserving mass $m$ to the fermions and studying the anomaly in the infrared PH-Pfaffian theory $\O(2)_{2,1}$ ---  our theory can have the 
$\upnu \in \Z_{16}$ anomaly of $CT$ symmetry with
$\upnu=0$ for PH-Pfaffian$_{+}$
and $\upnu=8$  for PH-Pfaffian$_{-}$
\cite{Metlitski:2014xqa, Metlitski1510.05663, WangLevin1610.04624, Tachikawa:2016cha,Cordova:2017kue}.\footnote{There 
are two versions PH-Pfaffian$_{\pm}$ depending on how $(CT)^2$ assigns to
the odd $q=1,3,5,7$ anyons of $\U(1)_8$, see 
\cite{Metlitski:2014xqa, Metlitski1510.05663, WangLevin1610.04624}
and \Sec{sec:quantum-numbers}.
The two choices are related by shifting the background gauge field for the $\mathbb{Z}_2$ subgroup one-form symmetry (generated by a fermion line) by $w_1^2$, see also \cite{Cordova:2017kue}. 
Readers should beware that 
we use $\upnu$ as topological classification index while $\nu$ as filling fraction.
} %\fixme{/JW.PH}

\item $\mathbb{Z}_4$ one-form global symmetry \cite{Gaiotto:2014kfa} from the $\O(2)$ Chern-Simons theory: The $\mathbb{Z}_2$ subgroup is generated by the $\O(2)$ Wilson line in the determinant sign representation.
The $\mathbb{Z}_4$ one-form symmetry has an 't Hooft anomaly, characterized by the spin $\frac{1}{4}$ statistics of the generating line \cite{Hsin:2018vcg}. %\fixme{/PH} 
The 't Hooft anomaly of the one-form symmetry is related to the fractional part of the Hall conductivity; see Appendix \ref{app:hall-conductivity}
We will not focus on this global symmetry in this work.

\item $\mathbb{Z}_2$ magnetic 0-form symmetry of the $\O(2)$ gauge field \cite{Cordova:2017vab}. We will not discuss this symmetry in this work.

\end{enumerate}
\item Gauge sector\footnote{{Readers may be curious about 
how the semi-direct product gauge structure of
$[\O(2)] = [\SO(2) \rtimes \mathbb{Z}_2^{\cal C}]$ in $\O(2)_{2,L}$
can be related to the direct product
gauge structure of
$[\U(1)]$ and $[\mathbb{Z}_2]$ 
in $\frac{\U(1)_8\times T_{L}}{\mathbb{Z}_2}$ CS with 
$T_{L}$ as some $L\in \Z_8$ class of
$[\mathbb{Z}_2]$ gauge theory.
The answer is that there is a duality between the two gauge theories
at the level 2 of $\O(2)_{2,L}$,
see \cite{Cordova:2017vab} and Appendix \ref{app:O2}. 
} %
}  ---
\begin{multline}
\O(2)_{2,1}\;\text{CS}\cong \frac{\U(1)_8\times T_{L=1}}{\mathbb{Z}_2}\;\text{CS}
\cong
\frac{\U(1)_8\times (\overline{\text{Ising}}\times \text{(spin-Ising)}
)}{\mathbb{Z}_2}
\;\text{CS}
\\
\cong
\frac{\U(1)_8\times ({\frac{\SU(2)_{-2}\times \U(1)_{4}}{\Z_2}}\times (\SO(3)_{1}\times \U(1)_{-1})
)}{\mathbb{Z}_2}
\;\text{CS}.
\end{multline}
The second line rewrites Ising and spin-Ising TQFTs as CS theories.\footnote{The Ising TQFT can be expressed as a non-abelian CS theory with a gauge group 
$\U(2)_{2,-4}\cong (\SU(2)_2\times \U(1)_{-4})/\Z_2$
from \cite{1602.04251SW}. 
By $\SO(3)_1$, we denoted the spin-CS theory with 
the level normalized such that the states on a 2-torus 
$T^2$ are subset of $\SU(2)_2$ states corresponding to odd
$\SU(2)$ representations ($\mathbf{1}$ and $\mathbf{3}$).
The spin-Ising TQFT is given by the $(\SO(3)_{1}\times \U(1)_{-1})$ CS.
}
For background information on this sector, see 
\Sec{app:O2} around \eq{eq:O2LtoU18TLmodZ2}.
\end{itemize}

In the following subsections, we discuss several deformations of our theory (\ref{eqn:GNY}):
\begin{enumerate}
\item In \Sec{Sec:PreservingDeformation}, 
we consider a $CT$-preserving mass deformation, but the $CT$-symmetry turns out to be spontaneously broken.  The
deformation explicitly breaks $\rm{PO}(4)=\O(4)/\Z_2$ down to $(\O(2) \times \O(2))/\Z_2$.
\item In \Sec{sec:CT-breaking}, we add an odd polynomial potential in $\phi$ to our action, which explicitly breaks 
$CT$-symmetry, but preserves the $(\O(2) \times \O(2))/\Z_2$ symmetry (or $\PO(4)$ if $m = 0$).
\item In \Sec{sec:moreK=8and113}, we add additional Majorana mass terms that break the entire $\PO(4)$ symmetry, but preserve the $CT$-symmetry. %\fixme{JW}
\end{enumerate}

\subsection{Physical arguments supporting the EFT}

Here we provide some arguments and intuition to support our EFT from simple physical considerations, following the setup in  \Rf{Lian:2018xep}, see \Fig{fig:BdG}.

\subsubsection{From gapless or gapped Fermi surfaces to four gapless Majorana nodes}
The first question to ask about our EFT \eq{eqn:GNY} is: why do we introduce
two Dirac fermions (or four Majorana fermions)? This can be understood from solving the 
Bogoliubov-de Gennes (BdG) equation 
\cite{Lian:2018xep}, which allows us to analyze the gap function $\Delta({\vec{k}, \vec{r}})$ around the domain wall between
the Pf and APf region. We remind ourselves that the Pf and APf
(and other related topological orders) can be obtained by
the superconductivity (SC) pairing of composite fermions (CF) in the 
Halperin-Lee-Read (HLR) theory \cite{halperin1993} 
or the superconductivity pairing of composite Dirac fermions (CDF) in Son's theory:
\bea \label{eq:pairing}
\begin{array}{rcc}
    & \text{{CF (HLR)}} &  \text{{CDF (Son)}}\\[2mm]
\text{Pf}: & p\text{-wave},& d\text{-wave}.\\[2mm]
\text{$K=8$ state}:& s\text{-wave}, & p\text{-wave}.\\[2mm]
\text{PHPf}:& p^*\text{-wave}, & s\text{-wave}.\\[2mm]
\text{$113$ state}:& d^*\text{-wave}, & p^*\text{-wave}.\\[2mm]
\text{APf}:& f^*\text{-wave} & d^*\text{-wave}.
\end{array}
\eea
\begin{figure}[!t]
\centering
(a) 
 \includegraphics[width=.40\textwidth]{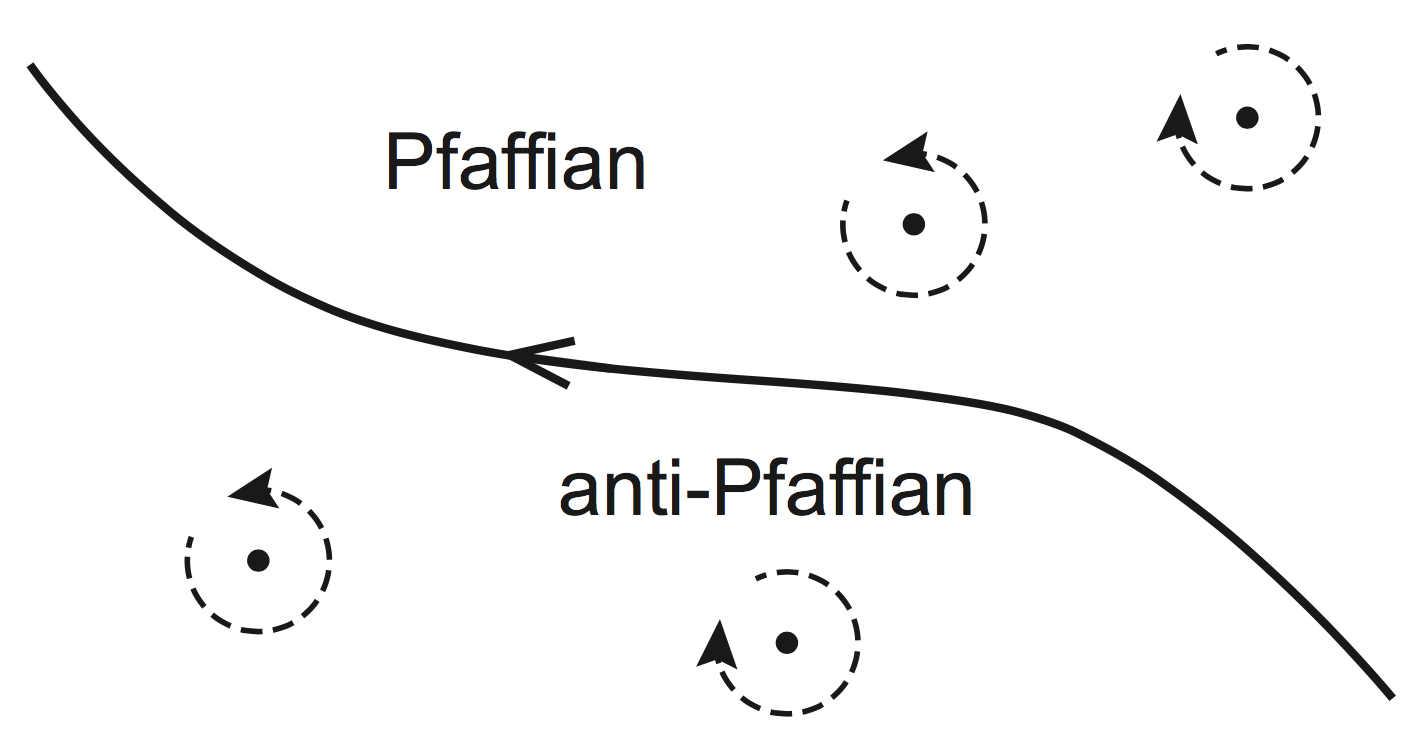}
\;\;\;
(b)\includegraphics[width=.40\textwidth]{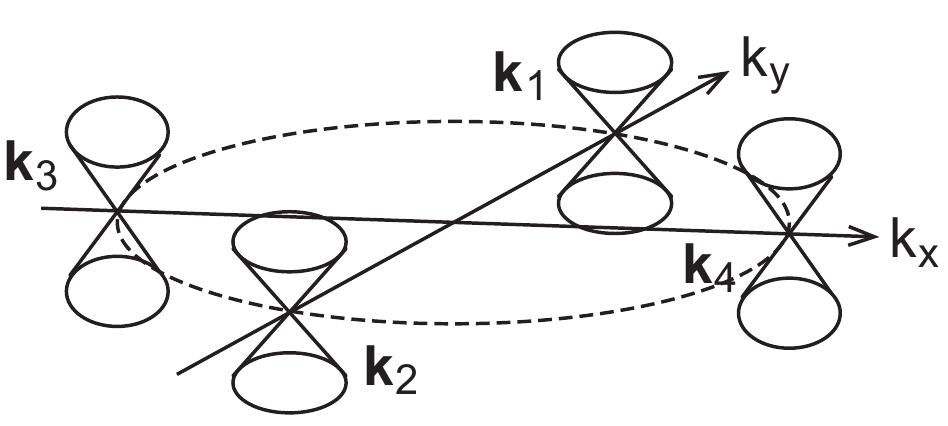}
\caption{Follow the setup in \Rf{Lian:2018xep}: 
(a) By solving the BdG equation in the 
mean-field, single-particle, semi-classical quantum mechanical manner
\cite{Lian:2018xep}, we can analyze the gap function $\Delta({\vec{k}, \vec{r}})$ around the domain wall between
the Pf and APf region. There are also $\pm \pi$ vortices shown in dashed circles.
(b)  Four Dirac nodes solved from
the BdG equation on Pf$\mid$APf domain wall 
in a momentum $\vec{k}$-space. Due to BdG Nambu space double counting, we only have physical degrees of freedom
of two Dirac nodes or four Majorana nodes. The four Majorana nodes are precisely the fermion nodes that we need in our EFT that we study in the real space. 
 }
 \label{fig:BdG}
\end{figure}
The HLR and Son theories describe \emph{gapless} theories with 
infinitely many gapless modes along
a continuous Fermi surface 
(more precisely, a $1d$ Fermi circle for a $2+1d$ theory). However, we can gap the Fermi surface and go to a \emph{gapped} theory
by introducing the superconductivity pairing to CF or CDF as \eq{eq:pairing} above.
Here we present the pairing gap function $\Delta(\vec{k}) \propto (k_x + \ii k_y)^{L_z}$
to obtain the five topological orders in the phase diagram \Fig{fig:Phase-Pf-APf-phase},
where $s,p,d,f$-wave pairing has
the $z$-directional angular momentum $L_z=0,1,2,3$ respectively, while the complex conjugate
$p^*,d^*,f^*$-wave pairing has the opposite sign of the angular momentum.

We can find that in the CF (HLR's) picture, the pairing function is
$\Delta({\vec{k}, \vec{r}})=\Delta_{\text{Pf}}(\mathbf{r})e^{\ii \theta_k} - \Delta_{\text{APf}}(\mathbf{r})e^{-\ii 3\theta_k}
\propto |\Delta| e^{-\ii\theta_k} \sin({2 \theta_k})
$;
while in the CDF (Son's) picture, 
the pairing function is
$\Delta({\vec{k}, \vec{r}})=\Delta_{\text{Pf}}(\mathbf{r})e^{\ii 2\theta_k}-\Delta_{\text{APf}}(\mathbf{r})e^{-\ii 2\theta_k}
\propto |\Delta|  \sin({2 \theta_k})$.\footnote{In general 
$\vec{k} = -i \frac{\partial}{\partial \vec{r}}$ is 
a differential operator in a disordered system, since 
$[\vec{k} = -i \frac{\partial}{\partial \vec{r}},H(\vec{k}, \vec{r})]\neq 0$ and $\vec{k}$ is not a good quantum number globally. See the detailed analysis in \cite{Lian:2018xep}.}
In either case,
the  $\sin({2 \theta_k})$ gives four gapless nodes around the otherwise 
fully gapped Fermi surface at $\theta_k=0,\pi/2,\pi,3\pi/2$, 
when we are spatially near the Pf$\mid$APf domain wall.
Thus, \Rf{Lian:2018xep} finds four Dirac nodes solved from
the BdG equation on Pf$\mid$APf domain wall 
in a momentum $\vec{k}$-space. Due to BdG Nambu space double counting degrees of freedom
at $\vec{k}$ and $-\vec{k}$, we only have physical degrees of freedom
of two Dirac nodes or equivalently four Majorana nodes. The four Majorana nodes explain precisely the origin of the fermions that we need in our EFT \eq{eqn:GNY} that we study in the real space. 
Furthermore, it will become clear in the later subsections 
how the additional gauge theory sectors and deformations can help to span the full phase diagram \Fig{fig:BdG}. This physical picture helps to motivate our EFT.

{Our EFT, including the deformations, can describe both the gapped TQFT phases
and gapless topological quantum phase transitions predicted in the phase diagram \Fig{fig:Phase-Pf-APf-phase}, similarly to the percolating phases and transitions in \Rf{Lian:2018xep}. For condensed matter purposes, we remark that 
the $2+1d$ gapless phases in our Lorentz invariant EFT have four interacting Majorana fermions ($2+1d$ Majorana cones in a 
momentum $\vec{k}$-space, 
in the non-interacting band theory limit in \Fig{fig:BdG} (b))
but without a Fermi surface 
--- the Fermi surface is gapped and left with only isolated gapless nodes.
In other words, we emphasize that 
the $2+1d$ gapless phase transitions in our EFT are similar to that
of a \emph{semi-metallic} phase transition with isolated gapless nodes,
instead of a \emph{metallic} phase transition with a gapless continuous Fermi surface.}

\subsubsection{More on energy and length scales, and emergent symmetries}
\label{Sec:energy-length-scale}

Before we dive into the detailed phase diagrams of our EFT in the next subsections,
we first summarize what we expect from the story in \Rf{Lian:2018xep}, about the 
energy and length scales (see also footnote \ref{ft:length-scale}), 
and emergent symmetries of the system.

We take the limit of no Landau-level mixing (LLM),
so the energy gap between Landau levels from the
cyclotron frequency 
$\omega_c = B/m \gg \Lambda_2 = e^2/\epsilon\ell_B \sim \sqrt{B}$
is assumed to be much larger than the Coulomb energy scale, which we set to be $\Lambda_2$, the disorder energy scale in \Fig{fig:Phase-Pf-APf-phase}.
The $\ell_B=\sqrt{\hbar c/eB}$ is the magnetic length scale.
There is yet another length scale set by the domain wall width w,
which is microscopically related to the phase-coherence length $\ell_{\varphi}$
of superconducting pairing fluctuations 
in the composite fermion picture of \eq{eq:pairing}.

\Rf{Lian:2018xep} analyzes the relation between energy/length scales and emergent symmetries of the $1+1d$ domain wall, showing that:
\begin{itemize}

\item The disorder energy $\Lambda \sim 1/\ell$ (the vertical axis of \Fig{fig:Phase-Pf-APf-phase})
is  related to the domain length scale $\ell$ which 
controls the Pf and APf puddle sizes.

\item The energy scale  $\Lambda_2$ is defined by
the inverse of magnetic length scale $\ell_B=\sqrt{\hbar c/eB}$. 
The $\Lambda_2$ is around the Fermi energy of the composite fermion.

\item
The energy scale $\Lambda_1$ is defined by the inverse of  
the correlation length $\ell_{\varphi}$ of superconducting 
phase pairing fluctuations 
in the composite fermion picture of Pf and APf,
which is also related to 
the inverse of the domain wall width w.

\item
When the disorder energy scale $\Lambda < \Lambda_1$,
we have weaker disorder and hence larger Pf and APf puddle sizes, so
the domain length scale $\ell$ is large. For large $\ell$, the 
$1+1d$ four Majorana edge modes running on the domain wall can be mixed together 
via scattering along the domain wall, 
which induces an emergent O(4) symmetry and a 
uniform velocity.\footnote{\Rf{Lian:2018xep} also 
uses a BKT-type perturbative analysis to show that,
regardless of spatial fluctuations (from impurity) or temporal fluctuations
(from the SC pairing phase), the velocity fluctuation correlation function 
$\langle\delta v_I(x)\delta v_I(x')\rangle=W_v\delta(x-x')$
has an irrelevant perturbation
driven by the phase fluctuation. This means the 
the $\langle\delta v_I(x)\delta v_I(x')\rangle$ flows to zero.
Therefore, with weak disorder $\Lambda < \Lambda_1$,
either at zero temperature or some small finite temperature
{(the experiment is performed around $10 \sim 30$ milli Kelvin (mK) \cite{Banerjee:2018qtz}),
} we
have an emergent O(4) symmetry.
}

\item When the disorder energy $\Lambda$ scale sits at $\Lambda_2 > \Lambda > \Lambda_1$,
then $\Lambda$ is below the Coulomb energy $e^2/\epsilon\ell_B$ and the Fermi energy $\bar{v} k_F$ 
set by $\Lambda_2$.
The $\Lambda$ is also below some factor of the magnitude of SC gap size $|\Delta|$. This implies that, from \Fig{fig:BdG} (b), the two physical 
Dirac nodes solved from BdG, have internal symmetries $\O(2)\times \O(2)$,
where each $\O(2)$ rotates the two Majorana nodes of a give Dirac node.

\item
When the disorder energy scale $\Lambda > \Lambda_2$,
we have stronger disorder  $\Lambda \sim 1/\ell$, hence smaller Pf and APf puddle sizes, so
the domain length scale $\ell$ is smaller. For smaller $\ell$, it is difficult to mix the 
$1+1d$ four Majorana edge modes running  on the domain wall, so we expect only 
the fermion parity symmetry when $\Lambda > \Lambda_2$.
The $\O(2)\times \O(2)$ symmetry will be broken when  $\Lambda > \Lambda_2$, because
the disorder is strong enough to exceed the gap size $|\Delta|$ or even the Fermi energy $\bar{v} k_F$,
so the Dirac nodes in \Fig{fig:BdG} (b) fluctuate and their dispersion and 
energy spectra 
can overlap with each other.  Because of this, we can no longer make sense of the two internal 
rotational symmetries.

\end{itemize}

{To summarize all the length scales, we have
\begin{multline}
\text{
lattice cutoff $a_{\text{lattice}}$ $<$ magnetic length $\ell_B$
$<$ 
domain wall width w 
  }\cr
\text{
 $\simeq$ phase-coherence length $\ell_{\varphi}$ 
$\lesssim$ \text{our EFT typical length scale} $\xi$
$<$
sample size $L$.
}
\end{multline}
}
The energy scales are given by the inverse of length scales:
\bea
a_{\text{lattice}}^{-1} > \Lambda_2= \ell_B^{-1}
 > \Lambda_1=\ell_{\varphi}^{-1} 
 \simeq {\text{w}}^{-1} 
 \gtrsim  \xi^{-1}
>
L^{-1}.
\eea

\subsection{Particle-hole (time-reversal $CT$)-preserving deformation}
\label{Sec:PreservingDeformation}

Let us turn on the deformation $m$ in \eq{eqn:GNY}. 

\paragraph{${0 \leq m < gv:}$} When $m$ increases from zero to $gv$, the theory has the following phases in the two vacua:
\begin{itemize}
\item At the vacuum $\langle\phi\rangle=-v$, $\Psi^1$ has mass $m-gv$, while $\Psi^2$ has mass $-m-gv$.
For $m$ below $gv$, at low energies we can integrate out both negative mass Dirac fermions, and the theory becomes the gapped TQFT 
\begin{equation} \label{eq:PfaffianO2}
\text{Pfaffian}:\quad \O(2)_{2,-1} \;\text{CS} -5\text{CS}_\text{grav}~.
\end{equation}
The O(2)$_{2,L}$ Chern-Simons gauge theory (see Appendix \ref{app:O2}) contains a $\U(1)_8$
Chern-Simons theory that contributes a net chiral central charge $c_-=c_L-c_R=1$, while the matter sectors do not contribute any 
net chiral central charge.
The theory has Hall conductivity $\sigma_{xy}$ and thermal Hall conductivity $\kappa_{xy}$ matching those of the Pfaffian state:\footnote{The spin gravitational Chern-Simons term
has chiral edge modes contributing to the thermal Hall conductivity $\kappa_{xy}$ by a chiral central charge $c_-=-1/2$; see Appendix
\ref{app:gravCS}.
}
$$
\sigma_{xy}=5/2,\quad \kappa_{xy}=1+5/2=7/2.
$$

\item At the vacuum $\langle\phi\rangle=+ v$,
$\Psi^1$ has mass $m+gv$,
 while $\Psi^2$ has mass $-m+gv$.
For $m$ below $gv$, then at low energies we can integrate out the two positive mass Dirac fermions, and the theory becomes the gapped TQFT 
\begin{equation} \label{eq:antiPfaffianO2}
\text{anti-Pfaffian}:\quad \O(2)_{2,3} \;\text{CS} -\text{CS}_\text{grav}~.
\end{equation}
The theory has Hall conductivity $\sigma_{xy}$ and thermal Hall conductivity $\kappa_{xy}$
$$
\sigma_{xy}=5/2,\quad \kappa_{xy}=1+1/2=3/2.
$$

The TQFT becomes, under level/rank duality \cite{Hsin2016blu1607.07457, Cordova:2017vab}, $\O(2)_{2,3}\leftrightarrow \O(2)_{-2,1}$ up to
$4\text{CS}_\text{grav}$, and thus it is the time-reversal image of the Pfaffian theory.
\end{itemize}

\begin{figure}[t]
  \centering
    \includegraphics[width=1\textwidth]{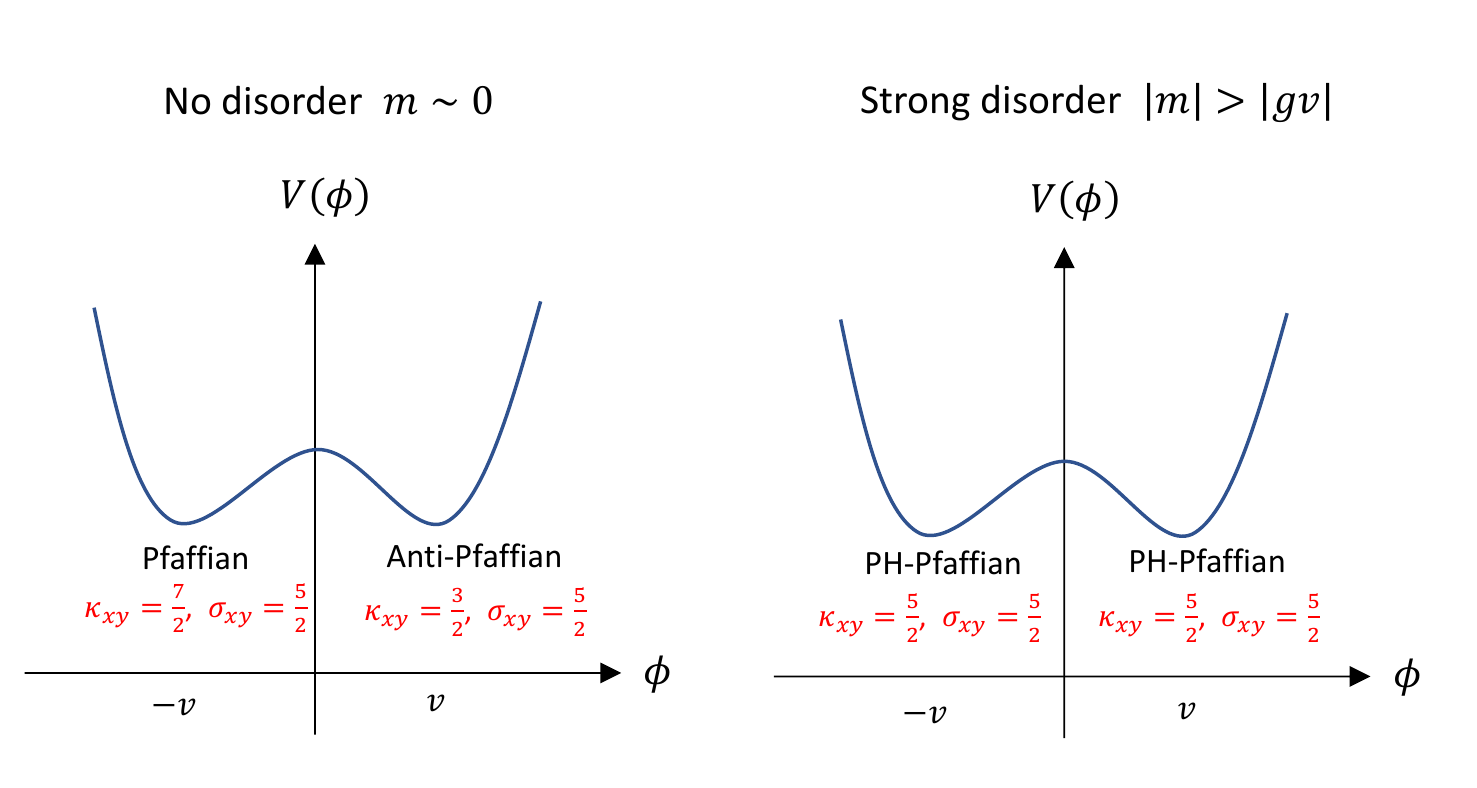}
      \caption{The theories comprising the vacua of the system (\ref{eqn:GNY}) depend on the deformation $m$.}\label{fig:vacua}
\end{figure}

The two different regimes capturing our time-reversal-symmetric deformations are depicted in Figure \ref{fig:vacua}.

\paragraph{${ m = gv:}$}
 When $m=gv=g|\langle\phi\rangle|$, one of the Dirac fermions becomes massless, and the theories are\footnote{For $m=gv$, when $\langle\phi\rangle=-v$, the fermion $\Psi^1$ becomes massless; 
  when $\langle\phi\rangle=+v$,
  $\Psi^2$ becomes massless.}
\begin{align}
&\langle\phi\rangle=-v:\quad
\O(2)_{2,0}\;\text{CS}+\Psi^1 \text{ in }\mathbf{1}_\text{odd}-4\text{CS}_\text{grav}~.
\cr
&\langle\phi\rangle=+v:\quad
\O(2)_{2,2}\;\text{CS}+\Psi^2 \text{ in }\mathbf{1}_\text{odd}-2\text{CS}_\text{grav}~.\label{eq:massless-critical}
\end{align}

\paragraph{${m > gv:}$} When $m>gv$, the two Dirac fermions acquire masses of {\it opposite} signs, and the two vacua become the \emph{same} gapped TQFT
\begin{equation}  \label{eq:PHPfaffianO2}
\text{PH-Pfaffian}:\quad \O(2)_{2,1}\;\text{CS}-3\text{CS}_\text{grav}~.
\end{equation}
The theory has the Hall conductivity $\sigma_{xy}$ and thermal Hall conductivity $\kappa_{xy}$
$$
\sigma_{xy}=5/2,\quad \kappa_{xy}=1+3/2=5/2.
$$

With $m$ treated as a proxy for disorder strength (the precise relation is discussed in Section \ref{sec:disorderm}),
the gapped phases in the above discussion are precisely those that appear in the scenario of \cite{Wang:2018WVH,Lian:2018xep}:
for small disorder strength, the microscopic theory is at a first order-like phase transition with coexisting Pfaffian and anti-Pfaffian phases, while increasing the disorder strength produces the PH-Pfaffian phase.
From the above discussion, it is thus natural to identify the parameter $m$ (or its magnitude $|m|$) in the effective phenomenological theory with the disorder strength in the microscopic material.

We remark that the first-order phase transition with distinct gapped vacua persists for a range of the parameter $m\in [0,gv)$, which is consistent with the phase diagram proposed in \cite{Wang:2018WVH,Lian:2018xep}. We may identify $\Lambda_1$ in \cite{Wang:2018WVH,Lian:2018xep} with $gv$, with $v$ controlled by the scalar mass $\mu$ in the effective theory. When $gv$ is small, the phase diagram approaches that described in \cite{Mross:2018MOSMH,Wang:2018WVH}.

\subsection{Particle-hole (time-reversal $CT $)-breaking deformation}
\label{sec:CT-breaking}

In this section, we investigate the effect of adding a time-reversal-breaking deformation that preserves the $\frac{\O(2) \times \O(2)}{\Z_2}$ symmetry ($\frac{\O(4)}{\Z_2}$ when $m = 0$). In the experiment, this corresponds to applying an additional time-reversal-breaking magnetic field that changes the filling fraction $\nu$ slightly.
In this discussion, we set the Yukawa coupling to $g=1$ for simplicity.

Since $\phi$ is a time-reversal-odd pseudoscalar field, we consider the simple time-reversal-breaking deformation given by an odd polynomial of $\phi$.
The most relevant deformation will be $\delta V(\phi) \propto -m_{odd}\,\phi$.
This modifies the scalar potential and lifts the degenerate vacua.

\begin{figure}[!t]
  \centering
    \includegraphics[width=0.7\textwidth]{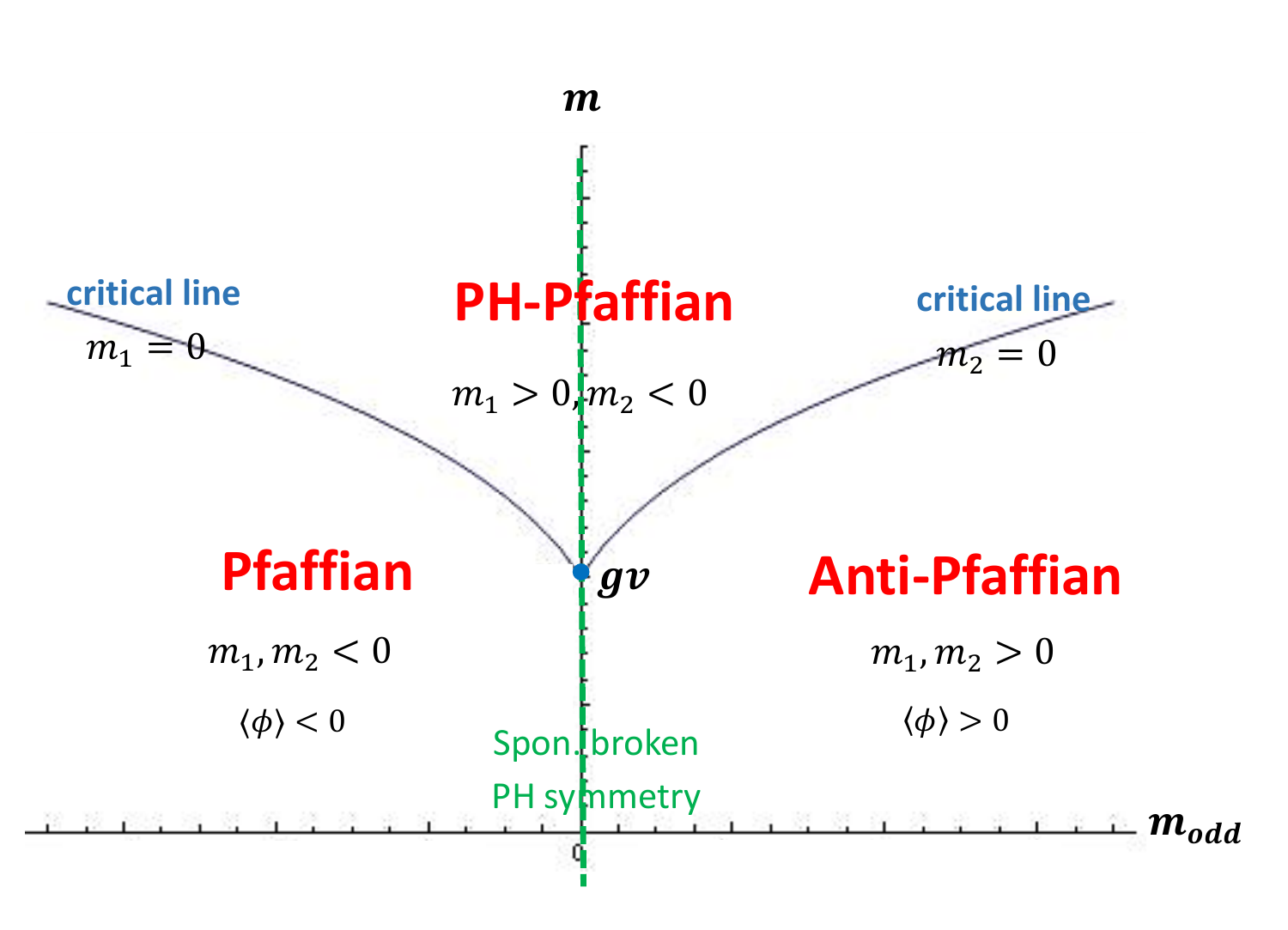}
      \caption{Phase diagram of the theory (\ref{eqn:GNY}) deformed by the PH-symmetry breaking scalar potential $\delta V(\phi) = -m_{odd} \, \phi$.
      Phase boundaries are plotted \emph{analytically} by Mathematica. 
      Here the parameter $m$ is from the Dirac mass term
      in (\ref{eqn:GNY}). The
      $m_1$ and $m_2$ are the induced masses in the IR for the ground
      states: 
     $m_1=m+g\langle\phi\rangle$, $m_2=-m+g\langle\phi\rangle$.
      The blue lines are critical lines where one of the Dirac fermion becomes massless, joined by blue dots where both fermions are massless at a critical 
      $m=g v = g |\langle \phi \rangle|$.
The green  
line in the middle represents a first order phase transition with spontaneously broken time-reversal symmetry ($i.e.$, anti-unitary particle-hole symmetry) that gives rise to domain wall excitations that interpolate between the Pfaffian and anti-Pfaffian phases.      
The phase diagram is in qualitative agreement with the schematic phase diagram discussed in \cite{Wang:2018WVH,Lian:2018xep} with the time-reversal breaking deformation $m_{odd}$ identified with the external magnetic field in the experiment and the time-reversal preserving deformation $|m|$ identified with the microscopic disorder strength. In experiment, it is so far undetermined whether Pfaffian or anti-Pfaffian
is favored at $\nu<\nu_c\simeq 5/2$ (or 
$\nu>\nu_c\simeq 5/2$); we can easily flip our 
phase diagram by
defining the sign of $m_{odd}$ to match the
tuning parameter for the filling fraction $\nu$.
}\label{fig:GNYphaseO(2)}
\end{figure}

In the lowest-order approximation, we can take the effect to be such that the original vacua shift to the locations $\langle\phi\rangle=v+m_{odd}+{\cal O}(m_{odd}^2)$ and $-v+m_{odd}+{\cal O}(m_{odd}^2)$.
Depending on the sign of $m_{odd}$, one of the above is the true vacuum: for $m_{odd}>0$ it is the former and for $m_{odd}<0$ it is the latter. In other words, the
true vacuum has a vev:
$$\langle \phi \rangle= v\text{ sgn}(m_{odd})+m_{odd}+{\cal O}(m_{odd}^2)
=(v+|m_{odd}|) \text{ sgn}(m_{odd})+{\cal O}(m_{odd}^2)
.$$ The two Dirac fermions then have masses given by (with $g=1$) 
\begin{align}
&
m_1=m+\langle\phi\rangle=m+m_{odd}+ v\text{ sgn}(m_{odd})+{\cal O}(m_{odd}^2)~, \nn\\
&m_2=-m+\langle\phi\rangle=-m+m_{odd}+ v\text{ sgn}(m_{odd})+{\cal O}(m_{odd}^2)~. \label{eq:m1-m2-mass}
\end{align}
There are critical lines when any of the fermions become massless.\footnote{The situation is similar to
\eq{eq:massless-critical}. 
One might worry that the critical line can receive a quantum correction; however, since the scalar field has a mass of the order
$m_\phi \propto {\mu}$ around the vacuum, in the vicinity of the critical line with distance less than $m_\phi$ there is a light fermion.}

The phase diagram whose coordinates are our two parameters
$(m,m_{odd})$ for a fixed $v$ is given in Figure \ref{fig:GNYphaseO(2)}.
Given by the mass deformation formula (\ref{eq:m1-m2-mass}), the left critical line in Figure \ref{fig:GNYphaseO(2)} has $m_2 \simeq 0$, while the right critical line in Figure \ref{fig:GNYphaseO(2)} has $m_1 \simeq 0$.
It is in qualitative agreement with the schematic phase diagram discussed in \cite{Wang:2018WVH,Lian:2018xep}, and suggests that the corresponding Pf$\vert$PHPF and APf$\vert$PHPf phase boundaries are given by second-order phase transitions.

\subsection{$K=8$ and $113$ states from $\frac{\mathrm{O}(2) \times \mathrm{O}(2)}{\Z_2}$-breaking masses}
\label{sec:moreK=8and113}

In our earlier discussion, we mainly focused on mass deformations preserving the $\frac{\mathrm{O}(2) \times \mathrm{O}(2)}{\Z_2}$ symmetry that transforms the two Dirac fermions. If we allow Majorana masses that break the $\frac{\mathrm{O}(2) \times \mathrm{O}(2)}{\Z_2}$ symmetry, the effective theory can also describe the $K=8$ state and the 113 state (the two states are related to one another by the particle-hole $CT$ symmetry).
Denote the four Majorana fermions by $\eta_{ia}$ where $i=1,2$ labels the Dirac fermions and $a=1,2$ labels the Majorana components. Consider the Majorana mass deformation:\footnote{We take $(\gamma^0)_{\alpha \beta}=({}\ii \sigma^y)_{\alpha \beta}={}\epsilon_{\alpha \beta}$ with the spinor indices ${\alpha, \beta}$.
Note that the Dirac mass term $\bar\Psi^j \Psi^j = \Psi^{j\dagger} \gamma^0 \Psi^j$. In the Majorana basis, we write 
$\Psi^j=\eta_{j1} +\ii \eta_{j2}$, and
$\bar\Psi^j \Psi^j =
\epsilon_{\alpha\beta}(\eta_{j1,\alpha}\eta_{j1,\beta}+
\eta_{j2,\alpha}\eta_{j2,\beta})\equiv
(\eta^2_{j1}+\eta^2_{j2})$.
We define the Majorana mass term as
$\epsilon_{\alpha\beta}\eta_{\alpha}\eta_{\beta}
\equiv  \eta^2$.
}
\begin{eqnarray}
-M(m)\left(\eta_{11}^2-\eta_{12}^2-\eta_{21}^2+\eta_{22}^2\right),\quad M(m)=\epsilon(m-m^*)\Theta(m-m^*)~,
\end{eqnarray}
where $\epsilon$ is a small number, $\Theta$ is the step function: $\Theta(x)=0$ for $x\leq 0$ and $\Theta(x)=1$ for $x>0$.
Then the deformation is only nonzero when $m>m^*$, where we take $m^*>gv$.
The deformation preserves the time-reversal $CT$ symmetry.
The four Majorana fermions have masses
\begin{align}
m_{11} &= m + g\langle\phi\rangle + M(m),\cr
m_{12} &= m + g\langle\phi\rangle - M(m),\cr
m_{21} &= - m+ g\langle\phi\rangle - M(m),\cr
m_{22} &= - m+ g\langle\phi\rangle + M(m),
\end{align}
where the vev $\langle\phi\rangle= v\text{ sgn}(m_{odd})+m_{odd}+{\cal O}(m_{odd}^2)$ depends on the CT symmetry-breaking deformation $m_{odd}$.

What becomes of the phase diagram under the deformation?
For $m\leq m^*$ it is the same as before, while for $m>m^*$ there are new gapped phases:
\begin{itemize}
    \item $m_{21},m_{22}<0$, and $m_{12}<0,m_{11}>0$:
    the theory flows to
    \begin{equation} \label{eq:K=8}
        K=8\;\text{ state}: \quad \O(2)_{2,0}\;\text{CS}-4\text{CS}_\text{grav}~,
    \end{equation}
    as a $\U(1)_8$ CS theory or equivalently the abelian $K=8$  
    $K$-matrix CS theory.
    If we write the $\U(1)_8$ CS 1-form gauge field as $b$, and the $\U(1)_{\text{EM}}$ gauge field 
    as $A$, then the action is
    $\int_{} \frac{8}{4\pi}b\dd b+\frac{2b}{2\pi}\dd A -4\text{CS}_\text{grav}$,
    up to a trivial spin-TQFT to represent a fermionic gapped sector
    (see Appendix A of \cite{Lian:2018xep}).
It has quantum Hall conductivity 
and thermal Hall conductivity 
$$
\sigma_{xy}=5/2,\quad \kappa_{xy}=1+2=3.
$$

    \item $m_{11},m_{12}>0$, and $m_{21}<0,m_{22}>0$:
    the theory flows to
    \begin{equation} \label{eq:113}
    113\;\text{ state}: \quad
        \O(2)_{2,2}\;\text{CS}-2\text{CS}_\text{grav}\leftrightarrow \U(1)_{-8}\;\text{CS}-6\text{CS}_\text{grav}~,
    \end{equation}
    where we used the duality $\O(2)_{2,2}\leftrightarrow \O(2)_{-2,0}-4\text{CS}_\text{grav}$ \cite{Cordova:2017vab} and $\O(2)_{2,0}\leftrightarrow \U(1)_8$. This phase is called the 113 state, since it can be described by the 3$d$ Abelian Chern-Simons theory action, with 1-form gauge field $b$, as
$$\frac{K_{IJ}}{4 \pi}\int b_I \dd b_J -4\text{CS}_\text{grav},
    \text{ with a $K$ matrix } \left(\begin{array}{cc}
         1 & 3   \\
         3 & 1
    \end{array}\right).$$
     It has quantum Hall conductivity and thermal Hall conductivity 
$$
\sigma_{xy}=5/2,\quad \kappa_{xy}=-1+1+2=2.
$$
    It is related to the previous phase by the anti-unitary particle-hole symmetry (up to an anomaly).

\end{itemize}

\begin{figure}[!t]
  \centering
    \includegraphics[width=0.8\textwidth]{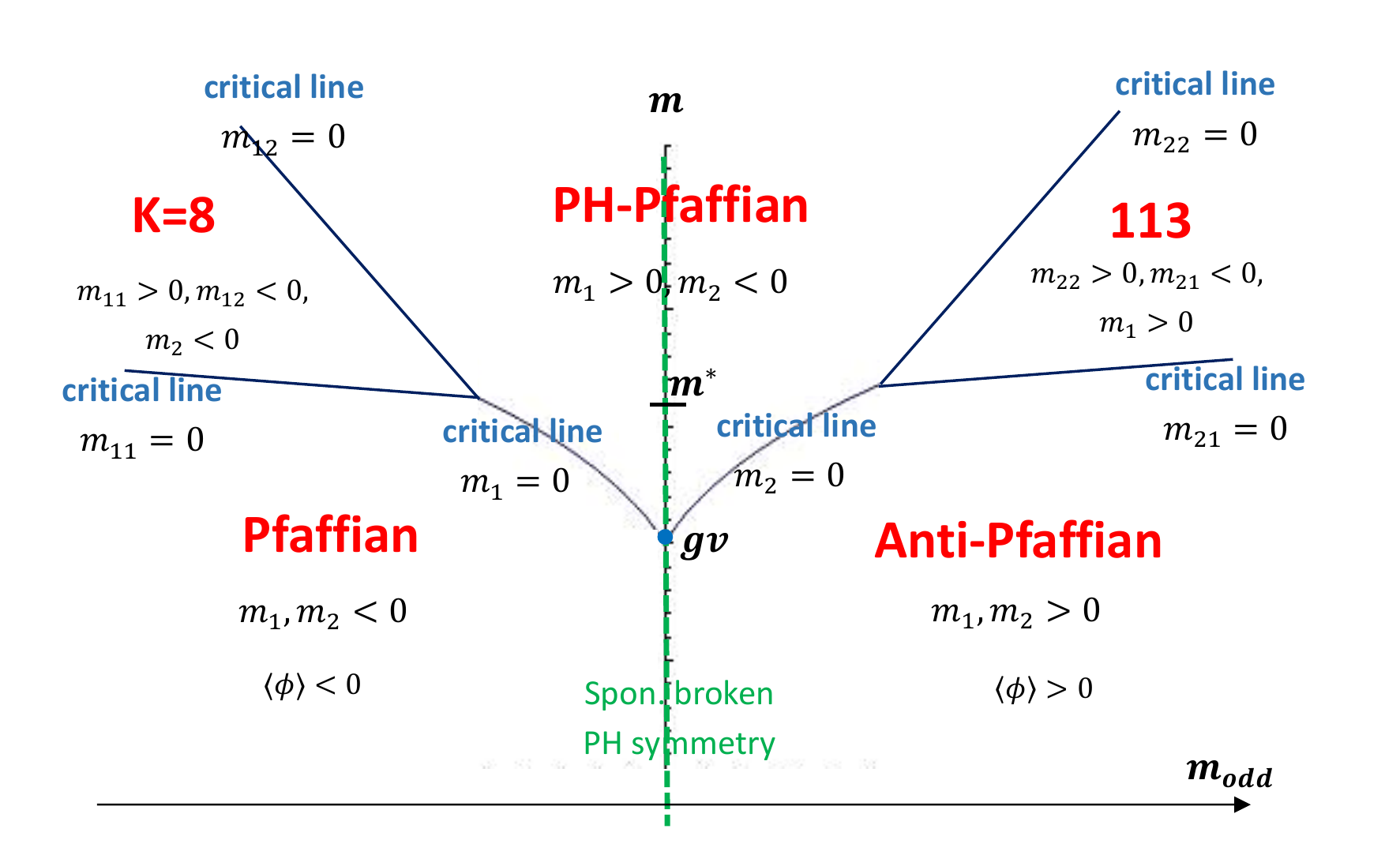}
      \caption{Phase diagram of the theory (\ref{eqn:GNY}) deformed by the PH-symmetry breaking term $\delta V(\phi)= -m_{odd}\phi$ and the
      Majorana mass $M(m)$.  Here the parameter $m$ is from the Dirac mass term
      in (\ref{eqn:GNY}). We abbreviate $m_{11},m_{12}<0$ as $m_1<0$ {\it etc}. 
      Phase boundaries can be plotted \emph{analytically} by Mathematica. 
      }\label{fig:phasemaj}
      
\end{figure}

In addition, there are critical lines separating the gapped phases where some of the fermions become massless.
The phase diagram is in Figure \ref{fig:phasemaj}.
In the rest of the discussion we will focus on the case without the deformation $M(m)$.

\subsection{Random coupling and the thermal metal phase}
\label{sec:disorderm}

In the theory (\ref{eqn:GNY}), we can further choose the parameter $m$ to be a random coupling with Gaussian distribution
\begin{equation}
\overline{m}=m_0,\quad\overline{m^2}=\delta^2.
\end{equation}
The theory depends on the average $m_0$ and the fluctuation $\delta$.
In \cite{Medvedyeva:2010MTB}, it is found that for strong fluctuations $\delta\rightarrow\infty$, the system of free Dirac fermions becomes a thermal metal.
We will set the magnitude of fluctuation to be
\begin{equation}
\delta=h(m_0)~
\end{equation}
for some non-negative, monotonically increasing function $h$ that grows faster than a linear function (for instance, $h(m_0)=m_0^2$). 
Then $m_0$ controls the disorder strength of the system.
At large enough $m_0$, {\it i.e.} strong disorder, the fluctuation becomes sufficiently strong and the model (\ref{eqn:GNY}) with random coupling $m$ enters a thermal metal phase.
Since in our model the electromagnetic background field only couples to the O(2) gauge field and does not couple to the fermions, the Hall conductivity does not depend on the mass of the fermions and remains the same value $\sigma_{xy}=\frac{5}{2}$. This is consistent with the proposal in \cite{Wang:2018WVH,Lian:2018xep}.

Near the critical lines of the phase diagram, the physical mass of one of the fermion becomes close to zero. If the disorder strength is nonzero for zero mass, $h(0)>0$, the disorder will cause the region sufficiently near the critical lines to have thermal metal behavior, which accommodates the behavior described in \cite{Wang:2018WVH} and illustrated in \Fig{fig:Phase-Pf-APf-phase}.

%\section{Anyonic excitations in the topological phases}
\section{Anyonic excitations and quantum observables from the EFT} %in the topological phases

\label{sec:quantum-numbers}

Let us spell out the key properties of the TFT phases and their anyonic excitations.\footnote{
The worldline of an anyon in quantum Hall liquids corresponds to a line operator in the low energy effective TQFT.
\label{line}}
We assume standard knowledge from the Chern-Simons (CS)  description of fQHE. 
We will delineate the following:
\begin{itemize}
\item Fractionalized anyon statistics, {\it i.e.} the spin or exchange statistics $\exp(\ii 2 \pi {\rm{s}})$ of anyons with spin ${\rm{s}}$.
\item Fractionalized $\U(1)_{\text{EM}}$ electromagnetic charge $Q/e$ ($e$ is the electron charge).  
\item Their PH-symmetry (time-reversal $CT$) transformation properties.
\end{itemize}
They are summarized in Tables \ref{table:Pf}, \ref{table:APf}, \ref{table:PHPf},  \ref{table:K=8}, and \ref{table:113},
for the Pfaffian, anti-Pfaffian, PH-Pfaffian,
$K=8$, and 113 states respectively, in the notation
\bea
(\exp(\ii 2 \pi {{\rm{s}}}), Q/e).
\eea
For the PH-Pfaffian, since it enjoys PH-symmetry (time-reversal $CT$), we also specify the $(CT)^2$ quantum number for the appropriate anyons, and write
\bea
(\exp(\ii 2 \pi {\rm{s}}), Q/e)_{{(CT)}^2}.
\eea

We will first examine the non-abelian states, {\it i.e.} the Pfaffian in \eq{eq:PfaffianO2},
anti-Pfaffian in \eq{eq:antiPfaffianO2}, and
PH-Pfaffian in \eq{eq:PHPfaffianO2}. They can be written as the following Chern-Simons theories (see Appendix A of 
\Rf{Lian:2018xep}, and Appendix \ref{app:O2}):
\bea
\text{Pfaffian}&:&\quad \frac{\U(1)_8\times {{\text{Ising}}}}{\mathbb{Z}_2}
-4\text{CS}_\text{grav},~\quad \; \;  c_-=1+1/2+4/2=7/2. \label{eq:1Pf}\\
\text{PH-Pfaffian}&:&\quad  \frac{\U(1)_8\times {\overline{\text{Ising}}}}{\mathbb{Z}_2}-4\text{CS}_\text{grav},~\quad \;  \;  c_-=1-1/2+4/2=5/2. \label{eq:2PHPf} \\
\text{anti-Pfaffian}&:&\quad \frac{\U(1)_8\times {\SU(2)_{-2}}}{\mathbb{Z}_2}-4\text{CS}_\text{grav},~\;  c_-=1-3/2+4/2=3/2.
 \label{eq:3APf}
\eea
with their chiral central charges $c_- = \kappa_{xy}$.
These TQFTs are obtained from gauging a diagonal one-form $\Z_2$ symmetry in the
($\U(1)_8$ CS theories) and the ($\nu \in \Z_8$-class spin-TQFTs)\footnote{Here
the  2+1$d$ $\nu \in \Z_8$-class spin-TQFTs
 are obtained from gauging the $\Z_2$ internal ``Ising'' symmetry of
the  2+1$d$ fermionic $\Z_2 \times \Z_2^f$-SPTs, with fermion parity symmetry $\Z_2^f$
\cite{1602.04251SW, Putrov2016qdo1612.09298PWY}. From this class of TQFTs, we will use the
Ising, $\overline{\text{Ising}}$, and ${\SU(2)_{-2}}$ cases.}
in 2+1$d$.
More generally, these TQFTs are
$ \frac{\U(1)_8\times {T_L}}{\mathbb{Z}_2}$ for $L=-1,+1,+3$, 
and one gauges a diagonal $\mathbb{Z}_{2}$ one-form symmetry generated by the composite line given by the tensor product of the charge 4 Wilson line of $\U(1)_8$ and a non-transparent fermion line in $T_L$.
 See Appendix \ref{app:O2} for details on the $T_L$ theories.
   When gauging a diagonal $\mathbb{Z}_{2,[1]}$  symmetry,
we identify their
 \emph{charged objects} (the line with odd $\U(1)$ charge in $\U(1)_8$ 
 and the  $\sigma$ line in $T_L$) 
 and their \emph{symmetry generators} or 
  \emph{charge operators}
  (the operator with $\U(1)$ charge 4
  in $\U(1)_8$  and the  $f$ line in $T_L$).
This reduces the 24 anyons in the quasi-excitation spectrum of
  ${\U(1)_8\times T_L}$ theory to
  the 12 anyons in the
$\frac{\U(1)_8\times T_L}{\mathbb{Z}_2}$ theory.

 \begin{enumerate}
\item
 {\bf Spin statistics.}  The spin of an anyon is given by
 \bea \label{eq:spin}
 \exp(\ii 2 \pi {\rm{s}})= \exp(\ii  2 \pi( {\rm{s}}_{\text{nab}}+ \frac{q^2}{2 K})),
 \eea
where $K$ is the level of abelian CS theory, and $q$ is the integer 
labeling the abelian anyon associated with the line operator $\e^{\ii q \oint b}$ of 
1-form gauge field $b$.\footnote{In the $K$-matrix CS theory, we replace $\frac{q^2}{2 K} \mapsto \frac{q^{{\rm T}} K^{-1} q}{2}$ where $q$ is a charge vector in the second expression.}
 Here, ${\rm{s}}_{\text{nab}}$ means the spin from the non-abelian sector of the TQFT.
 For the Ising, $\overline{\text{Ising}}$, and ${\SU(2)_{-2}}$ TQFTs
 in \eq{eq:1Pf}, (\ref{eq:2PHPf}) and (\ref{eq:3APf}), 
 their ${\rm{s}}_{\text{nab}}$ for the $(1,\sigma,f)$ anyons are
 given by the diagonal of the modular
 $T$ matrix: $(1,\e^{\ii \frac{\pi}{8}},-1)$,
 $(1,\e^{-\ii \frac{\pi}{8}},-1)$,
 and $(1,\e^{-\ii \frac{3\pi}{8}},-1)$ respectively. See, e.g., \cite{Lian:2018xep} for the data.

\item {\bf Electromagnetic charge.} 
 For the anyon's $\U(1)_{\text{EM}}$ charge $Q/e$,
we can look at the coupling $q$ of the electric current to the $\U(1)_{\text{EM}}$ gauge field $A$. The charge can be changed by an integer by tensoring the line with a classical Wilson line $\oint A$.\footnote{
If we demand the spin/charge relation with spin$^c$ connection $A$, then the isolated $\oint A$ is not well-defined and the transparent fermion line in all theories is charged under $\U(1)_\text{EM}$. Then the charge is instead taken modulo 2 from tensoring with $2\oint A$.
}
The $\U(1)_{\text{EM}}$ charge $Q$ and the Hall conductance $\sigma_{xy}$
can be computed via (see Appendix \ref{app:O2} for details):
 \bea \label{eq:charge-Q}
 Q/e= K^{-1} q,
\quad\quad\quad 
\sigma_{xy}=q K^{-1} q = q (Q/e).
 \eea
Based on the experimental constraint of $\sigma_{xy}=\nu=1/2$, 
we have to introduce the appropriate
$\U(1)_{\text{EM}}$ coupling $\int \frac{2b}{2\pi}\dd A
=\int\frac{1}{2\pi}(2b)\dd A$ to the action for the 
$\frac{\U(1)_8\times T_L}{\mathbb{Z}_2}$ theory, where the
$\U(1)_8$ CS theory action is $\int \frac{8}{4\pi}b\dd b$. This is a coupling with charge $q=2$.
Indeed, this gives half-filled $\nu=\sigma_{xy}=q K^{-1} q=2^2/8=1/2$. 
The anyon with $\U(1)$-charge 2 is identified with the non-abelian $\sigma$ anyon
in the gauged $\frac{\U(1)_8\times {T_L}}{\mathbb{Z}_2}$ CS theory.
This non-abelian anyon has $\U(1)_{\text{EM}}$ charge $Q/e= K^{-1} q=2/8=1/4$.
We can obtain all 12 anyons' $\U(1)_{\text{EM}}$ charges by the same argument, with the results shown
in Tables \ref{table:Pf}, \ref{table:APf}, and \ref{table:PHPf}.

\item {\bf PH-symmetry.} 
In the PH-Pfaffian theory, 
PH-symmetry (or time-reversal $CT$) is preserved, so to those anyons not permuted by the time-reversal symmetry, whose spin statistics $\exp(\ii 2 \pi {\rm{s}})$ are real-valued, \footnote{In other words, $\exp(\ii 2 \pi {\rm{s}})=\pm 1$ for such anyons, so they are self-bosonic or self-fermionic.} 
we can assign $(CT)^2 = \pm 1$ quantum numbers.
For those anyons $\alpha_{a}$ whose spin statistics $\exp(\ii 2 \pi {\rm{s}})$
are complex valued, the spin statistics are mapped to their complex conjugates $\exp(-\ii 2 \pi {\rm{s}})$ under the $CT$ transformation.  %\cred
{In fact, there are two versions of
PH-Pfaffian denoted as PH-Pfaffian$_{\pm}$ depending on how the $(CT)^2$ quantum number is assigned to the odd $q$ charge of $\U(1)_8$ anyons, which we elaborate in Table 
\ref{table:PHPf}.
}

On the other hand, the Pfaffian and anti-Pfaffian states do not have $CT$ symmetry. Instead, they map into each other under the $CT$ transformation as follows.
\begin{itemize}
\item
When the $\U(1)_8$ charge $q$ is even,
the abelian sector is paired with the abelian trivial anyon 1 or the fermionic anyon $f$, so under the $CT$ transformation:
$$
\text{Pf: } (q_{\text{even}}, f^n) \overset{CT}{\Longleftrightarrow}
\text{APf: } (q_{\text{even}}, f^{n+ \frac{q_{\text{even}}}{2}}),
$$
where $f^2=(f)^{\text{even}}=1$ and $n=0,1$. Namely,  
$$
\text{Pf: } (q_{\text{even}}, f^n) \overset{CT}{\Longleftrightarrow}
\text{APf: } (q_{\text{even}}, f^{n}), \text{ if 
$\frac{q_{\text{even}}}{2} \in$ even}.
$$
$$
\text{Pf: } (q_{\text{even}}, f^n) \overset{CT}{\Longleftrightarrow}
\text{APf: } (q_{\text{even}}, f^{n+1}), \text{ if 
$\frac{q_{\text{even}}}{2} \in$ odd}.
$$

\item When the $\U(1)_8$ charge $q$ is odd, the abelian sector is paired with the non-abelian 
$\sigma$ anyon, so under $CT$:
$$
\text{Pf: } (q_{\text{odd}}, \sigma) \overset{CT}{\Longleftrightarrow}
\text{APf: } (q_{\text{odd}}, \sigma).
$$
\end{itemize}

\end{enumerate}

The 12 anyons, and their spin statistics 
$\exp(\ii 2 \pi {\rm{s}})$, 
$\U(1)_{\text{EM}}$ charges, 
and $CT$ properties are organized
in Tables \ref{table:Pf}, \ref{table:APf}, and \ref{table:PHPf}.\footnote{Note that the sigma anyon $\sigma_n$ notation in our present work is actually the $\sigma_{-n}$ in  \Rf{Lian:2018xep}.} The list of anyons in the Tables contains not only \emph{quasiparticles} but also \emph{quasiholes} of quantum Hall liquids, to be explained in Appendix \ref{app:many-body-wavefunctions}.\footnote{As mentioned in footnote \ref{line},
 the line operator is a worldline of an anyon.
Moreover, the two open ends of a line
operator correspond to two anyons 
that can be fused to nothing (i.e. the open line can 
become a closed line after fusing two ends). Thus, the two open ends of a line operator correspond to a \emph{quasiparticle} and its \emph{quasihole} 
in the quantum Hall liquids of Appendix  \ref{app:many-body-wavefunctions}.
The entries in Tables \ref{table:Pf}, \ref{table:APf}, and \ref{table:PHPf} therefore contain data for anyons and their ``anti-particles''. The fusion of a quasiparticle and its quasihole
must include a trivial anyon 1 that carries zero global symmetry charges
and trivial spin statistics 
 $\exp(\ii 2 \pi {\rm{s}})=\pm 1$. 
 (More accurately, the spin statistics of the fusion outcome of two anyons contain not only the spin statistics of each individual anyon [from their modular $\cal{T}$ matrix], but also their mutual
 statistics from their relative angular momentum [from their modular $\cal{S}$ matrix]. Here spin-1/2 is allowed for intrinsically fermionic systems).
} Although there are 12 anyons, the number of ground states
on a spatial 2-torus $T^2$ known as the ground state degeneracy (GSD)
is only 6 for the Pf, APf, and PHPf states. The corresponding 6 ground states depend on the spin structure of the spin manifold $T^2$. 

%\newpage
\begin{table}[!t] %[!htbp]
\centering
\begin{tabular}{c|cccccccc}%{*9c}
\toprule
Pfaffian &  \multicolumn{7}{c}
 {$\U(1)_8$ {CS}} & \\
\midrule
{$T_{L=-1}$}   & 0 & 1 & 2 & 3 & 4 & 5 & 6 & 7 \\
\midrule
1    & $(+1, 0)$ &   & $(+\ii, \frac{1}{2})$ &   & $(+1,1)$ &   & $(+\ii, \frac{3}{2})$ & \\
$\sigma_{-1}$  &   &  $(\e^{\ii \frac{\pi}{4}}, \frac{1}{4})$ &   & $(-\e^{\ii \frac{\pi}{4}}, \frac{3}{4})$ &   &  $(-\e^{\ii \frac{\pi}{4}},\frac{5}{4})$ &   &  $(\e^{\ii \frac{\pi}{4}},\frac{7}{4})$ \\ 
$f$ & $(-1, 0)$&   & $(-\ii, \frac{1}{2})$&   & $(-1,1)$ &   & $(-\ii, \frac{3}{2})$ & \\
\bottomrule
\end{tabular}
\caption{Pfaffian data from $\frac{\U(1)_{8} \times {\text{Ising}}}{\Z_2}$ CS. 
We provide $(\exp(\ii 2 \pi {\rm{s}}), Q/e)$ for 12 anyons.
The  $\sigma_{-1}$ anyon has $\e^{\ii \frac{\pi}{8}}$ statistics.
}
\label{table:Pf}
\end{table}

\begin{table}[!h] %[!htbp]
\hspace{-14mm}
%\centering
\begin{tabular}{c|cccccccc} %{*9c}
\toprule
Anti-Pfaffian &  \multicolumn{7}{c}{$\U(1)_8$ {CS}} & \\
\midrule
{$T_{L=3}$}   & 0 & 1 & 2 & 3 & 4 & 5 & 6 & 7 \\
\midrule
1    & $(+1, 0)$ &   & $(+\ii, \frac{1}{2})$ &   & $(+1,1)$ &   & $(+\ii, \frac{3}{2})$ & \\
$\sigma_{3}$  &   &  $(\e^{-\ii \frac{\pi}{4}}, \frac{1}{4})$ &   & $(-\e^{-\ii \frac{\pi}{4}}, \frac{3}{4})$ &   &  $(-\e^{-\ii \frac{\pi}{4}},\frac{5}{4})$ &   &  $(\e^{-\ii \frac{\pi}{4}},\frac{7}{4})$ \\ 
$f$ & $(-1, 0)$&   & $(-\ii, \frac{1}{2})$&   & $(-1,1)$ &   & $(-\ii, \frac{3}{2})$ & \\
\bottomrule
\end{tabular}
\caption{Anti-Pfaffian data from the 
$\frac{\U(1)_{8} \times {\SU(2)_{-2}}}{\Z_2}$ CS.
We provide
$(\exp(\ii 2 \pi {\rm{s}}), Q/e)$ for 12 anyons.
The  $\sigma_{3}$ anyon has $\e^{-\ii \frac{3\pi}{8}}$ statistics.
}
\label{table:APf}
\end{table}

\begin{table}[!h] %[!htbp]
\hspace{-11mm}
%\centering
\begin{tabular}{c|cccccccc} %{*9c}
\toprule
PH-Pfaffian$_{\pm}$ &  \multicolumn{7}{c}{$\U(1)_8$ {CS}} & \\
\midrule
{$T_{L=1}$}   & 0 & 1 & 2 & 3 & 4 & 5 & 6 & 7 \\
\midrule
1    & $(+1, 0)_{1}$ &   & $(+\ii, \frac{1}{2})$ &   & $(+1,1)_{-1}$ &   &  $(+\ii, \frac{3}{2})$ & \\
$\sigma_1$  &   & $(1, \frac{1}{4})_{\pm 1}$ &   & $(-1, \frac{3}{4})_{\mp 1}$ &   & $(-1,\frac{5}{4})_{\mp 1}$ &   &  $(1,\frac{7}{4})_{\pm 1}$ 
\\ 
$f$ & $(-1, 0)_{1}$&   & $(-\ii, \frac{1}{2})$&   & $(-1,1)_{-1}$ &   & $(-\ii, \frac{3}{2})$ & \\
\bottomrule
\end{tabular}
\caption{PH-Pfaffian data from $\frac{\U(1)_{8} \times \overline{\text{Ising}}}{\Z_2}$ CS.
We provide $(\exp(\ii 2 \pi {\rm{s}}), Q/e)$ 
or $(\exp(\ii 2 \pi s), Q/e)_{(CT)^2}$ for 12 anyons.
The  $\sigma_{1}$ anyon has $\e^{-\ii \frac{\pi}{8}}$ statistics. In comparison, 
\Rf{Metlitski2014xqa1406.3032} represents related TQFT data in terms of
 $\frac{\U(1)_{-8} \times {\text{Ising}}}{\Z_2}$ CS. Moreover,
 there are two versions PH-Pfaffian$_{\pm}$ depending on how $(CT)^2$ assigns to
 the odd $q=1,3,5,7$ of $\U(1)_8$.
}
\label{table:PHPf}
\end{table}

Now we examine the abelian states. The $K=8$ state in \eq{eq:K=8}
has the action
    $\int_{} \frac{8}{4\pi}b\dd b+\frac{2b}{2\pi}\dd A -4\text{CS}_\text{grav}$
    plus a trivial spin-TQFT with $\{1,f \}$ (generated by a trivial line and a fermionic line).
    Note that the fermion
    $f$ does not couple to $\U(1)_{\text{EM}}$.
    
   The $113$ state in \eq{eq:113} has the action
   $\frac{K_{IJ}}{4 \pi}\int b_I \dd b_J +\frac{(q^{{\rm T}}_I \cdot  b_I)}{2\pi}\dd A
    -4\text{CS}_\text{grav},$
   where $q^{\rm{T}}$ denotes the transpose of the $q$ charge vector. There are two convenient expressions for this theory, related by a GL(2,$\mathbb{Z}$) transformation \cite{Lian:2018xep}:
   $$
K  =\left(\begin{array}{cc}
         1 & 3  \\
         3 & 1
    \end{array}\right), q^{\rm{T}}=(1,1)\; {\xleftrightarrow{\text{GL(2,$\mathbb{Z}$) transformation}}}\; 
    K  =\left(\begin{array}{cc}
         -8 & 0  \\
         0 & 1
    \end{array}\right), q^{\rm{T}}=(2,1).$$
    (We omit an electron charge normalization factor $e$.)

    The quantum numbers for the abelian states are shown in Tables \ref{table:K=8} and  Table \ref{table:113}.
    The spin statistics can be obtained from \eq{eq:spin} by dropping the
    ${\rm{s}}_{\text{nab}}$ part. The  $\U(1)_{\text{EM}}$
    charge can be determined from \eq{eq:charge-Q} as before.

\begin{table}[!t] %[!htbp]
\hspace{-16mm}
\begin{tabular}{c|cccccccc} %{*9c}
\toprule
$K=8$-state &  \multicolumn{7}{c}{$\U(1)_8$ {CS}} & \\
\midrule
{$T_{L=0}$}   & 0 & 1 & 2 & 3 & 4 & 5 & 6 & 7 \\
\midrule
1    & $(+1, 0)$ &$(\e^{\ii \frac{\pi}{8}},  \frac{1}{4})$ & $(\e^{\ii \frac{\pi}{2}},  \frac{1}{2})$   &  $(-\e^{\ii \frac{\pi}{8}},  \frac{3}{4})$ 
& $(+1, 1 )$ & $(-\e^{\ii \frac{\pi}{8}}, \frac{5}{4})$  &  $(\e^{\ii \frac{\pi}{2}},  \frac{3}{2})$ &
$(\e^{\ii \frac{\pi}{8}},  \frac{7}{4})$ \\
$f$ & $(-1, 0)$& $(-\e^{\ii \frac{\pi}{8}},  \frac{1}{4})$  & $(-\e^{\ii \frac{\pi}{2}},  \frac{1}{2})$  & $(\e^{\ii \frac{\pi}{8}},  \frac{3}{4})$   
& $(-1,  1)$ & $(\e^{\ii \frac{\pi}{8}}, \frac{5}{4})$  &  $(-\e^{\ii \frac{\pi}{2}},  \frac{3}{2})$ & 
$(-\e^{\ii \frac{\pi}{8}}, \frac{7}{4})$ \\
\bottomrule
\end{tabular}
\caption{Data for the $K=8$-state.
We provide
$(\exp(\ii 2 \pi {\rm{s}}), Q/e)$ for 16 anyons.
}
\label{table:K=8}
\end{table}

\begin{table}[!t] %[!htbp]
\hspace{-20mm}
\begin{tabular}{c|cccccccc} %{*9c}
\toprule
113-state &  \multicolumn{7}{c}{$\U(1)_{-8}$ {CS}} & \\
\midrule
{$T_{L=2}$}   & 0 & 1 & 2 & 3 & 4 & 5 & 6 & 7 \\
\midrule
1    & $(+1, 0)$ &$(\e^{-\ii \frac{\pi}{8}},  \frac{7}{4})$ & $(\e^{-\ii \frac{\pi}{2}},  \frac{3}{2})$   &  $(-\e^{-\ii \frac{\pi}{8}},    \frac{5}{4})$ & $(+1, 1)$ & 
$(-\e^{-\ii \frac{\pi}{8}},  \frac{3}{4})$  &  $(\e^{-\ii \frac{\pi}{2}},  \frac{1}{2})$ &
$(\e^{-\ii \frac{\pi}{8}},  \frac{1}{4})$ \\
$f$ & $(-1, 1)$& $(-\e^{-\ii \frac{\pi}{8}}, \frac{3}{4})$  & $(-\e^{-\ii \frac{\pi}{2}}, \frac{1}{2})$  & $(\e^{-\ii \frac{\pi}{8}}, \frac{1}{4})$   & $(-1, 0)$ & 
$(\e^{-\ii \frac{\pi}{8}}, \frac{7}{4})$  &  $(-\e^{-\ii \frac{\pi}{2}},\frac{3}{2} )$ & 
$(-\e^{-\ii \frac{\pi}{8}},  \frac{5}{4})$ \\
\bottomrule
\end{tabular}
\caption{Data for the 113-state. We provide
$(\exp(\ii 2 \pi {\rm{s}}), Q/e)$ for 16 anyons.
}
\label{table:113}
\end{table}

The $K=8$ and 113-states do not have $CT$ symmetry. 
Instead, they map into each other under the $CT$ transformation.
 Quantum numbers of their anyons are mapped as:
$$
K=8: \quad (\exp(\ii 2 \pi {\rm{s}}), Q/e  \mod 1) \overset{CT}{\Longleftrightarrow}
113: \quad (\exp(-\ii 2 \pi {\rm{s}}), Q/e \mod 1).
$$
The mod 1 comes from the freedom to tensor the anyons with the classical Wilson line $\oint A$.

Although there are 16 anyons in each the abelian state, the GSD
is only 8. The corresponding 8 ground states depend on the spin structure of the spin manifold 2-torus $T^2$.\footnote{Since all five theories are fermionic spin-TQFTs,
we can specify various spin structures on the $T^2$ to characterize the GSD.
There are 4
choices corresponding to the periodic (P) or anti-periodic (A) boundary
conditions along each of
two 1-cycles of  $T^2$: (P,P), (A,P), (P,A), and (A,A).
The Hilbert space up to an isomorphism only
depends on the fermionic parity $\Z_2^f$ (the $\Z_2$ value of the Arf invariant).
The fermionic parity $\Z_2^f$ is odd for (P,P), and  
the $\Z_2^f$ is even for (A,P), (P,A), (A,A).
We denote the corresponding spin 2-tori $T^2$ as $T^2_{\rm o}$ for odd and $T^2_{\rm e}$ for even.
The ground states on $T^2_{\rm o}$ or on $T^2_{\rm e}$ can come from different states.
The 6 ground states on $T^2$ in 
Table \ref{table:Pf}, 
\ref{table:APf} and \ref{table:PHPf}, 
depending on $T^2_{\rm o}$ or $T^2_{\rm e}$, are chosen
differently among 12 line operators.
The 8 ground states on $T^2$ in 
Table \ref{table:K=8} and \ref{table:113}, 
 depending on $T^2_{\rm o}$ or $T^2_{\rm e}$, are chosen 
differently among 16 line operators. 
In fact, rigorously speaking, only the (P,P) sector stays the same sector under the modular SL(2,$\Z$)'s $\CS$ and $\CT$ transformations,
while (A,P), (P,A), and (A,A) permute to each other under the modular $\CS$ and $\CT$.
The boundary conditions, P and AP, are also known as Ramond and Neveu-Schwarz sectors respectively in string theory. 
See more discussions about the spin structure dependence 
in \cite{Putrov2016qdo1612.09298PWY, 1801.05416WOP,Guo20181812.11959}.
}

\section{Domain wall theory and tension}
\label{sec:Domain-wall}

As reviewed in the introduction, the proposal of \cite{Lian:2018xep} suggests a percolation transition involving puddles of Pf and APf phases separated by domain walls. To this end, we consider the model \eqref{eqn:GNY} on the slice of parameter space with time-reversal symmetry preserved, {\it i.e.} $m_{odd}=0$. 
We would like to study some basic properties of the domain walls, from the EFT point of view, that result when time-reversal symmetry is spontaneously broken. 

Let us ignore the discrete gauge field which couples to the fermions, for now, and write the Lagrangian as (in the mostly positive Lorentzian signature)
\begin{equation}
\label{PS}
 {\cal L} = \sum_{i=1,2} \overline\Psi^i (\ii\slashed{D}- g \phi) \Psi^i + m ( \overline\Psi^1 \Psi^1 - \overline\Psi^2 \Psi^2 ) + {1\over2}(\partial_{\mu}\phi)(\partial^{\mu} \phi) - {1 \over 4}\lambda(\phi^2 - v^2)^2.
\end{equation} 
The vacua are doubly degenerate, with the vevs given by $\pm v$ where $v \equiv {\mu / \sqrt{\lambda}}$.
Throughout this section, we assume without loss of generality that $m, \, g \ge 0$.

The classical solution for a static domain wall is, as usual,
\begin{equation}\label{eq:scalar}
\phi_0(z) = {\mu \over \sqrt{\lambda}} \tanh{\mu (z- z_0) \over \sqrt{2}}
\end{equation} with $z_0$ the center-of-mass coordinate.\footnote{One also has an anti-domain wall of the opposite overall sign; we will focus on the properties of the domain wall.
}
We assume that the effective perturbative expansion parameter in the scalar sector, $\lambda / \mu$, is small to validate the semiclassical analysis that we perform presently. The classical action evaluated on the domain wall saddle is 
\begin{equation}
    S_{\phi}[\phi_0] = {2 \sqrt{2} \over 3} {\mu^3 \over \lambda} \int \dd^2x \, ,
\end{equation}
where $\dd^2x$ is over the parallel directions to the domain wall, and the transverse $z$ direction has already been integrated over. Divided by the area worldvolume area $\int \dd^2x$, this famously gives the classical domain wall tension \cite{Dashenetal1, Dashenetal2}
\begin{equation}
    \label{action0}
    \sigma_{cl} = {2 \sqrt{2} \over 3} {\mu^3 \over \lambda}.
\end{equation}

For nonzero fermion mass, the two vacua are gapped. At energies smaller than the $2+1d$ bulk gap, we have well-defined 1+1$d$ domain wall theories.  To derive the domain wall theories, we first analyze the fermionic zero modes (which survive the low energy limit) in section~\ref{Sec:ZeroModes}, and then proceed to quantize the zero modes to obtain the domain wall theories in section~\ref{Sec:DomainWallTheory}.  We then study another aspect of the domain walls -- their tension, and we do so at one-loop order.  

\subsection{Fermionic zero modes in the domain wall background}
\label{Sec:ZeroModes}

In the semiclassical approximation, the transverse profile of fermion modes solves the Dirac equation in the domain wall background:
\begin{equation}
(\epsilon_r \gamma^0 +  \ii \gamma^1 {\partial \over \partial z} \pm m - g \phi_0(z))\Psi^j_0(z)=0.
\end{equation}
We have, for the moment, suppressed dependence on the spatial direction parallel to the domain wall. 
We use the Majorana basis for $\Gamma$-matrices 
{$\gamma^0 = - \ii \sigma^y, \gamma^1 = \sigma^z, \gamma^2 = \sigma^x$}, and write the two-component spinors explicitly as $\Psi^j(z) = (u^j_r(z), v^j_r(z))^{{\rm T}}$, $j=1, 2$ (so the top component is of definite chirality and the bottom component has the opposite chirality). With these conventions, the above equation becomes
\begin{align*}
(-{\partial \over \partial z} \pm m - g \phi_0(z))v^j_r &= \epsilon_r u^j_r(z), \\
({\partial \over \partial z} \pm m- g \phi_0(z))u^j_r &= -\epsilon_r v^j_r(z).
\end{align*}
We are interested in the zero-modes, which survive the low energy limit. For $\epsilon_0=0$, we can solve these equations in the classical domain wall background: 
\begin{align}\label{eq:ferm1}
u^1_0(z) &= \psi^1_{0,+} \e^{-m (z-z_0)}\cosh\left((z-z_0){\mu \over \sqrt{2}}\right)^{\sqrt{2}g \over \sqrt{\lambda}},\\ 
v^1_0(z) &= \psi^1_{0,-}e^{m(z-z_0)} \cosh\left((z-z_0){\mu \over \sqrt{2}}\right)^{-\sqrt{2}g \over \sqrt{\lambda}},
\end{align}
and
\begin{align}\label{eq:ferm2}
u^2_0(z) &= \psi^2_{0,+} \e^{m(z-z_0)}\cosh\left((z-z_0){\mu \over \sqrt{2}}\right)^{\sqrt{2}g \over \sqrt{\lambda}},\\ 
v^2_0(z) &= \psi^2_{0,-} \e^{-m(z- z_0)}\cosh\left((z-z_0){\mu \over \sqrt{2}}\right)^{-\sqrt{2}g \over \sqrt{\lambda}}.
\end{align}

Let us discuss the properties of the zero modes in the Pfaffian/anti-Pfaffian regime $m < gv$ and the PH-Pfaffian regime $m > gv$. These properties will be the key in our subsequent determination of the respective domain wall theories in section~\ref{Sec:DomainWallTheory}.

When $m < g v$, since the solution for $u^j_0(z), j = 1, 2$ is not normalizable, we set both $\psi^j_{0,+} = 0$ and are therefore left with two complex parameters $\psi^j_{0,-}$, which constitute our expected four real fermionic zero modes of a single chirality (thus, they correspond to four chiral Majorana fermions). In the extreme limit of $m \ll g v$, the zero-modes satisfy $\bar{\Psi}\Psi \sim \Psi^{\dagger} \sigma_2 \Psi = 0$, and hence do not backreact on the scalar via the equations of motion.

When $m > g v$, the fermions delocalize and are essentially described by plane wave solutions. 
For each Dirac fermion, the normalizable edge modes of opposite chiralities survive on different sides of a half-space:
\begin{center}
    \begin{tabular}{c|c|c}
        Fermion & $z \ge z_0$ & $z \le z_0$ \\\hline
        $\Psi^1$ (mass $m > 0$) & $\psi^1_-$ & $\psi^1_+$ \\
        $\Psi^2$ (mass $-m < 0$) & $\psi^2_+$ & $\psi^2_-$ \\
    \end{tabular}
\end{center}
The semiclassical limit $\mu/\lambda \gg 1$ is also a ``hard-wall'' limit, in which the soliton solution tends towards a steep step-function at $z=z_0$ with an insurmountable height barrier.
Then we can indeed consider the normalizable edge modes on two half-spaces that can only interact via possible couplings on the interface. Among the relevant interactions, a 1+1$d$ Majorana mass term for each fermion species, induced from the bulk mass term, can survive precisely on the wall, and gaps out the fermionic degrees of freedom at low energies. This is rather analogous to wall-localized supersymmetric couplings that appear in \cite{DGP}.

\subsection{Domain wall worldvolume theory in $1+1d$}
\label{Sec:DomainWallTheory}

There is a natural proposal for the domain wall worldvolume theory following from simple anomaly considerations.  It is the O(2) WZW model coupled to two massless complex Dirac fermions by a common $\mathbb{Z}_2$ orbifold that acts as the charge conjugation in O(2). The chiral anomaly accounts for the relative shift of the Chern-Simons level in the two bulk vacua. Since the U(1) part of the gauge field is confined in 1+1$d$, the theory naturally flows to $\mathbb{Z}_2$ coupled to two complex fermions.  The domain wall theory has $\O(4)/\mathbb{Z}_2$ symmetry which rotates the four massless real fermions. This is consistent with the proposal in \cite{Lian:2018xep}. 

Let us now derive the domain wall worldvolume theory from first principles to verify this intuition. For the moment, we will ignore the presence of the discrete gauge field, and reinstate its effect at the end. The vacua, which spontaneously break the time-reversal invariance, occur at $\langle \phi \rangle = \pm v$. The fermions in each of these vacua have tree-level masses $\pm m +  g \langle \phi \rangle = \pm m \pm g v$.

To get the 1+1$d$ domain wall theories, we wish to quantize the zero-modes in the two regimes of interest, $m < gv$ and $m > gv$.  We first describe a sector of the worldvolume theory without fermions, and then describe the interesting fermionic sector alluded to above.  In the following, all quantities are the renormalized versions, as we imagine having already integrated out the bulk massive modes.

\paragraph{Goldstone mode}  Since the domain wall breaks translational invariance, there is an effective action for the bosonic Goldstone center-of-mass mode. It arises from promoting the modulus\footnote{Here, the term modulus refers to a massless scalar field with trivial potential (at least, at the order to which we are working in the derivative expansion; we discuss this more below). It has the geometric interpretation of being the center-of-mass coordinate of the domain wall.} $z_0$ $\in \R$ adiabatically to functions of the worldvolume directions $m \in (t, x)$. Integrating over $z$ and dropping a standard additive constant (hence our use of $\sim$ below) gives
\begin{align}
\label{TreeTension}
\mathcal{L}^{G}[z_0] = {1 \over 2} \int \dd z \left( (\partial_z \phi_0)^2 + (\partial_m z_0(x, t) \partial_z \phi_0 )^2 \right) \sim {\sigma_{cl} \over 2} (\partial_m z_0(x, t))^2,
\end{align}
where the bosonic tension is
\begin{align}
\sigma_{cl} = {2 \sqrt 2 \over 3} {\mu^3 \over \lambda},
\end{align}
in agreement with the tension \eqref{action0} derived from evaluating the classical (effective) action on the domain wall solution. We neglect irrelevant higher-derivative terms in the fluctuation $z_0(x, t)$.

\paragraph{Wess-Zumino-Witten (WZW) models}

Since the Chern-Simons sector of the bulk theory does not interact with the degrees of freedom on the wall except via the $\mathbb{Z}_2$ gauging of the fermions, the domain wall is transparent to the continuous gauge degrees of freedom. As is well known, the 1+1$d$ theory that furnishes a trivial interface for a Chern-Simons gauge field is the corresponding WZW theory. 

This bulk Chern-Simons term on the two sides of the wall contributes a diagonal CFT on the wall due to the opposite orientations with respect to the bulk. The theory on the wall can be constructed as follows.
First we start with the bulk theory $\SO(2)_2=\U(1)_2$ on both sides of the wall, so that the theory on the wall is naturally a compact boson at the self-dual radius. Then we deposit additional $L$ units of $\mathbb{Z}_2$ SPT phases in the bulk, which induce additional fermions on the wall.  The amount $L$ of $\mathbb{Z}_2$ SPT phases appropriate for each phase was discussed in Section~\ref{Sec:PreservingDeformation}, which we summarize here for the convenience of the reader:
\begin{center}
    \begin{tabular}{c|c}
        Phase & $L$ \\\hline
        PH-Pfaffian & $1$ \\
        Pfaffian & $-1$ \\
        Anti-Pfaffian & $3 \equiv -5 \mod 8$
    \end{tabular}
\end{center}
Finally, we  gauge the diagonal $\mathbb{Z}_2$ symmetry of the entire configuration that acts as charge conjugation on the $\SO(2)$ gauge field.
This introduces a single $\mathbb{Z}_2$ gauge field throughout the bulk and on the wall.
In other words, at the interface we identify the $\mathbb{Z}_2$ gauge field on the left side of the wall with the gauge field on the right.  
We may employ the relation among Chern-Simons theories
\begin{equation}
\SO(2)_2
\xrightarrow{\text{gauging $\mathbb{Z}_2$}}
\O(2)_{2, L} = {\U(1)_{8} \times T_L \over \mathbb{Z}_2}
\end{equation}
where the theories $T_L, ~ L \textrm{ mod }8$ are described in Appendix \ref{app:O2}.

In the PH-Pfaffian regime, we have $L=1$, and the contribution from the bulk on one side is given by gauging a diagonal $\mathbb{Z}_2$ symmetry in the product of a left-moving compact boson $\varphi$ at the self dual radius and a right-moving Majorana fermion $\zeta^R$. The contribution from the other side of the wall is the same with left exchanged with right. Of course, the chiral anomaly of this sector from both sides of the wall is trivial. 

In the Pfaffian/anti-Pfaffian regime, an interface interpolating between the Pfaffian and anti-Pfaffian WZW theories differs from this basic one ($L = 1$) by precisely four additional Majorana fermions of the same chirality. On the Pfaffian side of the wall we have a left-moving compact boson and a left-moving Majorana fermion, while on the anti-Pfaffian side we have a right-moving compact boson and five right-moving fermions as appropriate for the theories with $L = -1, (3 \equiv -5)$, respectively. Both sides are again gauged by a single $\mathbb{Z}_2$ gauge field. We denote the discrete $\mathbb{Z}_2$ gauge field below as $a$, which implements a projection on the spectrum --- on the $\Psi^j$ as well as $\varphi, \zeta$.

Therefore, the domain wall theory before the contribution of the SPT-induced fermions is
\begin{equation}
S^{G}[z_0] + S^{{WZW}}[a, \zeta, \varphi]
\end{equation} in the obvious notation, where the superscript $WZW$ denotes the appropriate WZW model for a given phase. The theory for the fermions that the SPT phases deposit on the wall will now be derived using our previous analysis of bulk fermionic zero modes in Section~\ref{Sec:ZeroModes}.

\paragraph{Fermionic sector}  
Let us study the Pfaffian/anti-Pfaffian regime $m < g v$ in the extreme limit of $m = 0$. We take the normalizable zero-modes $\psi_{0, -}$ and promote them to worldvolume fields. 
We substitute the corresponding solutions in terms of two complex Weyl fermions (\ref{eq:ferm1}, \ref{eq:ferm2}) into the Lagrangian \eqref{PS} to obtain
\begin{equation}\label{eq:Seffsmall}
\mathcal{L}^\text{Pf/APf}[a, \Psi] \sim \tilde{\sigma} \sum_{i=1,2} \overline\psi^i_{0, -} (x, t) ({\ii} \slashed{D}_{a})
\psi^i_{0, -}(x, t) ,
\end{equation}
where all derivatives only run over the worldvolume coordinates $k=x, t$, and we have used the superscript to indicate that this is the domain wall theory that interpolates between the Pfaffian and anti-Pfaffian vacua. The coefficient of the kinetic term $\tilde{\sigma}$ \footnote{Although we call this coefficient $\tilde{\sigma}$, due to its formal similarity with $\sigma$ as computed in eqn (\ref{TreeTension}), we stress that it is not to be confused with the tension. The fermionic contribution to the tension will be computed in later subsections.} is given by the integral 
\begin{equation}
    \tilde{\sigma} = \int_{-\infty}^{\infty} \dd z \cosh(\mu (z-z_0)/\sqrt{2})^{-{2 \sqrt{2} g \over \sqrt{\lambda}}} 
    = {\sqrt{2\pi} \over \mu} {\Gamma({\sqrt2 g \over \sqrt\lambda}) \over \Gamma({1\over2}+{\sqrt2 g \over \sqrt\lambda})},
\end{equation}
which in the limit of small Yukawa coupling becomes
\begin{equation}
    \tilde{\sigma} \sim {\sqrt{\lambda} \over \mu g}.
\end{equation}
The gauge field couples to the fermions on the wall exactly as it did in the bulk.  Note that the WZW sector is almost decoupled from the fermions except for the $\mathbb{Z}_2$ gauging.

In the PH-Pfaffian regime $m > g v$, let us set $g = 0$ for simplicity, and use the plane wave solutions on opposite sides of the walls. Doing the respective integrals for the surviving zero-modes over the two half-spaces ($z \leq z_0, z \geq z_0$) then gives
\begin{align}\label{eq:Sefflarge}
&\mathcal{L}^\text{PHPf}[a, \Psi] 
\nonumber
\\
&= {m \over \mu} \left[ \psi^1_{0, +}(x, t)\psi^1_{0, -}(x, t) - \psi^2_{0, +}(x, t)\psi^2_{0, -}(x, t) \right]
+ {1 \over 2m } \sum_{i=1, 2} \bar{\psi}^i_{0}(x, t) ({\ii} \slashed{D}_{a})\psi^i_{0}(x, t) \, ,
\end{align}
where now the superscript indicates that the domain wall theory is for the PH-Pfaffian phase.\footnote{The appearance of ${1 / \mu}$ is not only expected by dimensional analysis. Recall from Section~\ref{Sec:ZeroModes} that the plane wave solutions on opposite sides of the wall overlap in the vicinity of the wall, where a mass coupling is possible. On the domain wall, the mass term is therefore proportional to the width of the wall, which is $1 / \mu$.}
The mass term gaps out the fermions at low energies, hence only the Goldstone and WZW sectors of the domain wall theory survives on the wall.

The analysis of the zero modes in the two extreme regimes also suggests a natural candidate domain wall theory (in the universality class of the theory) that describes the wall's phase transition: a 1+1$d$ $\mathbb{Z}_2$-gauged Gross-Neveu-Yukawa theory (suppressing the dependence of the fields on the worldvolume coordinates $(x, t)$)\footnote{Analogous studies and proposals of domain wall worldvolume theories were made in the context of domain walls in four-dimensional QCD at $\theta = \pi $ \cite{GKS}, or four-dimensional SU(2) Yang-Mills gauge theory at $\theta = \pi$ \cite{Wang2019obe1910.14664WYZ}.}

\begin{equation}
\mathcal{L}^{wall} = - {1 \over 2}(\partial \varphi)^2 + g_2^2 \varphi^2 -{g_4 \over 4}\varphi^4 -g_3 \phi \left(\bar{\psi}^1 \psi^1 - \bar{\psi^2}\psi^2 \right)  +  \sum_j \bar{\psi}^j (%\cred{}
{\ii} \slashed{D}_a) \psi^j.
\end{equation} Here, the condensation of the scalar $\sigma$ as we tune the scalar mass term implements the phase transition between the two regimes. If we canonically normalize the fermions in $\mathcal{L}^\text{PHPf}$, then the coefficient of the mass term becomes ${m^2 \over \mu}$, so that we set ${m^2 \over \mu} \sim {g_3 g_2 \over \sqrt{g_4}}$, which naively suggests $g_4 \sim \mu^2, g_3 \sim g_2 \sim m$. We defer a more detailed analysis for future work.

\subsection{One-loop effective action and tension}

Let us return to our $2+1d$ bulk theory and study the (Euclidean) effective action and the domain wall tension from integrating out fermions at one-loop.\footnote{Y. Lin thanks Chi-Ming Chang and David Simmons-Duffin for useful discussions.}
We ignore the $\mathbb{Z}_2$ gauging and revisit its effect towards the end.

Consider expanding the theory in transverse fluctuations around a saddle $\phi = \phi_0 + \chi$, where $\phi_0$ could be either the vacuum saddle $\phi_0 = \pm v$ or the domain wall saddle \eqref{eq:scalar}.  The matter part of the action then takes the form
\begin{equation}
    S_{bulk} = {2 \sqrt{2} \over 3}{\mu^3 \over \lambda} + S_{fluct} + S_{ct},
\end{equation}
where (suppressing the $2+1$d spacetime dependence of the fields)
\begin{equation}
\begin{aligned}
\label{SFluct}
S_{fluct} &= \int \dd^3x \, {1 \over 2}\left\lbrace \chi\left(-{\partial^2 \over \partial z^2} - \mu^2 + 3 \lambda \phi_0^2 \right)\chi + \lambda \left(\phi_0\chi^3 + {1 \over 4}\chi^4\right)\right\rbrace \\
& + \sum_{i=1,2} \overline\Psi^i \ii\slashed{D}_{\cal C} \Psi^i + + m ( \overline\Psi^1 \Psi^1 - \overline\Psi^2 \Psi^2 ) + \int dz \left\lbrace- g \sum_{i=1,2} (\phi_0 + \chi)\bar{\Psi}^i \Psi^i \right\rbrace
\end{aligned}
\end{equation}
is the action for the fluctuations. We will study the counterterms $S_{ct}$ below.

\subsubsection{Effective action at $\phi_0 = v$}

At one-loop order around the vacuum saddle $\phi_0 = v$, there are terms in the fluctuation action that contribute to $\langle \chi \rangle$ via tadpole diagrams:
\begin{equation}
-g \chi \bar{\Psi}^i \Psi^i + \lambda v \chi^3.
\end{equation}
We need to include counterterms to cancel the tadpole so that the location of the vacuum remains fixed, $\langle \phi \rangle = v$.  Explicitly,
\begin{equation}
\begin{aligned}
\label{SCounter}
S_{ct} &= - {1 \over 2} \delta_b \mu^2 \int \dd^3x \, \phi^2 - {1 \over 2} \delta_f \mu^2 \int \dd^3x \, \phi^2 \, ,
\\
\delta_b \mu^2 &= \lambda v \int^{\Lambda}{\dd^3 k \over (2 \pi)^3}{1 \over k^2 + \mu^2} \, ,
\\
\delta_f \mu^2 &= 2 g \int^{\Lambda}{\dd^3 k \over (2 \pi)^3}\left[{g+m/v \over k^2 + (g v + m)^2} + {g-m/v \over k^2 + (g v - m)^2} \right] \, ,
\end{aligned} 
\end{equation}
where $\delta_b$ and $\delta_f$ denote counterterms that arise from consideration of bosonic $\chi$ and fermionic $\Psi^i$ loops, respectively, and $\Lambda$ is a UV cutoff.

\begin{figure}[!h]
\centering
\includegraphics[width=.325\textwidth]{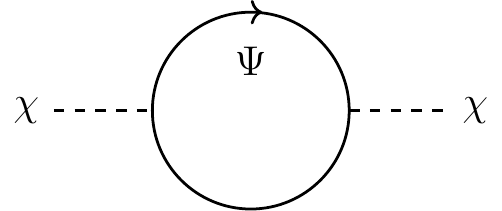}
%\\
%\begin{tikzpicture}
%\draw[thick,dashed] (1,0) -- (2,0) node [right] {$\chi$};
%\draw[thick,dashed] (-2,0) node [left] {$\chi$} -- (-1,0);
%\draw[thick,decoration={markings, mark=at position 0.25 with {\arrow{<}}}, postaction={decorate}] (0,0) circle (1);
%\node at (0,.5) {$\Psi$};
%\end{tikzpicture}
\caption{Fermionic one-loop renormalization of the fluctuating scalar mass.}
\label{Fig:One-Loop}
\end{figure}

The mass of the fluctuating $\chi$ field is given by $\sqrt2 \mu$ at tree level, but gets corrected at one-loop, with the Feynman diagram given in Figure~\ref{Fig:One-Loop}.  Since we want to focus on the effect of the fermions, let us ignore the bosonic loop corrections for now.  The one-loop effective action from integrating out two Dirac fermions with masses $\pm m$ and coupled to the scalar with Yukawa coupling $g$ is
\begin{equation}
  \delta_f {\cal L}_{eff} = \log { \det \mathbb{D}_{m, g}^{\phi = v + \chi} \over \det \mathbb{D}_{m, g}^{\phi = v} } { \det \mathbb{D}_{-m, g}^{\phi = v + \chi} \over \det \mathbb{D}_{-m, g}^{\phi = v} } - {1 \over 2} \delta_f \mu^2 \left[ (v + \chi)^2 - v^2 \right] \, ,
\end{equation}
where $\mathbb{D}_{m, g}^\phi$ is the effective Dirac operator
\begin{equation}
\label{Dirac}
\mathbb{D}_{m, g}^\phi \equiv \begin{pmatrix}
\displaystyle - \partial_1 + m + g v + g \chi & -\ii \partial_0 - \partial_2
\\
i \partial_0 - \partial_2 & \displaystyle \partial_1 + m + g v + g \chi & 
\end{pmatrix} \, .
\end{equation}
When $m \neq g v$, the leading terms in the derivative expansion amount to treating $\chi$ as a constant,
\begin{equation}
\begin{aligned}
\label{deltaLtadpole}
\delta_f {\cal L}_{eff} &= \int {\dd^3k \over (2\pi)^3} \Bigg\{ \log {k^2 + (m + g v + g \chi)^2 \over k^2 + (m + g v)^2} \, {k^2 + (m - g v - g \chi)^2 \over k^2 + (m - g v)^2} 
\\
& \hspace{.5in} - g [(v+\chi)^2 - v^2] \left[ {g + m/v \over k^2 + (m + g v)^2} + {g - m/v \over k^2 + (m - g v)^2} \right] \Bigg\}
\\
&= 
\begin{cases}
\displaystyle - {g^2 v^2 - m^2  \over 2\pi |g v|} (g \chi)^2
- {1 \over 3\pi} (g \chi)^3 & m < g v \, ,
\\
\displaystyle - {1 \over 6\pi} (g \chi)^3 & m = g v \, ,
\\
0 & m > g v \, .
\end{cases}
\end{aligned}
\end{equation}
We see that the mass of $\chi$ is renormalized as
\begin{equation}
\label{MassRenorm}
	2 \mu^2 = 2 \mu_{cl}^2 + {g^2 v^2 - m^2 \over \pi |g v|} g^2 \theta(g^2 v^2 - m^2) \, , \quad v = {\mu \over \sqrt\lambda} \, .
\end{equation}

In what regime can we trust this result?
Let us estimate this by computing some higher derivative terms in the one-loop effective action.  For a single Dirac fermion with mass $\bar m$, the first few higher derivative corrections quadratic in $\chi$ are
\begin{equation}\begin{aligned}
\delta_f S_{eff}^{(2)} &= - {g^2 \over 2} \chi \left( {1 \over 12 \pi |m-gv|} \partial^2 + {1 \over 240 \pi |m-gv|^3} \, \partial^4 + {\cal O}(\partial^6) \right) \chi + {\cal O}(\chi^3) \, ,
\end{aligned}\end{equation}
with some computational details given in Appendix~\ref{App:Determinant}.\footnote{To apply the results of Appendix~\ref{App:Determinant}, make the replacement $m \to m - gv$.}
Estimating the $\partial^2$ for bosonic fluctuations by the mass $\sqrt2 \mu$, we find that the higher derivative corrections are suppressed by factors of ${|m - g v| / \mu}$.
Thus our result is a reasonable approximation in the regime
\begin{equation}\begin{aligned}
\label{OneLoopRegime}
{|m - g v| \over \mu} \gtrsim {\cal O}(1) \, .
\end{aligned}\end{equation}
We will come back to this at the end of Section~\ref{Sec:Tension}.

\subsubsection{Domain wall tension}
\label{Sec:Tension}

To evaluate the one-loop corrections to the tension, we will closely follow the method of \cite{PY} (see also \cite{Rebhanetal}). First, we formulate the theory in a Euclidean box with half-length $L$ in the $z$ direction and area $V_{||}$ in the worldvolume directions, so that the energy density is given in terms of the effective action $\Gamma$ as 
\begin{equation}
\sigma = \textrm{lim}_{L, V_{||} \rightarrow \infty} {\Gamma(\phi_0) \over V_{||}} \, ,
\end{equation} 
under a scheme such that the expectation values of the vacua $\pm v$ are unrenormalized, and the effective action is normalized to vanish when evaluated on the vacua $\pm v$.

Formally, the full one-loop correction to this quantity receives contributions from the classical term $\sigma_{cl}$ (taking the form of \eqref{action0} with $\mu$ renormalized), 
the quantum correction $\sigma_{qu}$, and the counterterms $\sigma_{ct}$:
\begin{align}
\sigma &= \sigma_{cl} + \sigma_{qu} + \sigma_{ct} \\
&= \int \dd z \left(\mathcal{L}(\phi_0) - \mathcal{L}(v) \right) + \textrm{lim}_{L, V_{||} \rightarrow \infty} \left({1 \over 2 V_{||}}   \textrm{ln det} {\Delta \over \Delta^{(0)}}\right) \, .
\end{align}
The operators $\Delta, \Delta^{(0)}$ are the inverse propagators of a fluctuating field in the soliton background and in the vacuum (trivial background), respectively. The central idea of \cite{PY} is that the fluctuations are independent of the worldvolume coordinates and may therefore be partially diagonalized by a Fourier transform in those directions. Then, the ratio of functional determinants can be related to a ratio of solutions of ordinary differential equations, which is then (numerically) integrated over the transverse coordinates.\footnote{This bypasses numerous technical complications appearing in more traditional methods, and in particular provides a convenient way to deal with the regularization of sums of zero-point energies in different topological sectors. See, however, \cite{Dashenetal1, Shifman, CampbellLiao, Nastaseetal} for results in 1+1$d$ using analytic solutions of the fluctuation spectra and \cite{Grahametal1, Grahametal2, Grahametal3} for other approaches based on making successive Born approximations for scattering phase shifts.}

First, we express the one-loop tension in terms of the renormalized parameters of the theory. As is standard \cite{Shifman, Rajaraman, PY}, the renormalized scalar mass $\mu_{}$ can be related to the bare mass $\mu_{bare}$ by a one-loop computation in the perturbative sector of the fluctuation theory, {\it i.e.} in one of the two degenerate ground states. We follow \cite{PY} and use the $\bar{\textrm{MS}}$ scheme to fix the counterterms, and require that, as discussed above, the tadpole diagrams are cancelled by the counterterms. This coincides with the condition to fix the renormalized mass by requiring $\langle \phi \rangle = \pm \sqrt{ 6 \mu_{}^2 / \lambda} $.\footnote{The quartic coupling is only renormalized by a finite amount.}

The full one-loop tension can be broken up into a sum of the classical tension and the bosonic and fermionic one-loop contributions, of the form\footnote{The coefficient $\left[ {\delta_b\sigma_{qu} / \mu^2} \right]$ is also, in general, a function of the dimensionless scalar coupling $\lambda/\mu$. We take $\lambda/\mu$ to be small throughout our analysis for the semiclassical approximation, and just consider the leading order $\lambda$-independent contribution.}
\begin{equation}\label{FullTension}
\sigma = {2 \sqrt{2} \over 3} {\mu_{}^3 \over \lambda} + \left[{\delta_b\sigma_{qu} \over \mu^2}\right] \, \mu_{}^{2} + \left[{\delta_f\sigma_{qu} \over \mu^2}\right] 
\, \mu_{}^{2},
\end{equation} 
where $\mu$ is the {\it renormalized} scalar mass, $\left[{\delta_b\sigma_{qu} / \mu^2}\right]$ is a dimensionless constant, and $\left[{\delta_f\sigma_{qu} / \mu^2}\right]$ is a dimensionless quantity with the following functional dependence on a dimensionless fermion mass $w$ and a dimensionless Yukawa coupling $\gamma$,
\begin{equation}
\label{Dimensionless}
\left[ {\delta_f\sigma_{qu} \over \mu^2} \right] \left(w, \gamma \right), \quad w \equiv {m \over \mu}{\sqrt{\lambda} \over g}, \quad \gamma \equiv {g \over \sqrt{2\lambda}}.
\end{equation} 
The normalization of $w$ is chosen such that at $w=1$, the effective mass $m-gv$ of one of the Dirac fermions vanishes. The normalization of $\gamma$ is chosen so that $\gamma=1, \nu=0$ corresponds to the $\mathcal{N}=1$ supersymmetric point in the case of a theory with a single Majorana fermion.  

The classical piece of the domain wall tension in \eqref{FullTension}  with one-loop renormalized scalar mass is
\begin{equation}\label{ClassicalTension}\begin{aligned}
\sigma_{cl} &= {2\sqrt2 \over 3 \lambda} \left[\mu_0^2 + {g^2 v^2 - m^2 \over 2 \pi |g v|} g^2 \theta(g^2 v^2 - m^2) \right]^{3/2} \, ,
\end{aligned}\end{equation}
where the domain wall tension is renormalized at one-loop by the bosonic $\chi$ fluctuations alone. In relative terms, this correction is
\begin{equation}
    {\delta_f \sigma_{cl} \over \sigma_{cl}} \sim {g^3 v \over \mu^2} \sim {g^3 \over \mu \sqrt\lambda} \sim {\lambda \over \mu} \gamma^3 \, ,
\end{equation}
which is small in the semiclassical approximation with $\gamma \sim {\cal O}(1)$.
There is a first order transition in the domain wall tension at the critical point $m = g v$, but we expect it to be smoothed out by higher order corrections.

We can now determine $\left[ {\delta_b\sigma_{qu} / \mu^2} \right], \, \left[ {\delta_f\sigma_{qu} / \mu^2} \right]$ using the technology of \cite{PY}.  Since $\chi$ only self-interacts at one-loop, and since the relevant computation was performed in \cite{PY}, we can simply borrow their result, which was computed in dim-reg in terms of the analytically continued dimension $n$, and take $n \rightarrow 3$. The result is
\begin{equation}
\left[ {\delta_b\sigma_{qu} \over \mu^2} \right] = {3 \mu^2 \over 16\pi}\left(\textrm{log}(3)-4 \right) \sim -0.17.
\end{equation}

It remains to determine the integral encapsulating the quantum fermionic contributions to the tension, following \cite{PY}. As always, we would like to keep $\lambda/\mu$ small. In addition, we also want the Yukawa coupling to be small to suppress large backreaction by the fermions, but we can keep the ratio $g^2/\lambda$ finite.  Of course, when $g=0$, $\delta_f\sigma_{qu} = 0$.\footnote{An analogous computation performed in a supersymmetric theory with a single Majorana fermion and $g = \sqrt{2 \lambda}$ in \cite{Rebhanetal} gives $\delta_b \sigma_{qu}^{SUSY} + \delta_f \sigma_{qu}^{SUSY} = - {\mu^2 / 4 \pi}$. We reproduce this result.}

\begin{figure}[H]
\centering
\includegraphics[width=.7\textwidth]{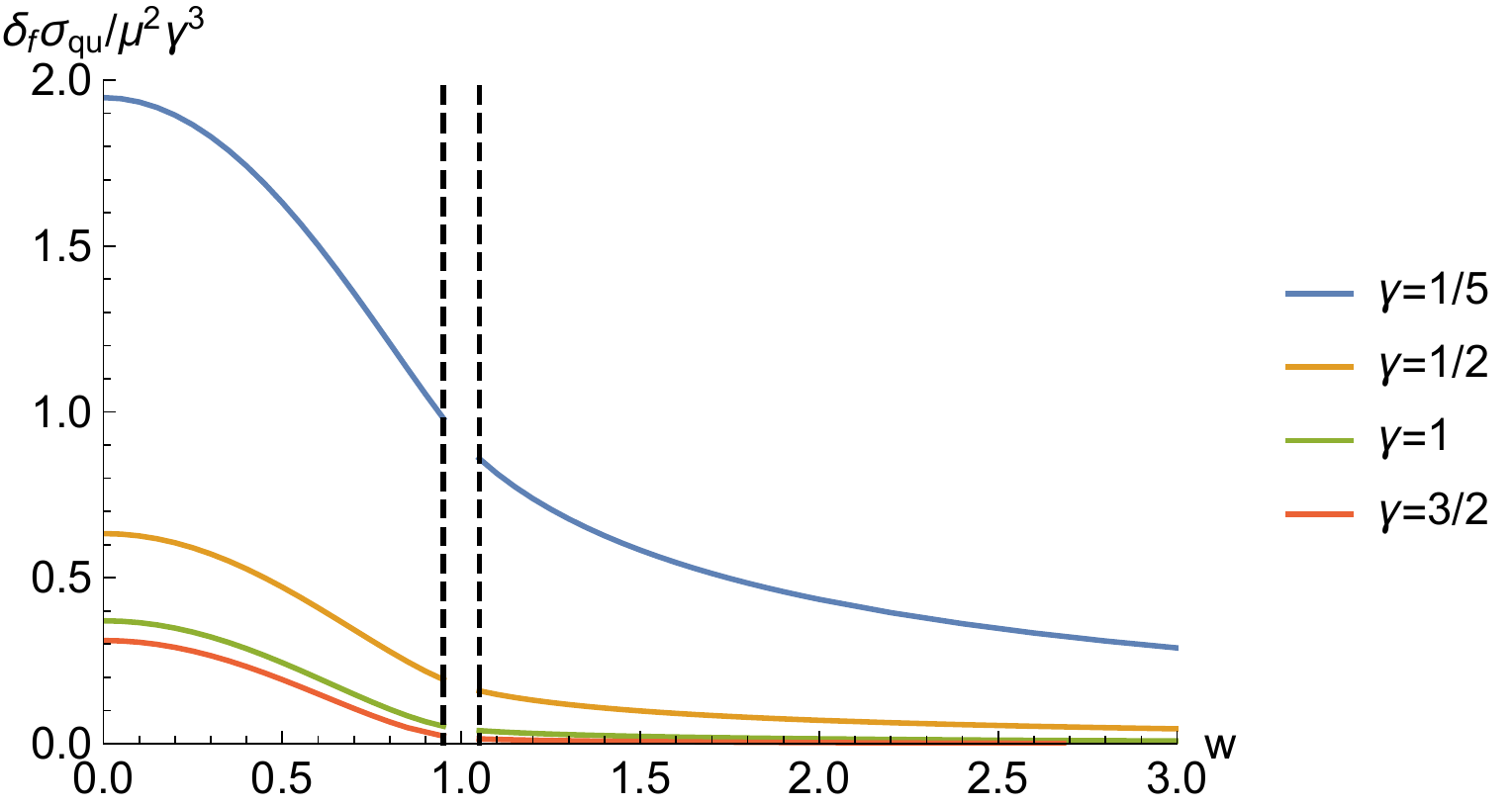}
\\
~
\\
\includegraphics[width=.72\textwidth]{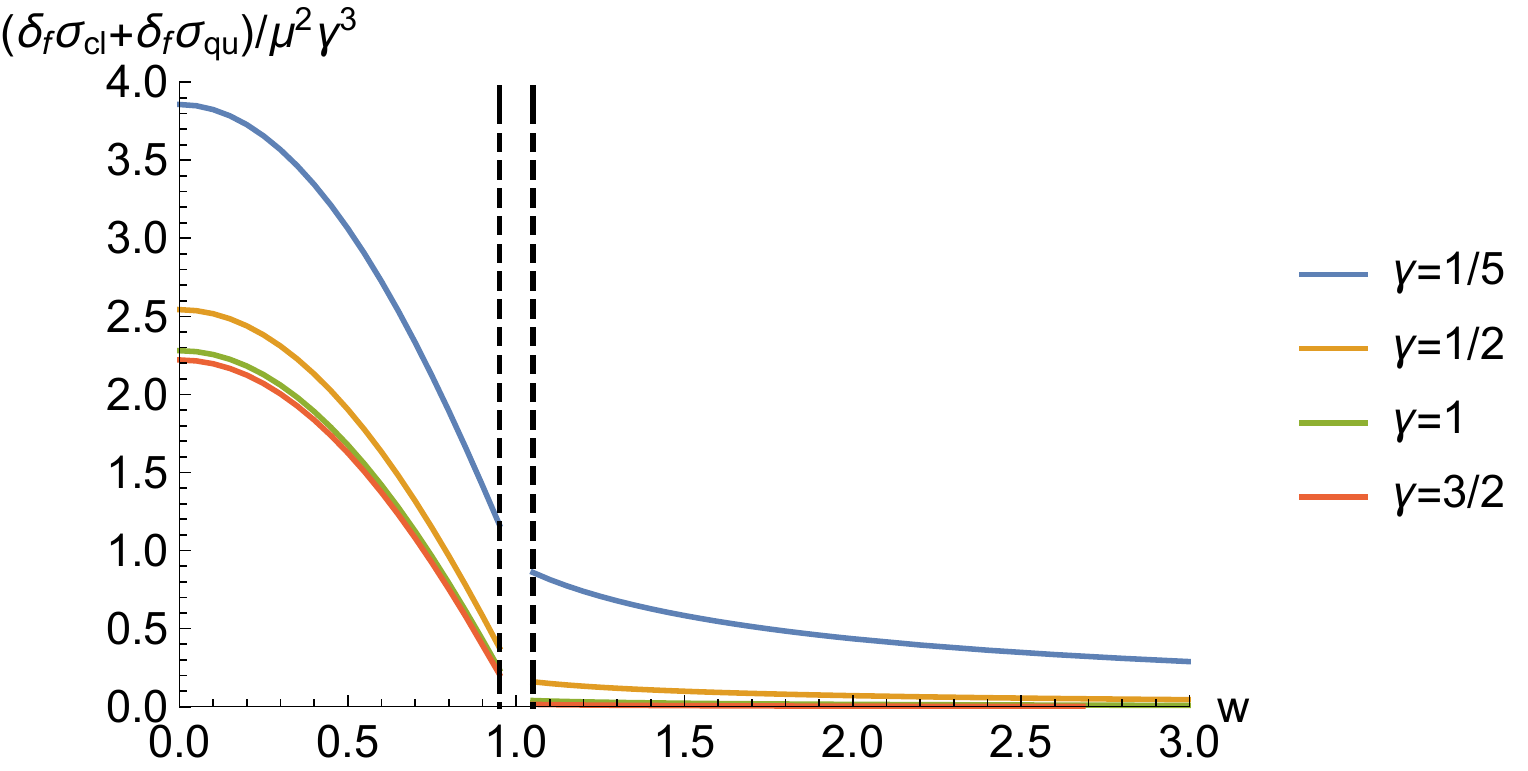}~~~~~
\caption{{\bf Top:} The fermionic {\it quantum} one-loop correction $\delta_f \sigma_{qu}$ to the domain wall tension. {\bf Bottom:} 
The mass renormalization and {\it quantum} one-loop correction combined.
{\bf Note:} The dimensionless fermion mass $w$ and Yukawa coupling $\gamma$ are defined in \eqref{Dimensionless}.  Notice that we have divided out by $\mu^2 \gamma^3$.  The region near $w = 1$ ($m = g v$) is blocked out by dashed lines because the fermions become light and higher derivative corrections become important.}
\label{Fig:Plot3d}
\end{figure}

Including the counterterm \eqref{SCounter}, the formula for the {\it quantum} one-loop tension from integrating out the fermions is
\begin{equation}\begin{aligned}
\label{deltaL}
\delta_f \sigma_{qu} &= - \log { \det \mathbb{D}_{m, g}^{\phi = \phi_0} \over \det \mathbb{D}_{m, g}^{\phi = v} } { \det \mathbb{D}_{-m, g}^{\phi = \phi_0} \over \det \mathbb{D}_{-m, g}^{\phi = v} }
\\
& \hspace{.5in} 
+ F \int^\Lambda {\dd^3k \over (2\pi)^3} \left[ {g(g+m/v) \over k^2 + (gv + m)^2} + {g(g-m/v) \over k^2 + (gv - m)^2} \right] \, ,
\end{aligned}\end{equation}
where
\begin{equation}
\mathbb{D}_{m, g}^{\phi=\phi_0} \equiv \begin{pmatrix}
\displaystyle - \partial_z + m + g \phi_0(z) & -\ii \partial_0 - \partial_2
\\
i \partial_0 - \partial_2 & \displaystyle \partial_z + m + g \phi_0(z) & 
\end{pmatrix} \, ,
\end{equation}
and
\begin{equation}\begin{aligned}
F = \int \dd z \left[ \phi_0(z)^2 - {\mu^2 \over \lambda} \right] = - 2\sqrt2 \, {\mu \over \lambda} \, .
\end{aligned}\end{equation}
This formal expression \eqref{deltaL} can be evaluated explicitly as outlined in Appendix~\ref{Sec:Numerics}.  The results are shown in Figure~\ref{Fig:Plot3d}, expressed in terms of the dimensionless variables $\nu$ and $\gamma$ defined in \eqref{Dimensionless}.  Some notable features are
\begin{itemize}
    \item The {\it quantum} one-loop correction to the tension $\delta_f \sigma_{qu}$ is of the same order as the correction from one-loop mass renormalization, namely: $\delta_f \sigma_{qu} \sim \delta_f \sigma_{cl} \sim {\cal O}(\mu^2 \gamma^3)$.  
    \item  Both are monotonically decreasing with respect to the fermion mass $m$.
    \item The effect diminishes rapidly once the mass $m$ increases past the critical point $m = g v$.  Note that there is no mass renormalization at all for $m > g v$.
\end{itemize}

\begin{figure}[!h]
\centering
\includegraphics[width=.6\textwidth]{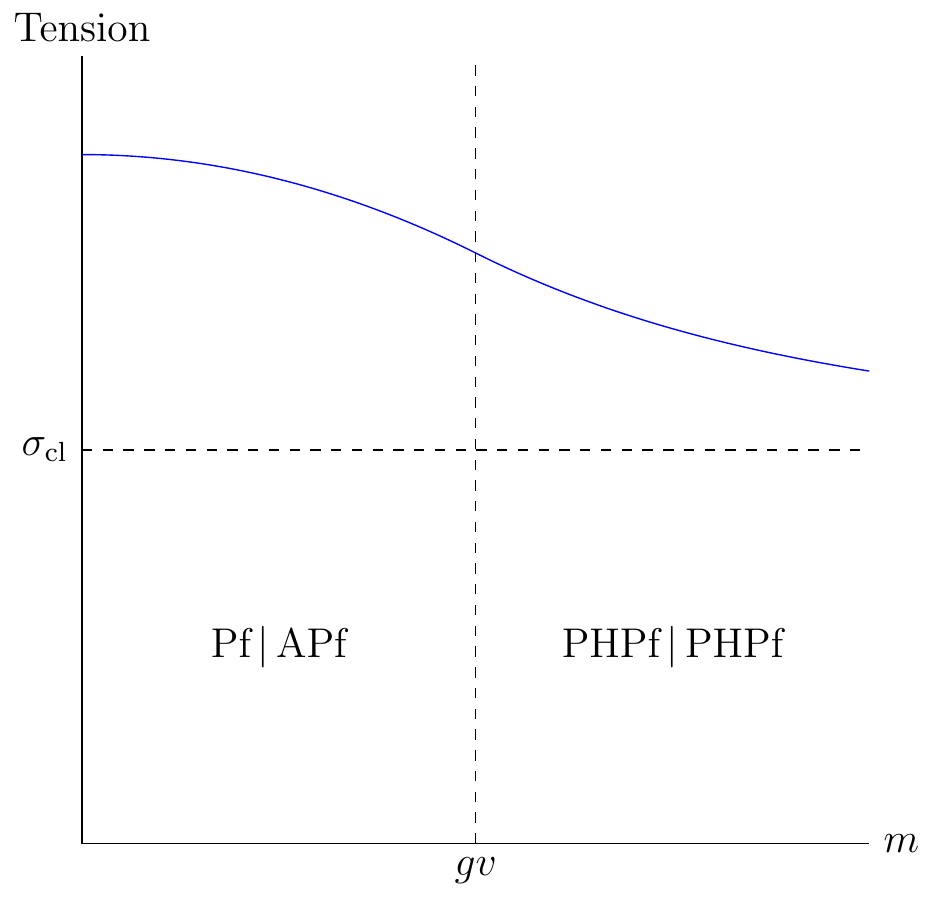}
%\\
%\begin{tikzpicture}[scale=4]
%\node at (.5,.5) {Pf\,$\mid$\,APf};
%\node at (1.5,.5) {PHPf\,$\mid$\,PHPf};
%\draw (0,0) -- (2,0) node [right] {$m$};
%\draw (0,0) -- (0,2) node [above] {Tension};
%\draw [dashed] (1,0) node [below] {$gv$} -- (1,2);
%\draw [dashed] (0,1) node [left] {$\sigma_\text{cl}$} -- (2,1);
%\draw [scale=1,domain=0:1,smooth,variable=\x,blue] plot ({\x},{1.75-.25*\x*\x});
%\draw [scale=1,domain=1:2,smooth,variable=\x,blue] plot ({\x},{1+1/(1+\x*\x});
%\end{tikzpicture}
\caption{Qualitative dependence of the domain wall tension on the fermion mass $m$, which is a proxy for the disorder strength $\Lambda$.  The critical line $m = gv$ separates the Pf/APf and PHPf regimes, and $\sigma_{cl}$ denotes the semi-classical tension in the absence of fermions.}
\label{Fig:Tension}
\end{figure}

The validity of the lowest order approximation in the derivative expansion was analyzed earlier, and the estimate \eqref{OneLoopRegime} translated to dimensionless quantities becomes
\begin{equation}\begin{aligned}
|w - 1| \gtrsim {{\cal O}(1) \over \gamma} \, .
\end{aligned}\end{equation}
As long as $\gamma$ is ${\cal O}(1)$, the one-loop tension to leading order in the derivative expansion is a valid approximation when $m$ is sufficiently large.  Furthermore, if the Yukawa coupling $g$ is large enough relative to $\lambda$, then we can also trust our results in some neighborhood of small $m$.\footnote{This may seem paradoxical at first, since the fermions clearly have no effect when $g = 0$.  However, we are interested in the dependence of the tension on the fermion mass $m$.  When $g$ is small, this dependence is small, but the higher order corrections relative to the approximate dependence is large.
}  Finally, we have assumed that the couplings $\lambda$ and $g$ are small relative to the masses $w$ and $m$, therefore the higher loop corrections and corrections involving more powers of $\chi$ are suppressed.

\subsubsection{The effect of gauging}

Let us discuss the effect of the $\mathbb{Z}_2$ gauge field on the one-loop tension. Recall that the $\mathbb{Z}_2$ gauge field acts as $\Psi^j \rightarrow -\Psi^j$ and leaves the scalar untouched. 
In the path integral, having a discrete gauge field amounts to summing over its holonomies.  For a domain wall interpolating between two vacua, the Euclidean spacetime is $\mathbb{R}^3$ with no boundary or nontrivial cycle, so it is unclear whether the gauging has any effect on the tension at all, even non-perturbatively.  On a spacetime with nontrivial cycles, it is logically possible that the sum over holonomies introduces new saddles that dominate over the original saddle (trivial holonomy), but such effects go beyond perturbation theory.

In fact, for the sake of argument, let us imagine that the $\mathbb{Z}_2$ is a subgroup of a continuous $\U(1)$ gauge symmetry acting on the fermions as $\Psi^j \rightarrow \e^{\ii\alpha} \Psi^j$, with associated gauge connection $A$.  
In the one-loop effective action from integrating out the fermions, to lowest order in the derivative expansion there in principle is a coupling of the form $(\partial^\mu \chi) \, A_\mu$.  However, we find that the coefficient of this term is zero by an explicit computation in Appendix~\ref{App:Determinant}.  Thus the $\U(1)$ gauging has no effect at the order of our approximation.

\section{Conclusion and future directions}
\label{sec:conclusion}

In this work, we have presented an effective field theory that captures the qualitative features of the phase diagram proposed in \cite{Lian:2018xep} to describe the $\nu=5/2$ fractional Quantum Hall system. We also studied some simple properties of domain walls present at the time-reversal-symmetric locus (or, in condensed matter terminology, the particle-hole-symmetric locus), including their effective worldvolume theory and their tension, computed to one-loop order in a semiclassical approximation. The tension computed with the EFT is {\it lower} in the PH-Pfaffian phase than in the Pfaffian/Anti-Pfaffian phase, suggesting that the former phase may in fact be energetically favored over the latter in the presence of domain walls.  This may explain the percolation transition, and serve as a resolution of the dilemma between the experiment \cite{Banerjee:2018qtz} (favoring PH-Pfaffian) and bulk energetics studies \cite{morf1998, rezayi2000, peterson2008, feiguin2009, wangh2009Haldane, storni2010DasSarma, rezayi2011Simon, papic2012HaldaneRezayi, zaletel2015MongPollmannRezayi, pakrouski2015PetersonNayak}
(favoring Pfaffian/Anti-Pfaffian). We leave a more exhaustive and complete study of bulk/domain wall energetics to future work.

%\color{red}
We make some additional remarks related to the bulk and domain wall systems:
\begin{enumerate}

\item Pf$\mid$APf \emph{domain wall vs} PHPf$\mid$PHPf \emph{domain wall}:\\
The Pf$\mid$APf domain wall is well-defined when particle-hole (PH) symmetry is broken, with two different vacua on two sides of bulk. But what do we mean by
PHPf$\mid$PHPf domain wall since PHPf presumably 
has a PH-symmetry preserving bulk?\footnote{We thank Dam T.~Son for raising this question.} An answer is that in our EFT,
the two vacua of the Higgs potential shown in \Fig{fig:vacua} indeed both break the PH symmetry: 
\begin{itemize}

\item The left side of \Fig{fig:vacua} gives vacua of the PH-symmetry breaking phases Pf and APf.

\item
The right side of \Fig{fig:vacua} gives two vacua both in the PHPf phase, but the two PHPf vacua are exchanged by PH-symmetry. We could have alternatively considered the Higgs potential with
the sign of the $\phi^2$ term flipped when we are in the PHPf phase --- if so, we would have a single PH-symmetry preserving vacuum. But to impose that the $\phi^2$ term flips sign \footnote{For fine-tuning, one can let $\frac{\mu^2}{2}\phi^2$ such that
$\mu^2 >0$ when $m<gv$, while
$\mu^2 <0$ when $m>gv$.} 
around the energy scale $m=gv$ would require a less natural fine-tuning on our EFT. However, the fine tuning could potentially be avoided in the following way. Consider the region near $m=gv$, where the $\phi^2$ term sign does not flip ``by hand'',
but rather the sign flip is induced by the following mechanism.
The PH-symmetry breaking Pf$\mid$APf domain wall
can transition to the PH-symmetry breaking PHPf$\mid$PHPf domain wall,
but the $1+1d$ PHPf$\mid$PHPf domain wall percolation may then induce a $2+1d$ bulk transition
to the PHPf phase, which would restore the PH-symmetry dynamically. We propose that this mechanism indeed occurs, while our proposal requires future study.
\end{itemize}

\item We contrast \emph{the order of quantum phase transitions} 
from two perspectives, (i) the domain wall percolating picture
 \cite{Lian:2018xep} versus (ii) our EFT:\\ 
$\bullet$ At zero disorder $\Lambda=0$, both (i) and (ii) have a first order discontinuous phase transition due to PH (or $CT$) breaking and the discontinuity jump at domain wall.\\
$\bullet$ At nonzero but small disorder at $\Lambda < \Lambda_1$ along the vertical axis $\nu=\nu_c$, 
the case (ii) gives a first order transition, while the case (i) can have a second order transition,
or a first order transition effected by disorder broadening the phase boundary (which can result
in a new second order transition within the broadening region). \\
$\bullet$ Away from the axis $\nu \simeq \nu_c$ but within 
topological orders in the phase diagram \Fig{fig:Phase-Pf-APf-phase},
we have second order or continuous transitions derived in our EFT for the case (ii) due to the continuous deformation of mass sign flipping. 
In the case (i),
the percolation phase transition can be indeed a second order transition,
at least in the free fermion limit --- 
chiral fermions running on the percolation domain walls where the
length scale of puddle \emph{diverges} at the transition.\\
$\bullet$ The upper phase boundary $\Lambda > \Lambda_2$
from the topological orders to the thermal metal is also
a second order or continuous transition, for both cases (i) and (ii).
Physically this transition is similar to an insulator-metal transition driven by strong disorder
known as \emph{Anderson localization-delocalization transition}, which is a second order continuous transition.

\item Our EFT should encode the \emph{universality class of gapless topological quantum phase transitions}. In our case, we have a second-order continuous 
phase transition controlled by a 
free CFT given by 2+1d Dirac or Majorana fermions whose masses flip sign. Naively, the O(2) gauge sector does not directly affect the dynamics and universality class. It will be important to explore further the nature of the phase transitions.

\item We may contrast the emergent global symmetries on the $1+1d$ domain wall (from lower to higher disorder: 
$\O(4) \to \O(2) \times  \O(2) \to \Z_2^F$ \cite{Lian:2018xep})
with the $2+1d$ bulk global symmetries of our $2+1d$ EFT (from lower to higher disorder: $ \frac{\O(4)}{\Z_2} \to \frac{\O(2) \times  \O(2)}{\Z_2} \to \Z_2^F$).
The two global symmetry patterns are almost equivalent, but differ by finite group sectors. It turns out that we can also formulate an alternative EFT (by modifying the deformation parameters of the original QFT) to exactly match the bulk and domain wall global symmetry patterns. This will be left for further exploration.

\item 
Our EFT can choose either 
two versions PH-Pfaffian$_{\pm}$, depending on how $(CT)^2$ assigns to
the anyons (they are the same topological order, but different symmetry-enriched
topological orders). PH-Pfaffian$_{+/-}$ have different anomaly,
with an anomaly index $\upnu =0$ or $8\in \Z_{16}$,  
of $CT$ symmetry. If the $\nu=5/2$ system in the lab has
 PH-Pfaffian$_{+/-}$ order, then we have less/more IR constraints from the $\upnu =0$ or $8$ anomaly. Moreover, other than this pure $CT$-anomaly, there is also a $CT$-PO(4) mixed anomaly.
 The 't Hooft anomaly implies that the associated global symmetry (such as $CT$) is not strictly 
 local and onsite, but it is an \emph{emergent global symmetry} at low energy and long distances.

\end{enumerate}

%\color{black}

We conclude with an incomplete list of additional  questions and future directions raised by this study. Of course, most interesting is whether the proposal of \cite{Lian:2018xep} indeed provides the correct microscopic description of the $\nu=5/2$ state. If so, we hope our EFT provides a useful conceptual framework for studying aspects of this system. 

\begin{enumerate}

\setcounter{enumi}{5}

\item Our effective field theory is a standard relativistic QFT, though various non-relativistic EFTs have been proposed to study quantum Hall systems (see {\it e.g.} \cite{son2015, halperin1993}).
Is there a useful non-relativistic bulk EFT description of this system?

It is worthwhile to note that our EFT is a (super-)renormalizable QFT in $2+1d$, and it is UV complete by itself. Although
our EFT does not require a further UV completion at higher energy, it may still be helpful to understand how this relativistic EFT can be obtained from RG flow from a non-relativistic EFT, the electron wavefunctions, or a lattice model 
at the condensed matter UV cutoff scale.

\item We computed the tension of the domain walls in an approximation where $\lambda/\mu <<1$. Roughly speaking, $\lambda$ and $\mu$ respectively govern the height and width of the domain walls, so that the limit corresponds to studying rigid and thick walls. It would be interesting to determine if the domain walls, assuming they are indeed realized in the $\nu =5/2$ system, actually satisfy this limit so that our tension result can be used reliably to understand energetics of the system.

\item It may be that the particle-hole symmetry is explicitly, weakly broken in the experimental setup. If so, the domain walls would be metastable. It would be instructive to compute the decay rate for these walls in our EFT when one turns on our small but non-vanishing $m_{odd}$ deformation.

\item It would be very instructive to compute the spin-structure-dependent ground state degeneracy by performing an explicit path integral in our EFT. Such a quantity could potentially be measured in a real experimental setting if one fixes the boundary conditions of the lab sample, {\it i.e.} periodic or anti-periodic boundary conditions, similar to those on a spatial 2-torus.

\end{enumerate}

\section{Acknowledgements}

The authors are listed in the alphabetical order by a standard convention.
We thank Bert Halperin for a conversation,
and Nathan Seiberg and Xiao-Gang Wen for comments on the manuscript.
JW thanks Biao Lian for a previous collaboration on \cite{Lian:2018xep} and especially acknowledges helpful comments from Jie Wang, Yizhuang You, and Yunqin Zheng. JW also thanks e-mail correspondences from David Mross and Chong Wang, and the feedback from the seminar attendees \cite{JW-seminar1}.
YL and NP are each supported by a Sherman Fairchild Postdoctoral Fellowship. JW was supported by
NSF Grant PHY-1606531 and Institute for Advanced Study.
This work is also supported by 
NSF Grant DMS-1607871 ``Analysis, Geometry and Mathematical Physics'' 
and Center for Mathematical Sciences and Applications at Harvard University. This material is based upon work supported by the U.S. Department of Energy, Office of Science, Office of High Energy Physics, under Award Number DE- SC0011632, and by the Simons Foundation through the Simons Investigator Award.

\appendix

\section{Gravitational Chern-Simons term and thermal Hall response}

\label{app:gravCS}

Any 3-manifold has a spin connection $\omega$.\footnote{
Explicitly, in terms of the frame metric $\eta_{ab}=g_{\mu\nu}e^\mu_a e^\nu_b$ and the coframe $\bar e^a_\mu e^\mu_b=\delta^a_b$
\begin{equation}
\omega^i_{j\mu}=\bar e^{i}_\nu\Gamma^{\nu}_{\lambda\mu}e^\lambda_j+\bar e^i_\nu \partial_\mu e^\nu_j~,
\end{equation}
where $\Gamma$ is the Christoffel symbol.
} The fermion spinor field of spin $1/2$ couples to the spin connection as $\nabla=\partial-\frac{1}{2}\omega$. Integrating out one massive Majorana fermion $\psi$ gives the gravitational spin Chern-Simons term for positive mass $m$ compared to negative mass:
\begin{equation}
\frac{Z_{\psi,m\gg0}}{Z_{\psi,m\ll 0}}=
\exp\left(\ii\int_{M_3}\text{CS}_\text{grav}{\dd^3x}\right)~.
\end{equation}
More explicitly,
\begin{equation}
\text{CS}_\text{grav}{\dd^3x}=\frac{1}{192\pi}\text{Tr }\left(\omega d\omega+\frac{2}{3}\omega^3\right)~.
\end{equation}
The spin gravitational Chern-Simons term contributes to the thermal Hall conductivity by a chiral central charge $c_-=-1/2$.\footnote{
This can be understood from the fact that the invertible TQFT $U(1)_{-1}$ has partition function $\e^{2\ii\int \text{CS}_\text{grav}}$ \cite{1602.04251SW}, and thus $2\text{CS}_\text{grav}$ has $c=-1$, so $\text{CS}_\text{grav}$ has $c=-1/2$.}

\section{$\mathbb{Z}_2$ gauge theory in $2+1d$}
\label{app:Z2-gauge}

Fermionic SPT phases with an internal unitary $\mathbb{Z}_2$ symmetry are known to be classified by $\Omega^\text{spin}_3=\mathbb{Z}_8$.
Denote the background $\mathbb{Z}_2$ gauge field by $B\in H^1(M,\mathbb{Z}_2)$ for the spacetime $M$. Then the partition function for the $\mathbb{Z}_2$ SPT phases can be described using the invertible fermionic TQFT $\SO(L)_1$ with a special orthogonal $\SO(L)$ gauge group as follows:
\begin{equation}
\e^{\ii f_L(B)}=\left( Z_{\SO(L)_1}[B]\right)^*Z_{\SO(L)_1}~,
\end{equation}
where $Z_{\SO(L)_1}[B]$ denotes the partition function of $\SO(L)_1$ coupled to $B$ by the magnetic symmetry $\pi\int w_2(\SO(L))\cup B$, while $\left( Z_{\SO(L)_1}[B]\right)^*$ is its complex conjugate.
Since $\SO(L)_1$ is an invertible spin TQFT, the right hand side is a phase that depends on $B$, which gives the SPT phase $f_L(B)$ on the left hand side. By the property $\SO(L)_1\times \SO(L')_{1}\leftrightarrow \SO(L+L')_1$, the phase can be written as 
\begin{equation}
f_L(B)=L\, f(B)~
\end{equation}
for some $f(B)$.

For an even $L$, we can use the property $\SO(2)_1=\U(1)_1$ to express $Lf(B)=(L/2)\cdot 2f(B)$ as the  $\U(1)\times \U(1)$ Chern-Simons term $-\frac{L/2}{4\pi}B\dd B+\frac{2}{2\pi}B\dd u$ with $u$ constrains $B$ to be a $\mathbb{Z}_2$ gauge field.
By the field redefinition $u\rightarrow u+B$, we find that for $L=8$ the SPT phase is the same as $L=0$.
This reproduces the $\mathbb{Z}_8$ classification of the SPT phases
\begin{equation}
L\sim L+8~.
\end{equation}

Gauging the $\mathbb{Z}_2$ symmetry with a dynamical gauge field by summing over $B$ gives rise to 8 different $\mathbb{Z}_2$ gauge theories. For $L=0$ it is the untwisted $\mathbb{Z}_2$ gauge theory (the $\mathbb{Z}_2$ toric code), while for $L=4$ it is the Dijkgraaf-Witten twisted $\mathbb{Z}_2$ gauge theory (the so-called double semion theory). See the list of 8 different $\mathbb{Z}_2$ gauge theories (where $L \in$ even yields an abelian TQFT and
$L \in$ odd yields a non-abelian TQFT)
in Table 2 of \cite{Putrov2016qdo1612.09298PWY}.

\section{Fermion path integral and counterterms}
\label{app:FermionPathIntegral}

Consider a $2+1d$ Majorana fermion coupled to a $\mathbb{Z}_2$ gauge field  $B\in H^1(M,\mathbb{Z}_2)$, and give it a large mass. The fermion path integral depends on the sign of the mass, given by
\begin{equation}\label{eqn:fermionpathintegral}
Z[B]_{m>0}=|Z|\exp\left(\frac{\pi \ii}{2}\eta(B)\right),\quad Z[B]_{m<0}=1~.
\end{equation}
The Atiyah-Patodi-Singer (APS) index theorem relates the exponential of the eta invariant to the topological actions
\begin{equation}\label{eqn:apsindex}
\exp\left(\frac{\pi \ii}{2}\eta(B)\right)=\exp\left(\ii f(B)+\ii\int\text{CS}_\text{grav}{\dd^3x}\right)~,
\end{equation}
where $f(B)$ is the basic fermionic $\mathbb{Z}_2$ SPT phase with a  $\mathbb{Z}_2$ background $B$, and $\int_{\partial Y}\text{CS}_\text{grav}{\dd^3x}=\frac{1}{192\pi}\int_Y\text{Tr}(R\wedge R)$ is the gravitational Chern-Simons term.
There are 8 fermionic SPT phases with $\mathbb{Z}_2$ symmetry $8f[B]\sim 0$ mod $2\pi$, and they correspond to the 8 pure $\mathbb{Z}_2$ gauge theories in $2+1d$ (some of them need a spin structure). We will call them the 8 levels of $2+1d$ $\mathbb{Z}_2$ gauge theories; see Appendix \ref{app:Z2-gauge}.
In our convention, the $\U(1)_4\times \U(1)_{-1}$ theory corresponds to the $6$th class.

The $\O(2)$ Chern-Simons gauge theory has two levels: $\O(2)_{K,L}$ with the level $K\in\mathbb{Z}$ associated with the instanton number in $4d$, while $L$ represents $8$ $\mathbb{Z}_2$ gauge theories $Lf(w_1^{\O(2)})$, where $w_1^{\O(2)}$ is the $\mathbb{Z}_2$-valued first Stiefel Whitney class of the $\O(2)$ bundle.

For massless Majorana fermions in the one-dimensional representation odd under $\mathbb{Z}_2$ charge conjugation, we will write the theory using the effective Chern-Simons levels
\begin{equation}
\O(2)_{K,L}\;\text{CS}+N_f\psi\text{ in }\mathbf{1}_\text{odd}+M\text{CS}_\text{grav}~,
\end{equation}
where $M,L$ are integers if $N_f$ is even, and half-integers if $N_f$ is odd.
Integrating out a massive $2+1d$ Majorana fermion shifts the effective Chern-Simons level to be
\begin{equation}
m>0:\; \O(2)_{K,L+\frac{1}{2}}\;\text{CS}+(M+\frac{1}{2})\text{CS}_\text{grav},\qquad
m<0:\; \O(2)_{K,L-\frac{1}{2}}\;\text{CS}+(M-\frac{1}{2})\text{CS}_\text{grav}~.
\end{equation}
The difference between the shifts for different signs is given by (\ref{eqn:fermionpathintegral}) and (\ref{eqn:apsindex}).

\section{Gauging one-form symmetry in $2+1d$ TQFT}
\label{app:gaugingoneform}

Here we review some rules for gauging a one-form symmetry in 2+1$d$ TQFT.
For gauging a $\mathbb{Z}_2$ one-form symmetry generated by the symmetry generator charge line $a$ of integer spin, the rules are (see {\it e.g.} \cite{Moore:1988ss,Moore:1989yh,Bais:2008ni,Hsin:2018vcg}):
\begin{itemize}
\item Discard the lines that transform non-trivially under the one-form symmetry. These are the lines (of objects charged under the one-form symmetry) that braid non-trivially with $a$.
\item Identify every remaining line $W$ with its fusion with $a$: $W\sim W\cdot a$.
\item For the remaining lines that are fixed points under fusion with $a$, there are two copies of the line.
\end{itemize}
In the corresponding chiral algebra, the procedure is equivalent to extending the chiral algebra by a simple current that obeys, together with the identity, the $\mathbb{Z}_2$ fusion algebra.

%\fixme{PH}
For instance, $\U(1)_{8k}$ Chern-Simons theory has a $\mathbb{Z}_2$ subgroup one-form symmetry generated by the Wilson line of charge $4k$ with integer spin. The above procedure produces $\U(1)_{2k}$ Chern-Simons theory after gauging one-form $\mathbb{Z}_2$ symmetry. Another way to obtain the result is that gauging the one-form symmetry makes the original $\U(1)$ gauge field $b$ no longer well-defined, but $b'=2b$ is a well-define $\U(1)$ gauge field. Expressing the original Chern-Simons term $\frac{8k}{4\pi}b\dd b$ in terms of the $\U(1)$ gauge field $b'$ gives $\U(1)_{2k}$.

% \cred{For example, gauging a one-form $\Z_k$ symmetry in $2+1d$ $\U(1)_{k^2 n}$ CS gives rise to a new gauge theory $\U(1)_{n}$ CS.}

\section{O(2)$_{2,L}$ Chern-Simons theories}
\label{app:O2}

In this Appendix, we summarize the $2+1d$
O(2)$_{2,L}$ gauge theories,
which
are fermionic spin 
Chern-Simons theories definable on spin manifolds.
For the zero level of  $2+1d$ $\mathbb{Z}_2$ gauge theories (written in terms of $\O(2)_{K,L}$ gauge theories with $L \in \mathbb{Z}_8$ levels in the previous Appendix), the $\O(2)_{2,0}$ gauge theory has the same chiral algebra as the $\U(1)_8$ gauge theory, and thus we have
\begin{equation}
\O(2)_{2,0}\;\text{CS}\leftrightarrow \U(1)_8\;\text{CS}~.
\end{equation}
In general, we denote the $L$th $\mathbb{Z}_2$ gauge theory by $T_L$ (with an action $Lf[B]$ for the $\mathbb{Z}_2$ gauge field $B$),
and have the equivalence
\begin{equation} \label{eq:O2LtoU18TLmodZ2}
\O(2)_{2,L}\;\text{CS}\leftrightarrow \frac{\U(1)_8\times T_L}{\mathbb{Z}_2}\;\text{CS}~,
\end{equation}
where the quotient denotes gauging a diagonal $\mathbb{Z}_2$ one-form symmetry generated by the composite line of the tensor product of the charge 4 Wilson line in $\U(1)_8$ and the non-transparent fermion line in $T_L$ (if we express $T_L\leftrightarrow \Spin(L)_{-1}\times \SO(L)_1$, it is the Wilson line in the vector representation of $\Spin(L)$).\footnote{
Here $T_L$ can be written as
another $\mathbb{Z}_8$ class of fermionic spin TQFTs \cite{1602.04251SW, Putrov2016qdo1612.09298PWY}
 as the $\Spin(L)_{-1}\times SO(L)_1$ Chern-Simons gauge theory in 2+1$d$ \cite{Cordova:2017vab}.
Explicitly, we can
express the relations with the following CS theories:
\begin{eqnarray}
T_1  &\leftrightarrow& \overline{\text{Ising}}\times \text{(spin-Ising)},\;\,\\
T_2 &\leftrightarrow& \U(1)_{-4}\times \U(1)_1 \simeq \text{K-matrix $\left(
\begin{array}{cc}
 0 & 2 \\
 2 & 1 \\
\end{array}
\right)$ CS},\;\,\\
T_3 &\leftrightarrow& \SU(2)_{-2}\times \SO(3)_1,\;\,\\
T_4 &\leftrightarrow& \SU(2)_{-1}\times \SU(2)_{-1}\times \SO(4)_{1}
\simeq \text{K-matrix $\left(
\begin{array}{cc}
 0 & 2 \\
 2 & 2 \\
\end{array}
\right)$ CS $\times \{1,f\}$}
,\\
T_5 &\leftrightarrow& \SU(2)_{2}\times \SO(3)_{-1},\;\,\\
T_6 &\leftrightarrow& \U(1)_{4}\times \U(1)_{-1}
 \simeq \text{K-matrix $\left(
\begin{array}{cc}
 0 & 2 \\
 2 & -1 \\
\end{array}
\right)$ CS}
,\;\,\\
T_7 &\leftrightarrow& {\text{Ising}}\times \overline{\text{(spin-Ising)}},\;\,\\
T_8 = T_0  &\leftrightarrow&  \text{untwisted  $\mathbb{Z}_2$ gauge theory $\times \{1,f\}$}
\simeq \text{K-matrix $\left(
\begin{array}{cc}
 0 & 2 \\
 2 & 0 \\
\end{array}
\right)$ CS $\times \{1,f\}$}. 
\end{eqnarray}
and $T_{-L} =\overline{T_{L}}$ where bar denotes its time-reversal $CT$ ($i.e.$, particle-hole conjugate) image. The $\Spin(L)_{-1} \times \SO(L)_1$ theories have a net zero chiral central charge $c_-=c_L-c_R=0$, and they are equivalent to 2+1$d$ Kitaev spin liquids \cite{KITAEV20062} tensored with suitable invertible spin TQFTs (with only $\{1,f\}$,
a trivial operator and a transparent 
spin-1/2 
fermionic line operator) to cancel the chiral central charge.
Here the K-matrix CS theories 
have a gauge group given by products of $\U(1)\times \U(1) \times \dots$   groups,
with a symmetric-bilinear integer matrix $K$. 
In our case,  only for an even integer $L$,
we have the $K$ matrix = 
$\left(
\begin{array}{cc}
 0 & 2 \\
 2 & L/2 \mod 4 \\
\end{array}
\right)$
which corresponds to an abelian CS theory.
The $\mathbb{Z}_8$ class of fermionic spin TQFTs can be obtained by gauging the $\mathbb{Z}_2$-internal symmetry of fermionic symmetry-protected topological states generated by
the spin bordism group $\Omega_3^{\Spin}(B\mathbb{Z}_2)=\mathbb{Z}_8$.
Their $\mathbb{Z}_8$ class 
bordism invariant as an 
invertible TQFT can be also written
schematically as 
$$
\e^{\ii S[a, s]}=\e^{\frac{2 \pi \ii\nu}{8}\text{ABK}[\text{PD}(a), \;\; s|_{\text{PD}(a)}]}
$$
with $\nu \in \Z_8$,
a spin structure
$s \in \text{Spin}({M^{3}})$, 
and the background $\Z_2$ gauge connection 
$a\in H^1({M^{3}},\Z_2)$. 
Here $\text{ABK}[\ldots]$ denotes the $\Z_8$ valued Arf-Brown-Kervaire invariant of Pin$^-$ 2-manifold
from the Poincar\'e dual (PD) of $a$ \cite{Putrov2016qdo1612.09298PWY}.
The $s|_{\text{PD}(a)}$ is the $\text{Pin}^-$ structure on $\text{PD}(a)$
For more details, see Table 2 of \cite{Putrov2016qdo1612.09298PWY}.

Moreover, if we disregard the thermal Hall conductance (the chiral central charge $c_-$) difference,
the $T_L$ can also be related to the $\Spin(L)_{-1}$ Chern-Simons gauge theory in 2+1$d$,
which is a bosonic non-spin TQFT with a
$\Spin(L)$ gauge group at the level $-1$ \cite{Cordova:2017vab}
with a chiral central charge $c_-=-L/2 \mod 4$.

{The relation of the $\Z_8$ class here
and the Kitaev's $\Z_{16}$ class \cite{KITAEV20062}
is also examined in other recent work on non-abelian 
fractional quantum Hall states, see \cite{Ma2019nynFeldman190402798,Ma2020pjs2003.05954}.
We also thank Greg Moore for pointing out 
another discussion on the
$\Z_8$ classes of 3d spin CS theory from the
symmetric bilinear $K$-matrix abelian 
CS theory perspective \cite{Belov2005zeMoore0505235}.}
}

To compare our present work to Ref.\cite{Lian:2018xep},
we note that Ref.\cite{Lian:2018xep} writes the TQFTs for
Pfaffian, PH-Pfaffian, and anti-Pfaffian states as:
\bea
\text{Pfaffian}&:&\quad \frac{\U(1)_8\times {{\text{Ising}}}}{\mathbb{Z}_2}
-4\text{CS}_\text{grav},~\quad \; \;  c_-=1+1/2+4/2=7/2.\\
\text{PH-Pfaffian}&:&\quad  \frac{\U(1)_8\times {\overline{\text{Ising}}}}{\mathbb{Z}_2}-4\text{CS}_\text{grav},~\quad \;  \;  c_-=1-1/2+4/2=5/2.\\
\text{anti-Pfaffian}&:&\quad \frac{\U(1)_8\times {\SU(2)_{-2}}}{\mathbb{Z}_2}-4\text{CS}_\text{grav},~\;  c_-=1-3/2+4/2=3/2.
\eea
whereas here we write
\bea
\text{Pfaffian}&:&\quad \O(2)_{2,-1}\;\text{CS}-5\text{CS}_\text{grav},~\quad c_-=1+5/2=7/2.\\
\text{PH-Pfaffian}&:&\quad \O(2)_{2,1}\;\text{CS}-3\text{CS}_\text{grav},~\quad \;\; c_-=1+3/2=5/2.\\
\text{anti-Pfaffian}&:&\quad \O(2)_{2,3}\;\text{CS}-\text{CS}_\text{grav},~\quad \quad \, c_-=1+1/2=3/2.
\eea

In $T_L$, the $\mathbb{Z}_{2,[1]}$ symmetry acts on the magnetically charged line, which is the line operator associated to the so-called $\sigma$ anyon. In the condensed matter terminology, this is called  the \emph{magnetic vortex line} or \emph{vison loop}. $\mathbb{Z}_{2,[1]}$ takes this line to minus itself ($-\sigma$). In the $\U(1)_8$ Chern-Simons theory, the $\mathbb{Z}_{2,[1]}$ symmetry also transforms the lines with \emph{odd} $\U(1)$ charges to minus themselves. 
In other words, the line operators of the $\sigma$ anyon in $T_L$
and of odd $\U(1)$ charge in $\U(1)_8$
are both charged objects under the diagonal $\mathbb{Z}_{2,[1]}$ 1-form symmetry. As already mentioned in the main text, the $\mathbb{Z}_{2,[1]}$ 
 symmetry generators are line operators in the $\U(1)_8$ theory with $\U(1)$ charge 4 and the fermionic line $f$ in $T_L$. 
 
Another way to think of the $\mathbb{Z}_{2,[1]}$ symmetry transformation is that it arises when the charged lines are linked with the symmetry generator lines. Then, the path integral picks up an extra $(-1)$ sign. The link configurations resulting in this sign are
 \begin{itemize}
\item

 In the $\U(1)_8$ Chern-Simons theory, when the 
odd $\U(1)$ charge line links with the $\U(1)$ charge 4 line,
we get a statistical Berry phase $\exp(\frac{2 \pi \ii}{8} \mathbb{Z}_{odd} \cdot 4)=(-1)$.

\item
 In the $T_L$ theory, when 
the $\sigma$ line links with the the fermionic $f$ line,
we get a statistical Berry phase $(-1)$.

 \end{itemize}
 
We reviewed in Appendix \ref{app:gaugingoneform} what it means to gauge a $\mathbb{Z}_{2,[1]}$  symmetry. By gauging the diagonal $\mathbb{Z}_{2,[1]}$  symmetry described above, we reduce the 24 line operators in the
  ${\U(1)_8\times T_L}$ theory to
  the 12 line operators in the
$ \frac{\U(1)_8\times T_L}{\mathbb{Z}_2}$ theory. 
See Tables \ref{table:Pf},
\ref{table:APf} and 
\ref{table:PHPf}.

\subsection{Hall conductivity}
\label{app:hall-conductivity}

The theory also has a $\mathbb{Z}_4$ one-form global symmetry generated by the line with $\U(1)$ charge 2 in the $\U(1)_8$ Chern-Simons theory.
One can use this one-form symmetry to couple to a background electromagnetic $\U(1)$ gauge field $A$ at level-1 as $\frac{1}{2\pi}(2b)\dd A=\frac{2b}{2\pi}\dd A$. The $U(1)_8$ action, including the coupling to the probe electromagnetic background, is:
\begin{equation}
\frac{8}{4\pi}b\dd b+\frac{2b}{2\pi}\dd A~,
\end{equation}
where $b$ is the gauge field of $\U(1)_8$
Chern-Simons theory, and $A$ couples to the properly quantized $\U(1)$ gauge field $2b$.
This system gives the Hall conductance
$$
\sigma_{xy}=2^2/8=1/2.
$$
In addition, we can add a Chern-Simons action for the background gauge field $A$
\begin{equation}
 \label{eq:probe-U1EFT}
\int_{3d} \Big(\frac{8}{4\pi}b\dd b+\frac{2b}{2\pi}\dd A \Big)+\frac{r}{4\pi}A\dd A~,
\end{equation}
where the coefficient $r\in\mathbb{Z}$ is quantized to be an integer in fermionic systems. 
Then the system has a quantum Hall conductance
\begin{equation}
\sigma_{xy}=\frac{1}{2}+r.
\end{equation}
measured in units of $e^2/h$.
In the application to the experiment with quantum Hall conductivity $\sigma_{xy}={5\over 2}$ (see \cite{Lian:2018xep} and the references therein), we take $$r=2.$$ Here the $r=2$ corresponds to the lowest (zeroth) Landau levels with both spin-up and spin-down complex fermions, which contribute $\sigma_{xy}=2$ quantum Hall conductance.
The first Landau level contributes
another $\sigma_{xy}=\frac{1}{2}$
from the half-filled first Landau level with spin-polarized fermions.

The discussion does not change when there are fermions in the nontrivial one-dimensional representation of $\mathbb{Z}_2$ ({\it i.e.} they are coupled to $T_L$ but not $\U(1)_8$) that do not couple to the background $A$.

There is another way to see the quantum Hall conductance using the $\mathbb{Z}_4$ one-form symmetry. The line generating that symmetry has spin $1/4$, and thus the one-form symmetry has the anomaly 
\begin{equation}
\frac{8}{4\pi}\int_{4d}B_2 B_2
\end{equation}
where $B_2$ is the two-form background field of the $\mathbb{Z}_4$ one-form symmetry. The anomaly then implies that the coupling to $A$ by fixing the value $B_2=\frac{1}{4}\dd A$ has a  half-integer Hall conductance (see Appendix E of \cite{Benini:2018reh}). 
The one-form symmetry is present in massive or massless theories with the same 't Hooft anomaly, since the one-form symmetry is preserved by mass deformations, and thus the quantum Hall conductance is the same across the phase diagram.

\subsection{Quantum numbers of quasi-excitations}
\label{sec:Quantum-numbers-quasi-excitations}

Using (\ref{eq:O2LtoU18TLmodZ2}) and  (\ref{eq:probe-U1EFT}), we see that a line with odd charge in the $\U(1)_8$ theory is identified with the $\sigma$ line in $T_L$ theory. It is therefore the line operator of a non-abelian anyon (when $L$ is odd)
with quantum dimension 2.
Let us label this line operator
as
$$
(\text{odd},\sigma).
$$
Moreover, we can determine the U(1) electromagnetic charge of this anyonic
quasi-excitation from the level $K=8$
and the charge vector ${q}=2$ in (\ref{eq:probe-U1EFT}) via
$$
Q_{(\text{odd},\sigma)}= \frac{1}{K}{q}=\frac{1}{8}2=\frac{1}{4}.
$$
This means our theory has fractional $\U(1)$ 
electromagnetic charges $\pm \frac{1}{4}$ from 
quasiparticle ${(\text{odd},\sigma)}$ 
and quasihole excitations.

Two such non-abelian anyons ${(\text{odd},\sigma)}$ fuse to abelian anyons, also called semions, with quantum 
dimension 1:
$$
(\text{even},s).
$$
The semions have fractional spin statistics with spin $\frac{1}{4}$, and also have fractional electromagnetic charges:
$$
Q_{(\text{even},{s})}= 
\frac{1}{K}q=\frac{1}{8}2\cdot2=\frac{1}{2}.
$$
Therefore, there are fractional $\U(1)$
electromagnetic charges in the theory $\pm \frac{1}{2}$ from the  
quasiparticle semion ${(\text{even},s)}$ 
and the corresponding
quasihole. 

\section{Many-body wavefunctions}
\label{app:many-body-wavefunctions}
In this Appendix, we recall and examine
the electron wavefunctions for the non-abelian 
Pf/PHPf/APf states and also abelian states. In contrast to the EFT language used in the bulk of our work (which uses the second-quantization language), this section is formulated in a many-body quantum mechanics picture (the first quantization language). This Appendix can be
a companion to the \Sec{sec:quantum-numbers}.

\subsection{Pfaffian state for  $\kappa_{xy}=7/2$}
%\begin{enumerate}
%\item
The {\bf Pfaffian} state wavefunction
 $\Psi_{\text{Pf}}$ was introduced 
by Moore-Read \cite{moore1991, read2000Green}  
for a $\nu=1/2$ fractional quantum Hall state in the zeroth Landau level. It is a rotationally invariant state. The wavefunction is
\begin{equation} \label{eq:psi-Pf}
\Psi_{\text{Pf}}(\{z_i\})={\rm{Pf}}\left(\frac{1}{z_i-z_j}\right)\big( \prod_{1=i<j}^{N}(z_i-z_j)^k \big)\e^{-\sum_{i=1}^{N} |z_i|^2/4\ell_B^2}\ .
\end{equation}
In particular, we look at $k=2$,
for $N$ electrons, and for a magnetic length $\ell_B=\sqrt{\hbar c/(|e| B)}$ for magnetic field $B$. Here $z_i^* \equiv \bar{z_i}$ is the complex conjugate of the coordinate ${z_i}=x_i+\ii y_i \in \mathbb{C}$. 
The Pf is the Pfaffian of the rank-$N$ antisymmetric matrix $M_{ij}=1/(z_i-z_j)$, so 
$({\rm Pf}(M_{ij}))^2 = \det(M_{ij})$. 
Namely,
for an even positive $N$, 
we have a degree-$N/2$ polynomial
$$
{\rm Pf} (M_{ij}) = {1 \over 2^{N/2} (N/2)!} \sum_{\sigma \in S_N}
{\rm sgn} \sigma \prod_{l=1}^{N/2} M_{\sigma(2l-1)\sigma(2l)},
$$
with the symmetry group $S_N$
and ${\rm sgn} \sigma = \pm 1$ the signature of the element $\sigma \in S_N$,
so
$$
{\rm Pf} (\frac{1}{z_i-z_j}) = {1 \over 2^{N/2} (N/2)!} \sum_{\sigma \in S_N}
{\rm sgn} \sigma \prod_{l=1}^{N/2}\frac{1}{z_{\sigma(2l-1)}-z_{\sigma(2l)}}  .
$$
The ${\rm{Pf}}\left(\frac{1}{z_i-z_j}\right)$ factor is crucial to obtain an antisymmetric wavefunction, as appropriate for a fermionic electron system.
The Laughlin-like factor $\big( \prod_{1=i<j}^{N}(z_i-z_j)^2 \big)$ with second-order zeros dictates that there are repulsive interactions between electrons.
The ${\rm{Pf}}\left(\frac{1}{z_i-z_j}\right)$ factor cancels some of the zeros present in the Laughlin-like factor $\big( \prod_{1=i<j}^{N}(z_i-z_j)^2 \big)e^{-\sum_{i=1}^{N} |z_i|^2/4\ell_B^2}$, making the electrons less repulsive on net. This  implies that electrons in $\Psi_{\text{Pf}}$ 
are closer together than those in a purely Laughlin-like state.
All the electrons are spin polarized in the $\Psi_{\text{Pf}}$ state.

 {\bf Filling fraction}:
To determine the filling fraction $\nu$ of $\Psi_{\text{Pf}}$,
we compute the angular momentum operator $L_{z_i}=\hbar({z_i}\partial_{z_i} - {z_i^*}\partial_{{z_i^*}})$ acting on the $i$-th electron.
The highest power of $z_i$ in $\Psi_{\text{Pf}}$
is $z_i^{k (N-1)-1}$ where $k (N-1)-1$ is from the Laughlin factor
and $-1$ is from the Pf factor. This
gives rise to the angular momentum $k  \hbar$ for the 
$i$-th electron, which encircles the larger
area of the droplet with a radius $r_k=\sqrt{2 (k (N-1)-1)} \ell_B$ (at the location
where the wavefunction density is maximal). 
Recall $\Phi_0=2 \pi (\ell_B)^2 B= h c/|e|$, so we verify that $\Psi_{\text{Pf}}$ has the 
$$\nu=\frac{\text{number of particles}}{\text{{number of flux quanta}}}
=
\frac{N}{\Phi_B/\Phi_0}= 
\frac{N}{(\pi (r_k)^2)/(2 \pi (\ell_B)^2)}
\simeq 1/k, \quad \text{as } N \to \infty,$$ 
(i.e., $\nu=1/2$ and $\sigma_{xy}=1/2$ for $k=2$ for the Moore-Read  Pfaffian).
To employ this wavefunction to the study of $\nu=5/2$, we employ the Pf state for the first half-filled, spin-polarized Landau level, while we also include spin up and down electrons
 fully occupying the zeroth Landau levels. This gives a total filling fraction of $\nu=5/2$; also, $\sigma_{xy}=5/2$.
The interaction produces an energy gap the order of the Coulomb interaction energy
$e^2/\ell_B$, so this state is \emph{incompressible}.

{\bf Quasi-excitations}: We can obtain a quasihole by adding a hole excitation in a complex coordinate $\zeta$,
\bea
&&\Psi_{\text{Pf}}^{\text{hole}}(\zeta;\{z_i\})\propto
\big( \prod_{i'=1}^{N}(\zeta-z_{i'}) \big)\Psi_{\text{Pf}}(\{z_i\}) \label{eq:Pf-hole}\\
&&={\rm{Pf}}\left(\frac{(\zeta-z_i)(\zeta-z_j)+(\zeta-z_j)(\zeta-z_i)}{z_i-z_j}\right)\big( \prod_{1=i<j}^{N}(z_i-z_j)^k \big)\e^{-\sum_{i=1}^{N} |z_i|^2/4\ell_B^2}.\nn
\eea
We could view the $z_i$ as dynamical variables (that should be integrated over to obtain the density), while viewing $\zeta$ as a background, or probe, parameter.
Because the additional factor $\big( \prod_{i'=1}^{N}(\zeta-z_{i'}) \big)$ introduces more zeros into the wavefunction, the system becomes less repulsive, so also less dense --- this is a hallmark of a hole excitation.
If $\zeta$ is instead a dynamical variable, then the 
$(\zeta-z_{i'})^k$ factor introduces an electron at position $\zeta$. For $\zeta$ a background parameter, this has the interpretation of
removing an electron at $\zeta$.
Putting this together, a $k$-fold factor $(\zeta-z_{i'})^k$ removes an electron at $\zeta$ and so, given the electron charge $-|e|$, we have produced a quasihole of charge $|e|/k$. The second line in \eq{eq:Pf-hole} is a rewriting of the first line, by absorbing the quasihole into the Pf factor:
${\rm{Pf}}\left(\frac{(\zeta_1-z_i)(\zeta_2-z_j)+(\zeta_2-z_j)(\zeta_1-z_i)}{z_i-z_j}\right)\vert_{\zeta_1=\zeta_2=\zeta}$
which can be regarded as the quasihole splitting  into two 
further fractional quasiholes at $\zeta_1$ and $\zeta_2$, each 
with charge $|e|/(2k)$.
For the Moore-Read Pfaffian at $k=2$, we have a quasihole of charge $|e|/2$ which further fractionate to a quasiholes of charge $|e|/4$.

For each quasihole, there is a corresponding quasiparticle excitation with opposite global symmetry quantum numbers, but with the same spin statistics. The quasiparticles/quasiholes may be regarded as vortices/anti-vortices because the phase of the wavefunction winds when a particle 
winds around the quasiexcitation at $\zeta$. For example, the 
fractionalized charge $|e|/4$ or $-|e|/4$ excitations are in fact the
$\pm \pi$-vortices, which we shall identify as the non-abelian $\sigma$ anyons in our EFT and TQFTs.

 {\bf Chiral central charge} $c_-= c_L-c_R$ (the degrees of freedom of 1+1$d$ left-moving minus right-moving edge modes) can be determined from two parts of the wavefunctions: first, the Laughlin sector $(z_i-z_j)^k$ corresponds to U(1)$_k$ CS theory. It has an edge theory which can be described as a complex chiral boson or fermion, which yields $c_-=1$. (The readers can find a systematic description of the $1+1d$ edge theory in Ref.~\cite{Lian:2018xep}.)
 Second, the Pf factor corresponds to the angular momentum $L_z=1$
 between the composite fermion with a chiral p-wave ($p_x + \ii p_y$) pairing \cite{read2000Green}. This gives rise to 
  $c_-=1/2$ corresponding to an edge theory given by a real-valued chiral Majorana mode. 
  The total $c_-$ for the Pfaffian state \eq{eq:psi-Pf} is $c_-=3/2$.

{\bf Composite fermion pairing}: The above discussion is consistent with the fact that the Ising TQFT contributes $c_-=1/2$ in \eq{eq:1Pf}, which can be induced from the $(p_x+ \ii p_y)$-wave pairing of composite fermions (CF), with angular momentum $\propto
(k_x+ \ii k_y)$ for $L_z=1$. 
In the Dirac composite Fermi liquid (CFL) picture,
the Dirac CF gains a $\pi$-Berry phase around the Fermi surface.
For Dirac CF, the pairing becomes the $(d_x+ \ii d_y)$-wave pairing 
 $\propto
(k_x+ \ii k_y)^2$ with $L_z=2$. 

\subsection{Anti-Pfaffian state for  $\kappa_{xy}=3/2$}

%\item
{\bf Anti-Pfaffian} (APf) state wavefunction:
 The bulk system for the Pfaffian state does not have a time-reversal ($CT$)/particle-hole symmetry \cite{Greiter:1991raWenWilczekPRL}, but  \Rf{lee2007} considered the \eq{eq:psi-Pf}'s particle-hole conjugate wavefunction, dictated by the particle-hole transformation \cite{PhysRevB.29.6012Girvin1984}, and named it the anti-Pfaffian state:
\begin{multline}\label{eq:psi-APf}
\Psi_{\text{APf}}(\{z_i\})=
\int 
 (\prod_{{i'}=1}^{N}
\dd \xi_{i'}
\dd \xi_{i'}^*
)
\prod_{i,j'=1}^{N}(z_i-\xi_{j'})
\cdot
\prod_{1=i'<j'}^{N}(\xi_{i'}-\xi_{j'})
\e^{-\sum_{j'=1}^{N} |\xi_{j'}|^2/4\ell_B^2} \cdot \Psi_{\text{Pf}}(\{\xi_{i'}^*\})
\\
\cdot \prod_{1=i<j}^{N}(z_i-z_j) 
 \cdot \e^{-\sum_{i=1}^{N}  |z_i|^2/4\ell_B^2}
\ .
\end{multline}
We can break down the $\Psi_{\text{APf}}$ state we are interested in \cite{levin2007,lee2007}
 as a combination of two component pieces. 
The first piece is the $\nu=1/2$ $\Psi_{\text{APf}}$ with respect to the
$\nu=1$ IQH state. The second piece can be viewed as 
a $\nu=1$ integer quantum Hall state (the IQH state with respect to the $\nu=0$ vacuum). The first part is nothing but the
particle-hole conjugate of the 
$\nu=1/2$ $\Psi_{\text{Pf}}$ with respect to the
$\nu=0$ vacuum. Indeed, the first line in \eq{eq:psi-APf}
corresponds to the first line, while 
the second line in \eq{eq:psi-APf}
corresponds to the second part.
 From this description, we see that the filling fraction is $\nu=1/2$ by construction
 and the contribution to the chiral central charge of APf from
 the first part is $c_-=-3/2$ and from the second part is $c_-=1$
 for a total of $c_-=-3/2+1=-1/2$.
 
 The $\SU(2)_{-2}$ TQFT contributes $c_-=-3/2$ in \eq{eq:3APf}, which can be induced from the $(f_x- \ii f_y)$-wave pairing of CF, 
with its angular momentum $\propto
(k_x- \ii k_y)^3$ for $L_z=-3$. 
For Dirac CF, the pairing becomes the $(d_x- \ii d_y)$-wave pairing with $L_z=-2$.

%\item
\subsection{Particle-Hole Pfaffian state for  $\kappa_{xy}=5/2$}

{\bf Particle-Hole Pfaffian} (PH-Pfaffian, or PHPf)
wavefunction \cite{ZuckerFeldman2016} (see also \cite{Balram2018Ajit1803.10427, 1804.01107PHPf} and other attempts \cite{yang2017particlehole1701.03562,
yang2020compressed2001.01915}) can be written as
\begin{multline} \label{eq:psi-PHPf}
\Psi_\mathrm{PHPf}(\{z_i\}) = \mathcal{P}_\mathrm{LLL}\left[\mathrm{Pf} \left(\frac{1}{z_i^* - z_j^*}\right) 
\big(\prod_{1=i<j}^{N}(z_i - z_j)^2 \big)
e^{-\sum_{i=1}^{N} |z_i|^2/4\ell_B^2}\right] \\
\simeq \int  (\prod_{{i'}=1}^{N}
\dd \xi_{i'}
\dd \xi_{i'}^*)\langle\{z_i\}|\{\xi_{i'}\}\rangle
\left[\mathrm{Pf} \left(\frac{1}{\xi_{i'}^* - \xi_{j'}^*}\right) 
\big(\prod_{1=i<j}^{N}(\xi_{i'} - \xi_{j'})^2 \big)
\e^{-\sum_{i'=1}^{N} |\xi_{i'}|^2/4\ell_B^2}\right]
\\
=
\int  (\prod_{{i'}=1}^{N}
\dd \xi_{i'}
\dd \xi_{i'}^*)
\exp\big(-(|\xi_{i'}|^2-2{\xi_{i'}^*}z_i+|z_i|^2)/(4l_B^2)\big)\\
\cdot \left[\mathrm{Pf} \left(\frac{1}{\xi_{i'}^* - \xi_{j'}^*}\right) 
\big(\prod_{1=i<j}^{N}(\xi_{i'} - \xi_{j'})^2 \big)
\e^{-\sum_{i'=1}^{N} |\xi_{i'}|^2/4\ell_B^2}\right].
\end{multline}
The $\mathcal{P}_\mathrm{LLL}$ is the projection onto the lowest Landau level (LLL).
From the first line in \eq{eq:psi-PHPf},
we can see that the filling fraction is still $\nu=1/2$, as one can read off from the Laughlin factor using the same reasoning from the Pfaffian case.
Moreover, the $\mathrm{Pf} \left(\frac{1}{z_i^* - z_j^*}\right) $ tells us the pairing of composite fermions possesses angular momentum $L_z=-1$
 between the composite fermion with an anti-chiral p-wave  
 ($p_x - \ii p_y$) pairing\cite{read2000Green}, which gives rise to 
  $c_-=1/2$. The total chiral central charge of the PH-Pfaffian \eq{eq:psi-APf} therefore has $c_-=1-1/2=1/2$.
  The second line in \eq{eq:psi-PHPf} rewrites the projection in terms of the coherent state projection so the wave function is projected into the LLL.
  
  The above discussion is consistent with the fact that 
the $\overline{\text{Ising}}$ TQFT contributes $c_-=-1/2$ in \eq{eq:2PHPf}, which can be induced from the $(p_x- \ii p_y)$-wave pairing of CF, 
with angular momentum $\propto
(k_x- \ii k_y)$ at $L_z=-1$. 
For a Dirac CF picture, the pairing becomes the $s$-wave pairing with $L_z=0$.

%\item
\subsection{$K=8$-state for  $\kappa_{xy}=3$}

{\bf $K=8$-state} wavefunction is a bosonic wavefunction but can be written as a 
fermionic wavefunction
by dressing it with
a fermionic tensor product state:
\begin{multline} \label{eq:psi-K=8}
\Psi_{{K=8}}(\{z_i\}\})=
\big( \prod_{1=i<j}^{N}(z_i-z_j)^{8} \big)
 \e^{-\sum_{i=1}^{N} |z_i|^2/4\ell_B^2}\cdot (\text{fermionic tensor product state})
\ .
\end{multline}
The filling fraction is $\nu=1/2$ with an appropriate charge coupling, 
when the charge $2e$ quasi-excitations are coupled to
the $\U(1)$ electromagnetic gauge field at level-1.
The chiral central charge is $c_-=1$, as always for a Laughlin wavefunction.
 The is consistent with $\U(1)_{8}$ TQFT with $c_-=1$.

%\item

\subsection{113-state for  $\kappa_{xy}=2$}

{\bf 113-state} wavefunction is a special case of the lmn wavefunction, known as the Halperin wavefunction 
(the multi-component generalization of Laughlin wavefunction)
with ${\rm l}=1$, ${\rm m}=1$, ${\rm n} =3$ for some $N+N'$ electron system:
\begin{multline} \label{eq:psi-113-state}
\Psi_{{113}}(\{z_i\},\{w_{i'}\})=
\big( \prod_{1=i<j}^{N}(z_i-z_j)^{\rm l} \big)
\big( \prod_{1=i'<j'}^{N'}(w_{i'}-w_{j'})^{\rm m} \big)
\big( \prod_{i}^{N} \prod_{j'}^{N'}(z_{i}-w_{j'})^{\rm n}  \big)\\
\left. \e^{-\sum_{i=1}^{N} |z_i|^2/4\ell_B^2}
e^{-\sum_{i'=1}^{N} |w_{i'}|^2/4\ell_B^2} \right|_{{\rm l}=1,\, {\rm m}=1,\, {\rm n} =3} \ .
\end{multline}
The filling fraction is $\nu=1/2$
with an appropriate charge coupling.
The chiral central charge is $c_-=1-1=0$ coming from two modes with
opposite chiralities.

%\end{enumerate}

In general, we expect that quasiparticles and quasiholes of the above many-body wavefunctions in this Appendix agree with the anyons (and their quantum numbers) of TQFTs shown in Table \ref{table:Pf}, 
\ref{table:APf}, \ref{table:PHPf},  \ref{table:K=8}, and \ref{table:113}
in \Sec{sec:quantum-numbers}.

\section{One-loop computations}
\label{App:One-loop}

\subsection{Fermionic functional determinant}
\label{App:Determinant}

We explicitly carry out the computation of some terms in the fermionic functional determinant given by integrating out a Dirac fermion $\Psi$ in the following Lagrangian in $d$ spacetime dimensions:
\begin{equation}
{\cal L} = \overline\Psi (\ii\slashed{D} + m - \slashed A - \chi) \Psi \, .
\end{equation}
The quadratic term in the functional determinant effective action is formally written as
\begin{equation}\begin{aligned}
& {1\over2} \, {\rm Tr} \, {\slashed p - m \over p^2 + m^2} \, (\chi + \slashed A) \, {\slashed p - m \over p^2 + m^2} \, (\chi + \slashed A) \, .
\end{aligned}\end{equation}
For the $\chi^2$ piece, we simplify by 
\begin{equation}\begin{aligned}
\label{TraceDerivatives}
& {1\over2} \, {\rm Tr} \, {\slashed p - m \over p^2 + m^2} \, \chi \, {\slashed p - m \over p^2 + m^2} \, \chi
\\
&= {1\over2} \, {\rm Tr} \, {\slashed p - m \over p^2 + m^2}  \, {\slashed p - \ii \slashed \partial - m \over (p - \ii \partial)^2 + m^2} \, \chi \, \chi
\\
&= {1\over2} \, {\rm Tr} \, {{-p^2 + m^2 - 2 m \slashed p +\ii(p \partial + m \slashed \partial) }\over {(p^2 + m^2)^2}} \sum_{n=0}^\infty \left( 
{{2\ii p \partial + \partial^2} \over {p^2 + m^2}} \right)^n \, \chi \, \chi \, ,
\end{aligned}\end{equation}
where the derivatives only act on the first $\chi$ and not the second.  The formal expression can be explicitly evaluated by the momentum space integral
\begin{equation}
    \int {\dd^dp \over (2\pi)^d} \, .
\end{equation}
The trace simply kills all slashed objects.  We will compute up to $\partial^4$ order, so we keep up to $n = 4$ in the sum.
By Lorentz invariance, we can perform the following replacements in the integrand:
\begin{equation}\begin{aligned}
(p \partial)^2 \to {1 \over d} p^2 \partial^2 \, , \quad (p \partial)^4 \to {3 \over d(d+2)} p^4 \partial^4 \, 
\end{aligned}\end{equation}
The result is\footnote{The first term is divergent and regularized by analytic continuation in spacetime dimension.}
\begin{equation}\begin{aligned}
{1\over2} \, \chi \left( {|m| \over \pi} - {1 \over 12 \pi |m|} \partial^2 - {1 \over 240 \pi |m|^3} \, \partial^4 + {\cal O}(\partial^6) \right) \chi \, .
\end{aligned}\end{equation}

Next, let us consider the $\chi A$ piece.  
\begin{equation}\begin{aligned}
& {1\over2} \, {\rm Tr} \, {\slashed p - m \over p^2 + m^2} \, \chi \, {\slashed p - m \over p^2 + m^2} \, \slashed A
\\
&= {1\over2} \, {\rm Tr} \, {\slashed p - m \over p^2 + m^2}  \, {\slashed p -\ii\slashed \partial - m \over (p -\ii\partial)^2 + m^2} \, \chi \slashed A
\\
&= {1\over2} \, {\rm Tr} \, {-p^2 + m^2 - 2 m \slashed p +\ii(p \partial + m \slashed \partial) \over (p^2 + m^2)^2} \sum_{n=0}^\infty \left( {2\ii p \partial + \partial^2 \over p^2 + m^2} \right)^n \, \chi \slashed A \, ,
\end{aligned}\end{equation}
where the derivatives only act on $\chi$ but not $A$.
Evaluating the trace gives
\begin{equation}\begin{aligned}
\label{chiA}
{2^{\lfloor d/2 \rfloor} \over 2} \, {2 m p^\mu - \ii m \partial^\mu \over (p^2 + m^2)^2} \sum_{n=0}^\infty \left( {2 \ii p \partial + \partial^2 \over p^2 + m^2} \right)^n  \, \chi A_\mu \, .
\end{aligned}\end{equation}
The pieces with only one derivative combine to
\begin{equation}\begin{aligned}
& {2^{\lfloor d/2 \rfloor} \over 2} {\ii  m} \int {\dd^dp \over (2\pi)^d} { - (p^2 + m^2) \delta^{\mu\nu} + 4 p^\mu p^\nu \over (p^2 + m^2)^3} \, (\partial_\nu \chi) \, A_\mu
\\
&= {2^{\lfloor d/2 \rfloor} \over 2} {\ii  m} \int {\dd^dp \over (2\pi)^d} \, {(4/d-1)p^2 - m^2 \over (p^2 + m^2)^3} (\partial^\mu \chi) \, A_\mu \, ,
\end{aligned}\end{equation}
whose coefficient in $d = 3$ evaluates to
\begin{equation}\begin{aligned}
\label{Coef3d}
{\ii  m \over 2\pi^2} \int_0^\infty \dd p \, p^2 \, {(p^2/3 - m^2) \over (p^2 + m^2)^3} = 0 \, .
\end{aligned}\end{equation}

As for the $A \chi$ piece,
\begin{equation}\begin{aligned}
& {1\over2} \, {\rm Tr} \, {\slashed p - m \over p^2 + m^2} \, \slashed A \, {\slashed p - m \over p^2 + m^2} \, \chi
\\
&= {1\over2} \, {\rm Tr} \, {\slashed p - m \over p^2 + m^2}  \, \gamma^\mu \, {\slashed p - \ii \slashed \partial - m \over (p - \ii \partial)^2 + m^2} \, A_\mu \chi
\\
&= {1\over2} \, {\rm Tr} \, {(\slashed p - m) \gamma^\mu (\slashed p - \ii \slashed \partial - m) \over (p^2 + m^2)^2} \sum_{n=0}^\infty \left( {2 \ii p \partial + \partial^2 \over p^2 + m^2} \right)^n \, A_\mu \chi
\\
&= {1\over2} \, {\rm Tr} \, {- \ii \slashed p \gamma^\mu \slashed \partial - m \gamma^\mu (\slashed p - \ii \slashed \partial) - m \slashed p \gamma^\mu \over (p^2 + m^2)^2} \sum_{n=0}^\infty \left( {2 \ii p \partial + \partial^2 \over p^2 + m^2} \right)^n \, A_\mu \chi
\\
&= {2^{\lfloor d/2 \rfloor} \over 2} \, {\ii  \varepsilon^{\nu \mu \sigma} p_\nu \partial_\sigma + m (2 p^\mu - \ii \partial^\mu) \over (p^2 + m^2)^2} \sum_{n=0}^\infty \left( {2 \ii p \partial + \partial^2 \over p^2 + m^2} \right)^n \, A_\mu \chi \, ,
\end{aligned}\end{equation}
where the derivatives act on $A$ but not $\chi$, and the parity odd piece with a Levi-Civita symbol is present only if $d = 3$.  At one derivative order, the parity odd piece vanishes upon integrating over $p$, and we are left with the same expression as the $\chi A$ term \eqref{chiA} except now the derivative acts on $A$ instead of $\chi$.  Thus, upon integration by parts, the $A \chi$ term is an identical contribution to the effective action as the $\chi A$ term.  When $d = 3$, the coefficient of $(\partial_\mu \chi) \, A^\mu$ in the effective Lagrangian vanishes, as we found in \eqref{Coef3d}.

\subsection{Domain wall tension}
\label{Sec:Numerics}

Let us discuss how to practically perform the computation of the fermionic one-loop contribution to the domain wall tension, given by the formula \eqref{deltaL}.  The log determinant of the first-order differential operator $\mathbb{D}_{m, g}^{\phi=\phi_0}$ can be related to those of second-order differential operators.  Let $\varepsilon$ be the Levi-Civita symbol.  
Formally,
\begin{equation}
\log \det \mathbb{D}_{m, g}^{\phi=\phi_0} = {1\over2} \log \det (\varepsilon \mathbb{D}_{m, g}^{\phi=\phi_0})^2 \, .
\end{equation}
Next, we write everything explicitly in transverse momentum space,
\begin{equation}
\log \det \mathbb{D}_{m, g}^{\phi=\phi_0} = \int {\dd^2k_\parallel \over (2\pi)^2} \log \det \mathbb{D}_{m, g; k_\parallel}^{\phi=\phi_0} = {1\over2} \int {\dd^2k_\parallel \over (2\pi)^2} \log \det (\varepsilon \mathbb{D}_{m, g; k_\parallel}^{\phi=\phi_0} )^2 \, ,
\end{equation}
where
\begin{equation}
\mathbb{D}_{m, g; k_\parallel}^{\phi=\phi_0} \equiv \begin{pmatrix}
\displaystyle - \partial_z + m + g \phi_0(z) & k_0 - \ii k_2
\\
- k_0 - \ii k_2 & \displaystyle \partial_z + m + g \phi_0(z) & 
\end{pmatrix} \, ,
\end{equation}
and
\begin{equation}
(\varepsilon \mathbb{D}_{m, g; k_\parallel}^{\phi=\phi_0})^2 = \begin{pmatrix}
- \partial_z^2 + ( g \phi_0(z) - m )^2 - g \phi'_0(z) + k_\parallel^2 & 0
\\
0 & - \partial_z^2 + ( g \phi_0(z) - m )^2 + g \phi'_0(z) + k_\parallel^2 
\end{pmatrix}
\end{equation}
is a diagonal matrix of second order differential operators.  If we define
\begin{equation}
\mathbb{M}_{m, g; k_\parallel}^{\phi=\phi_0} \equiv - \partial_z^2 + ( g \phi_0(z) - m )^2 - g \phi'_0(z) + k_\parallel^2 \, ,
\end{equation}
then
\begin{equation}
\label{LogDetDiracSquared}
\log \det \mathbb{D}_{m, g; k_\parallel}^{\phi=\phi_0} = {1\over2} ( \log \det \mathbb{M}_{m, g; k_\parallel}^{\phi=\phi_0} + \log \det \mathbb{M}_{-m, -g; k_\parallel}^{\phi=\phi_0} ) \, .
\end{equation}
Hence, the integral \eqref{deltaL} can written as
\begin{equation}
\begin{aligned}
\delta_f \sigma &= - {1\over2} \int {\dd^2k_\parallel \over (2\pi)^2} \Bigg[ \log { \det \mathbb{M}_{m, g; k_\parallel}^{\phi = \phi_0} \over \det \mathbb{M}_{m, g; k_\parallel}^{\phi = v} } { \det \mathbb{M}_{-m, g; k_\parallel}^{\phi = \phi_0} \over \det \mathbb{M}_{-m, g; k_\parallel}^{\phi = v} } { \det \mathbb{M}_{m, -g; k_\parallel}^{\phi = \phi_0} \over \det \mathbb{M}_{m, -g; k_\parallel}^{\phi = v} } { \det \mathbb{M}_{-m, -g; k_\parallel}^{\phi = \phi_0} \over \det \mathbb{M}_{-m, -g; k_\parallel}^{\phi = v} }
\\
& \hspace{1in} - {F g(g+m/v) \over (k_\parallel^2 + (gv + m)^2)^{1/2}} - {F g(g-m/v) \over (k_\parallel^2 + (gv - m)^2)^{1/2}} \Bigg] \, .
\end{aligned}
\end{equation}
To compute the log determinants in the integrand, we apply the Gel'fand-Yaglom theorem to relate $\log \det \mathbb{M}_{m, g; k_\parallel}^{\phi}$ to a boundary value problem for the second order differential operator $\mathbb{M}_{m, g; k_\parallel}^{\phi}$, and solve it numerically, as in \cite{PY}.

\bibliographystyle{utphys}
\bibliography{biblio,juven-Ref}

\providecommand{\href}[2]{#2}\begingroup\raggedright\begin{thebibliography}{10}

\bibitem{willett1987}
R.~Willett, J.~P. Eisenstein, H.~L. St\"ormer, D.~C. Tsui, A.~C. Gossard, and
  J.~H. English, ``Observation of an even-denominator quantum number in the
  fractional quantum hall effect,''
  \href{http://dx.doi.org/10.1103/PhysRevLett.59.1776}{{\em Phys. Rev. Lett.}
  {\bfseries 59} (Oct, 1987) 1776--1779}.
  \url{https://link.aps.org/doi/10.1103/PhysRevLett.59.1776}.

\bibitem{moore1991}
G.~Moore and N.~Read, ``Nonabelions in the fractional quantum hall effect,''
  \href{http://dx.doi.org/http://dx.doi.org/10.1016/0550-3213(91)90407-O}{{\em
  Nucl. Phys. B} {\bfseries 360} no.~2, (1991) 362 -- 396}.

\bibitem{1991Wen}
X.~G. Wen, ``Non-abelian statistics in the fractional quantum hall states,''
  \href{http://dx.doi.org/10.1103/PhysRevLett.66.802}{{\em Phys. Rev. Lett.}
  {\bfseries 66} (Feb, 1991) 802--805}.
  \url{https://link.aps.org/doi/10.1103/PhysRevLett.66.802}.

\bibitem{levin2007}
M.~Levin, B.~I. Halperin, and B.~Rosenow, ``Particle-hole symmetry and the
  pfaffian state,'' \href{http://dx.doi.org/10.1103/PhysRevLett.99.236806}{{\em
  Phys. Rev. Lett.} {\bfseries 99} (Dec, 2007) 236806}.
  \url{https://link.aps.org/doi/10.1103/PhysRevLett.99.236806}.

\bibitem{lee2007}
S.-S. Lee, S.~Ryu, C.~Nayak, and M.~P.~A. Fisher, ``Particle-hole symmetry and
  the $\ensuremath{\nu}=\frac{5}{2}$ quantum hall state,''
  \href{http://dx.doi.org/10.1103/PhysRevLett.99.236807}{{\em Phys. Rev. Lett.}
  {\bfseries 99} (Dec, 2007) 236807}.
  \url{https://link.aps.org/doi/10.1103/PhysRevLett.99.236807}.

\bibitem{son2015}
D.~T. Son, ``Is the composite fermion a dirac particle?,''
  \href{http://dx.doi.org/10.1103/PhysRevX.5.031027}{{\em Phys. Rev. X}
  {\bfseries 5} (Sep, 2015) 031027},
  \href{http://arxiv.org/abs/1502.03446}{{\ttfamily arXiv:1502.03446
  [cond-mat.mes-hall]}}.

\bibitem{halperin1993}
B.~I. Halperin, P.~A. Lee, and N.~Read, ``Theory of the half-filled landau
  level,'' \href{http://dx.doi.org/10.1103/PhysRevB.47.7312}{{\em Phys. Rev. B}
  {\bfseries 47} (Mar, 1993) 7312--7343}.

\bibitem{Jolicoeur200707051619}
T.~Jolicoeur, ``Non-abelian states with negative flux: A new series of quantum
  hall states,'' \href{http://dx.doi.org/10.1103/physrevlett.99.036805}{{\em
  Physical Review Letters} {\bfseries 99} no.~3, (Jul, 2007) },
  \href{http://arxiv.org/abs/0705.1619}{{\ttfamily arXiv:0705.1619
  [cond-mat.mes-hall]}}. \url{http://dx.doi.org/10.1103/physrevlett.99.036805}.

\bibitem{ZuckerFeldman2016}
P.~T. Zucker and D.~E. Feldman, ``Stabilization of the particle-hole pfaffian
  order by landau-level mixing and impurities that break particle-hole
  symmetry,'' \href{http://dx.doi.org/10.1103/PhysRevLett.117.096802}{{\em
  Phys. Rev. Lett.} {\bfseries 117} (Aug, 2016) 096802}.
  \url{https://link.aps.org/doi/10.1103/PhysRevLett.117.096802}.

\bibitem{Banerjee:2018qtz}
M.~Banerjee, M.~Heiblum, V.~Umansky, D.~E. Feldman, Y.~Oreg, and A.~Stern,
  ``{Observation of half-integer thermal Hall conductance},''
  \href{http://dx.doi.org/10.1038/s41586-018-0184-1}{{\em Nature} {\bfseries
  559} no.~7713, (2018) 205--210},
\href{http://arxiv.org/abs/1710.00492}{{\ttfamily arXiv:1710.00492
  [cond-mat.mes-hall]}}.
%%CITATION = ARXIV:1710.00492;%%.

\bibitem{kane1997}
C.~L. Kane and M.~P.~A. Fisher, ``Quantized thermal transport in the fractional
  quantum hall effect,''
  \href{http://dx.doi.org/10.1103/PhysRevB.55.15832}{{\em Phys. Rev. B}
  {\bfseries 55} (Jun, 1997) 15832--15837}.
  \url{https://link.aps.org/doi/10.1103/PhysRevB.55.15832}.

\bibitem{WenPhysRevLett.70.355}
X.-G. Wen, ``Topological order and edge structure of \ensuremath{\nu}=1/2
  quantum hall state,''
  \href{http://dx.doi.org/10.1103/PhysRevLett.70.355}{{\em Phys. Rev. Lett.}
  {\bfseries 70} (Jan, 1993) 355--358}.
  \url{https://link.aps.org/doi/10.1103/PhysRevLett.70.355}.

\bibitem{Mross:2018MOSMH}
D.~F. Mross, Y.~Oreg, A.~Stern, G.~Margalit, and M.~Heiblum, ``Theory of
  disorder-induced half-integer thermal hall conductance,''
  \href{http://dx.doi.org/10.1103/PhysRevLett.121.026801}{{\em Phys. Rev.
  Lett.} {\bfseries 121} (Jul, 2018) 026801}.
  \url{https://link.aps.org/doi/10.1103/PhysRevLett.121.026801}.

\bibitem{Wang:2018WVH}
C.~Wang, A.~Vishwanath, and B.~I. Halperin, ``Topological order from disorder
  and the quantized hall thermal metal: Possible applications to the
  $\ensuremath{\nu}=5/2$ state,''
  \href{http://dx.doi.org/10.1103/PhysRevB.98.045112}{{\em Phys. Rev. B}
  {\bfseries 98} (Jul, 2018) 045112}.
  \url{https://link.aps.org/doi/10.1103/PhysRevB.98.045112}.

\bibitem{Lian:2018xep}
B.~Lian and J.~Wang, ``{Theory of the disordered $\nu = \frac{5}{2}$ quantum
  thermal Hall state: Emergent symmetry and phase diagram},''
  \href{http://dx.doi.org/10.1103/PhysRevB.97.165124}{{\em Phys. Rev.}
  {\bfseries B97} no.~16, (2018) 165124},
\href{http://arxiv.org/abs/1801.10149}{{\ttfamily arXiv:1801.10149
  [cond-mat.mes-hall]}}.
%%CITATION = ARXIV:1801.10149;%%.

\bibitem{chen2014Fidkowski}
X.~Chen, L.~Fidkowski, and A.~Vishwanath, ``Symmetry enforced non-abelian
  topological order at the surface of a topological insulator,''
  \href{http://dx.doi.org/10.1103/PhysRevB.89.165132}{{\em Phys. Rev. B}
  {\bfseries 89} (Apr, 2014) 165132}.
  \url{https://link.aps.org/doi/10.1103/PhysRevB.89.165132}.

\bibitem{Metlitski2014xqa1406.3032}
M.~A. Metlitski, L.~Fidkowski, X.~Chen, and A.~Vishwanath, ``{Interaction
  effects on 3D topological superconductors: surface topological order from
  vortex condensation, the 16 fold way and fermionic Kramers doublets},''
\href{http://arxiv.org/abs/1406.3032}{{\ttfamily arXiv:1406.3032
  [cond-mat.str-el]}}.
%%CITATION = ARXIV:1406.3032;%%.

\bibitem{morf1998}
R.~H. Morf, ``Transition from quantum hall to compressible states in the second
  landau level: New light on the
  $\ensuremath{\nu}\phantom{\rule{0ex}{0ex}}=\phantom{\rule{0ex}{0ex}}5/2$
  enigma,'' \href{http://dx.doi.org/10.1103/PhysRevLett.80.1505}{{\em Phys.
  Rev. Lett.} {\bfseries 80} (Feb, 1998) 1505--1508}.
  \url{https://link.aps.org/doi/10.1103/PhysRevLett.80.1505}.

\bibitem{rezayi2000}
E.~H. Rezayi and F.~D.~M. Haldane, ``Incompressible paired hall state, stripe
  order, and the composite fermion liquid phase in half-filled landau levels,''
  \href{http://dx.doi.org/10.1103/PhysRevLett.84.4685}{{\em Phys. Rev. Lett.}
  {\bfseries 84} (May, 2000) 4685--4688}.
  \url{https://link.aps.org/doi/10.1103/PhysRevLett.84.4685}.

\bibitem{peterson2008}
M.~R. Peterson, T.~Jolicoeur, and S.~Das~Sarma, ``Finite-layer thickness
  stabilizes the pfaffian state for the 5/2 fractional quantum hall effect:
  Wave function overlap and topological degeneracy,''
  \href{http://dx.doi.org/10.1103/PhysRevLett.101.016807}{{\em Phys. Rev.
  Lett.} {\bfseries 101} (Jul, 2008) 016807}.
  \url{https://link.aps.org/doi/10.1103/PhysRevLett.101.016807}.

\bibitem{feiguin2009}
A.~E. Feiguin, E.~Rezayi, K.~Yang, C.~Nayak, and S.~Das~Sarma, ``Spin
  polarization of the $\ensuremath{\nu}=5/2$ quantum hall state,''
  \href{http://dx.doi.org/10.1103/PhysRevB.79.115322}{{\em Phys. Rev. B}
  {\bfseries 79} (Mar, 2009) 115322}.
  \url{https://link.aps.org/doi/10.1103/PhysRevB.79.115322}.

\bibitem{wangh2009Haldane}
H.~Wang, D.~N. Sheng, and F.~D.~M. Haldane, ``Particle-hole symmetry breaking
  and the $\ensuremath{\nu}=\frac{5}{2}$ fractional quantum hall effect,''
  \href{http://dx.doi.org/10.1103/PhysRevB.80.241311}{{\em Phys. Rev. B}
  {\bfseries 80} (Dec, 2009) 241311}.
  \url{https://link.aps.org/doi/10.1103/PhysRevB.80.241311}.

\bibitem{storni2010DasSarma}
M.~Storni, R.~H. Morf, and S.~Das~Sarma, ``Fractional quantum hall state at
  $\ensuremath{\nu}=\frac{5}{2}$ and the moore-read pfaffian,''
  \href{http://dx.doi.org/10.1103/PhysRevLett.104.076803}{{\em Phys. Rev.
  Lett.} {\bfseries 104} (Feb, 2010) 076803}.
  \url{https://link.aps.org/doi/10.1103/PhysRevLett.104.076803}.

\bibitem{rezayi2011Simon}
E.~H. Rezayi and S.~H. Simon, ``Breaking of particle-hole symmetry by landau
  level mixing in the $\ensuremath{\nu}=5/2$ quantized hall state,''
  \href{http://dx.doi.org/10.1103/PhysRevLett.106.116801}{{\em Phys. Rev.
  Lett.} {\bfseries 106} (Mar, 2011) 116801}.
  \url{https://link.aps.org/doi/10.1103/PhysRevLett.106.116801}.

\bibitem{papic2012HaldaneRezayi}
Z.~Papi\ifmmode~\acute{c}\else \'{c}\fi{}, F.~D.~M. Haldane, and E.~H. Rezayi,
  ``Quantum phase transitions and the $\ensuremath{\nu}\mathbf{=}5/2$
  fractional hall state in wide quantum wells,''
  \href{http://dx.doi.org/10.1103/PhysRevLett.109.266806}{{\em Phys. Rev.
  Lett.} {\bfseries 109} (Dec, 2012) 266806}.
  \url{https://link.aps.org/doi/10.1103/PhysRevLett.109.266806}.

\bibitem{zaletel2015MongPollmannRezayi}
M.~P. Zaletel, R.~S.~K. Mong, F.~Pollmann, and E.~H. Rezayi, ``Infinite density
  matrix renormalization group for multicomponent quantum hall systems,''
  \href{http://dx.doi.org/10.1103/PhysRevB.91.045115}{{\em Phys. Rev. B}
  {\bfseries 91} (Jan, 2015) 045115}.
  \url{https://link.aps.org/doi/10.1103/PhysRevB.91.045115}.

\bibitem{pakrouski2015PetersonNayak}
K.~Pakrouski, M.~R. Peterson, T.~Jolicoeur, V.~W. Scarola, C.~Nayak, and
  M.~Troyer, ``Phase diagram of the $\ensuremath{\nu}=5/2$ fractional quantum
  hall effect: Effects of landau-level mixing and nonzero width,''
  \href{http://dx.doi.org/10.1103/PhysRevX.5.021004}{{\em Phys. Rev. X}
  {\bfseries 5} (Apr, 2015) 021004}.
  \url{https://link.aps.org/doi/10.1103/PhysRevX.5.021004}.

\bibitem{Greiter:1991raWenWilczekPRL}
M.~Greiter, X.-G. Wen, and F.~Wilczek, ``{Paired Hall state at half filling},''
  \href{http://dx.doi.org/10.1103/PhysRevLett.66.3205}{{\em Phys. Rev. Lett.}
  {\bfseries 66} (1991) 3205--3208}.

\bibitem{imry1975Ma}
Y.~Imry and S.-k. Ma, ``Random-field instability of the ordered state of
  continuous symmetry,''
  \href{http://dx.doi.org/10.1103/PhysRevLett.35.1399}{{\em Phys. Rev. Lett.}
  {\bfseries 35} (Nov, 1975) 1399--1401}.
  \url{https://link.aps.org/doi/10.1103/PhysRevLett.35.1399}.

\bibitem{chalker1988}
J.~T. Chalker and P.~D. Coddington, ``Percolation, quantum tunnelling and the
  integer hall effect,'' {\em J. Phys. C} {\bfseries 21} no.~14, (1988) 2665.

\bibitem{Simon2018mhp180109687}
S.~H. Simon, ``{Interpretation of thermal conductance of the ?=5/2 edge},''
  \href{http://dx.doi.org/10.1103/PhysRevB.97.121406}{{\em Phys. Rev.}
  {\bfseries B97} no.~12, (2018) 121406},
\href{http://arxiv.org/abs/1801.09687}{{\ttfamily arXiv:1801.09687
  [cond-mat.mes-hall]}}.
%%CITATION = ARXIV:1801.09687;%%.

\bibitem{Ma2018xvjFeldman1809.05488}
K.~K. Ma and D.~Feldman, ``{Partial equilibration of integer and fractional
  edge channels in the thermal quantum Hall effect},''
  \href{http://dx.doi.org/10.1103/PhysRevB.99.085309}{{\em Phys. Rev. B}
  {\bfseries 99} no.~8, (2019) 085309},
  \href{http://arxiv.org/abs/1809.05488}{{\ttfamily arXiv:1809.05488
  [cond-mat.mes-hall]}}.

\bibitem{simon2019partial}
S.~H. Simon and B.~Rosenow, ``Partial equilibration of the anti-pfaffian edge
  due to majorana disorder,'' \href{http://arxiv.org/abs/1906.05294}{{\ttfamily
  arXiv:1906.05294 [cond-mat.str-el]}}.

\bibitem{Simon1909.12844Zaletel}
S.~H. Simon, M.~Ippoliti, M.~P. Zaletel, and E.~H. Rezayi, ``Energetics of
  pfaffian?anti-pfaffian domains,''
  \href{http://dx.doi.org/10.1103/physrevb.101.041302}{{\em Physical Review B}
  {\bfseries 101} no.~4, (Jan, 2020) },
  \href{http://arxiv.org/abs/1909.12844}{{\ttfamily arXiv:1909.12844
  [cond-mat.mes-hall]}}. \url{http://dx.doi.org/10.1103/physrevb.101.041302}.

\bibitem{Mulligan2004.04161}
H.~Asasi and M.~Mulligan, ``Partial equilibration of anti-pfaffian edge modes
  at $\nu=5/2$,'' \href{http://arxiv.org/abs/2004.04161}{{\ttfamily
  arXiv:2004.04161 [cond-mat.str-el]}}.

\bibitem{Feldman2018ssm180503204}
D.~E. Feldman, ``{Comment on ?Interpretation of thermal conductance of the
  ?=5/2 edge?},'' \href{http://dx.doi.org/10.1103/PhysRevB.98.167401}{{\em
  Phys. Rev.} {\bfseries B98} no.~16, (2018) 167401},
\href{http://arxiv.org/abs/1805.03204}{{\ttfamily arXiv:1805.03204
  [cond-mat.mes-hall]}}.
%%CITATION = ARXIV:1805.03204;%%.

\bibitem{Cordova:2017vab}
C.~C{\'o}rdova, P.-S. Hsin, and N.~Seiberg, ``{Global Symmetries, Counterterms,
  and Duality in Chern-Simons Matter Theories with Orthogonal Gauge Groups},''
  \href{http://dx.doi.org/10.21468/SciPostPhys.4.4.021}{{\em SciPost Phys.}
  {\bfseries 4} no.~4, (2018) 021},
\href{http://arxiv.org/abs/1711.10008}{{\ttfamily arXiv:1711.10008 [hep-th]}}.
%%CITATION = ARXIV:1711.10008;%%.

\bibitem{Hsin2016blu1607.07457}
P.-S. Hsin and N.~Seiberg, ``{Level/rank Duality and Chern-Simons-Matter
  Theories},'' \href{http://dx.doi.org/10.1007/JHEP09(2016)095}{{\em JHEP}
  {\bfseries 09} (2016) 095}, \href{http://arxiv.org/abs/1607.07457}{{\ttfamily
  arXiv:1607.07457 [hep-th]}}.

\bibitem{Metlitski:2014xqa}
M.~A. Metlitski, L.~Fidkowski, X.~Chen, and A.~Vishwanath, ``{Interaction
  effects on 3D topological superconductors: surface topological order from
  vortex condensation, the 16 fold way and fermionic Kramers doublets},''
  \href{http://arxiv.org/abs/1406.3032}{{\ttfamily arXiv:1406.3032
  [cond-mat.str-el]}}.

\bibitem{Metlitski1510.05663}
M.~A. Metlitski, ``{$S$-duality of $u(1)$ gauge theory with $\theta =\pi$ on
  non-orientable manifolds: Applications to topological insulators and
  superconductors},'' \href{http://arxiv.org/abs/1510.05663}{{\ttfamily
  arXiv:1510.05663 [hep-th]}}.

\bibitem{WangLevin1610.04624}
C.~Wang and M.~Levin, ``{Anomaly indicators for time-reversal symmetric
  topological orders},''
  \href{http://dx.doi.org/10.1103/PhysRevLett.119.136801}{{\em Phys. Rev.
  Lett.} {\bfseries 119} no.~13, (2017) 136801},
  \href{http://arxiv.org/abs/1610.04624}{{\ttfamily arXiv:1610.04624
  [cond-mat.str-el]}}.

\bibitem{Tachikawa:2016cha}
Y.~Tachikawa and K.~Yonekura, ``{On time-reversal anomaly of 2+1d topological
  phases},'' \href{http://dx.doi.org/10.1093/ptep/ptx010}{{\em PTEP} {\bfseries
  2017} no.~3, (2017) 033B04},
  \href{http://arxiv.org/abs/1610.07010}{{\ttfamily arXiv:1610.07010
  [hep-th]}}.

\bibitem{Cordova:2017kue}
C.~C\'ordova, P.-S. Hsin, and N.~Seiberg, ``{Time-Reversal Symmetry, Anomalies,
  and Dualities in (2+1)$d$},''
  \href{http://dx.doi.org/10.21468/SciPostPhys.5.1.006}{{\em SciPost Phys.}
  {\bfseries 5} no.~1, (2018) 006},
  \href{http://arxiv.org/abs/1712.08639}{{\ttfamily arXiv:1712.08639
  [cond-mat.str-el]}}.

\bibitem{Gaiotto:2014kfa}
D.~Gaiotto, A.~Kapustin, N.~Seiberg, and B.~Willett, ``{Generalized Global
  Symmetries},'' \href{http://dx.doi.org/10.1007/JHEP02(2015)172}{{\em JHEP}
  {\bfseries 02} (2015) 172}, \href{http://arxiv.org/abs/1412.5148}{{\ttfamily
  arXiv:1412.5148 [hep-th]}}.

\bibitem{Hsin:2018vcg}
P.-S. Hsin, H.~T. Lam, and N.~Seiberg, ``{Comments on One-Form Global
  Symmetries and Their Gauging in 3d and 4d},''
  \href{http://dx.doi.org/10.21468/SciPostPhys.6.3.039}{{\em SciPost Phys.}
  {\bfseries 6} no.~3, (2019) 039},
\href{http://arxiv.org/abs/1812.04716}{{\ttfamily arXiv:1812.04716 [hep-th]}}.
%%CITATION = ARXIV:1812.04716;%%.

\bibitem{1602.04251SW}
N.~Seiberg and E.~Witten, ``{Gapped Boundary Phases of Topological Insulators
  via Weak Coupling},'' \href{http://dx.doi.org/10.1093/ptep/ptw083}{{\em PTEP}
  {\bfseries 2016} no.~12, (2016) 12C101},
  \href{http://arxiv.org/abs/1602.04251}{{\ttfamily arXiv:1602.04251
  [cond-mat.str-el]}}.

\bibitem{Medvedyeva:2010MTB}
M.~V. Medvedyeva, J.~Tworzyd\l{}o, and C.~W.~J. Beenakker, ``Effective mass and
  tricritical point for lattice fermions localized by a random mass,''
  \href{http://dx.doi.org/10.1103/PhysRevB.81.214203}{{\em Phys. Rev. B}
  {\bfseries 81} (Jun, 2010) 214203}.
  \url{https://link.aps.org/doi/10.1103/PhysRevB.81.214203}.

\bibitem{Putrov2016qdo1612.09298PWY}
P.~Putrov, J.~Wang, and S.-T. Yau, ``{Braiding Statistics and Link Invariants
  of Bosonic/Fermionic Topological Quantum Matter in 2+1 and 3+1 dimensions},''
  \href{http://dx.doi.org/10.1016/j.aop.2017.06.019}{{\em Annals Phys.}
  {\bfseries 384} (2017) 254--287},
  \href{http://arxiv.org/abs/1612.09298}{{\ttfamily arXiv:1612.09298
  [cond-mat.str-el]}}.

\bibitem{1801.05416WOP}
J.~{Wang}, K.~{Ohmori}, P.~{Putrov}, Y.~{Zheng}, Z.~{Wan}, M.~{Guo}, H.~{Lin},
  P.~{Gao}, and S.-T. {Yau}, ``{Tunneling Topological Vacua via Extended
  Operators: (Spin-)TQFT Spectra and Boundary Deconfinement in Various
  Dimensions},'' {\em ArXiv e-prints} (Jan., 2018) ,
  \href{http://arxiv.org/abs/1801.05416}{{\ttfamily arXiv:1801.05416
  [cond-mat.str-el]}}.

\bibitem{Guo20181812.11959}
M.~Guo, K.~Ohmori, P.~Putrov, Z.~Wan, and J.~Wang, ``{Fermionic Finite-Group
  Gauge Theories and Interacting Symmetric/Crystalline Orders via
  Cobordisms},'' \href{http://dx.doi.org/10.1007/s00220-019-03671-6}{{\em
  Commun. Math. Phys.} (2020) 1--82},
  \href{http://arxiv.org/abs/1812.11959}{{\ttfamily arXiv:1812.11959
  [hep-th]}}.

\bibitem{Dashenetal1}
R.~F. Dashen, B.~Hasslacher, and A.~Neveu, ``{Nonperturbative Methods and
  Extended Hadron Models in Field Theory 1. Semiclassical Functional
  Methods},''
\href{http://dx.doi.org/10.1103/PhysRevD.10.4114}{{\em Phys. Rev.} {\bfseries
  D10} (1974) 4114}.
%%CITATION = PHRVA,D10,4114;%%.

\bibitem{Dashenetal2}
R.~F. Dashen, B.~Hasslacher, and A.~Neveu, ``{Nonperturbative Methods and
  Extended Hadron Models in Field Theory 2. Two-Dimensional Models and Extended
  Hadrons},''
\href{http://dx.doi.org/10.1103/PhysRevD.10.4130}{{\em Phys. Rev.} {\bfseries
  D10} (1974) 4130--4138}.
%%CITATION = PHRVA,D10,4130;%%.

\bibitem{DGP}
T.~Dimofte, D.~Gaiotto, and N.~M. Paquette, ``{Dual boundary conditions in 3d
  SCFT?s},'' \href{http://dx.doi.org/10.1007/JHEP05(2018)060}{{\em JHEP}
  {\bfseries 05} (2018) 060},
\href{http://arxiv.org/abs/1712.07654}{{\ttfamily arXiv:1712.07654 [hep-th]}}.
%%CITATION = ARXIV:1712.07654;%%.

\bibitem{GKS}
D.~Gaiotto, Z.~Komargodski, and N.~Seiberg, ``{Time-reversal breaking in
  QCD$_{4}$, walls, and dualities in 2 + 1 dimensions},''
  \href{http://dx.doi.org/10.1007/JHEP01(2018)110}{{\em JHEP} {\bfseries 01}
  (2018) 110},
\href{http://arxiv.org/abs/1708.06806}{{\ttfamily arXiv:1708.06806 [hep-th]}}.
%%CITATION = ARXIV:1708.06806;%%.

\bibitem{Wang2019obe1910.14664WYZ}
J.~Wang, Y.-Z. You, and Y.~Zheng, ``{Gauge Enhanced Quantum Criticality and
  Time Reversal Domain Wall: SU(2) Yang-Mills Dynamics with Topological
  Terms},'' \href{http://arxiv.org/abs/1910.14664}{{\ttfamily arXiv:1910.14664
  [cond-mat.str-el]}}.

\bibitem{PY}
A.~Parnachev and L.~G. Yaffe, ``{One loop quantum energy densities of domain
  wall field configurations},''
  \href{http://dx.doi.org/10.1103/PhysRevD.62.105034}{{\em Phys. Rev.}
  {\bfseries D62} (2000) 105034},
\href{http://arxiv.org/abs/hep-th/0005269}{{\ttfamily arXiv:hep-th/0005269
  [hep-th]}}.
%%CITATION = HEP-TH/0005269;%%.

\bibitem{Rebhanetal}
A.~Rebhan, P.~van Nieuwenhuizen, and R.~Wimmer, ``{One loop surface tensions of
  (supersymmetric) kink domain walls from dimensional regularization},''
  \href{http://dx.doi.org/10.1088/1367-2630/4/1/331}{{\em New J. Phys.}
  {\bfseries 4} (2002) 31},
\href{http://arxiv.org/abs/hep-th/0203137}{{\ttfamily arXiv:hep-th/0203137
  [hep-th]}}.
%%CITATION = HEP-TH/0203137;%%.

\bibitem{Shifman}
M.~Shifman, {\em {Advanced topics in quantum field theory.}}
\newblock Cambridge Univ. Press, Cambridge, UK, 2012.
\newblock
\url{http://www.cambridge.org/mw/academic/subjects/physics/theoretical-physics-and-mathematical-physics/advanced-topics-quantum-field-theory-lecture-course?format=AR}.
\newblock
%%CITATION = INSPIRE-1115926;%%.

\bibitem{CampbellLiao}
D.~K. Campbell and Y.-T. Liao, ``{A Semiclassical Analysis of Bound States in
  the Two-Dimensional Sigma Model},''
\href{http://dx.doi.org/10.1103/PhysRevD.14.2093}{{\em Phys. Rev.} {\bfseries
  D14} (1976) 2093}.
%%CITATION = PHRVA,D14,2093;%%.

\bibitem{Nastaseetal}
H.~Nastase, M.~A. Stephanov, P.~van Nieuwenhuizen, and A.~Rebhan,
  ``{Topological boundary conditions, the BPS bound, and elimination of
  ambiguities in the quantum mass of solitons},''
  \href{http://dx.doi.org/10.1016/S0550-3213(98)00773-1}{{\em Nucl. Phys.}
  {\bfseries B542} (1999) 471--514},
\href{http://arxiv.org/abs/hep-th/9802074}{{\ttfamily arXiv:hep-th/9802074
  [hep-th]}}.
%%CITATION = HEP-TH/9802074;%%.

\bibitem{Grahametal1}
E.~Farhi, N.~Graham, P.~Haagensen, and R.~L. Jaffe, ``{Finite quantum
  fluctuations about static field configurations},''
  \href{http://dx.doi.org/10.1016/S0370-2693(98)00354-2}{{\em Phys. Lett.}
  {\bfseries B427} (1998) 334--342},
\href{http://arxiv.org/abs/hep-th/9802015}{{\ttfamily arXiv:hep-th/9802015
  [hep-th]}}.
%%CITATION = HEP-TH/9802015;%%.

\bibitem{Grahametal2}
N.~Graham and R.~L. Jaffe, ``{Unambiguous one loop quantum energies of
  (1+1)-dimensional bosonic field configurations},''
  \href{http://dx.doi.org/10.1016/S0370-2693(98)00795-3}{{\em Phys. Lett.}
  {\bfseries B435} (1998) 145--151},
\href{http://arxiv.org/abs/hep-th/9805150}{{\ttfamily arXiv:hep-th/9805150
  [hep-th]}}.
%%CITATION = HEP-TH/9805150;%%.

\bibitem{Grahametal3}
N.~Graham and R.~L. Jaffe, ``{Fermionic one loop corrections to soliton
  energies in (1+1)-dimensions},''
  \href{http://dx.doi.org/10.1016/S0550-3213(99)00148-0}{{\em Nucl. Phys.}
  {\bfseries B549} (1999) 516--526},
\href{http://arxiv.org/abs/hep-th/9901023}{{\ttfamily arXiv:hep-th/9901023
  [hep-th]}}.
%%CITATION = HEP-TH/9901023;%%.

\bibitem{Rajaraman}
R.~Rajaraman, {\em {SOLITONS AND INSTANTONS. AN INTRODUCTION TO SOLITONS AND
  INSTANTONS IN QUANTUM FIELD THEORY}}.
\newblock
1982.
\newblock
%%CITATION = INSPIRE-181162;%%.

\bibitem{JW-seminar1}
J.~Wang, ``{``Mother Effective Field Theory for Fractional Quantum Hall Systems
  near $\nu= 5/2$'' {
  \href{https://www.youtube.com/watch?v=FoXUR7K9QmE}{https://www.youtube.com/watch?v=FoXUR7K9QmE}.
  \href{https://www.youtube.com/watch?v=nkEf65XK07I}{https://www.youtube.com/watch?v=nkEf65XK07I}.}},''
  {\em (Seminar Talk at Ultra Quantum Matter Simons Collaboration May 26th,
  2020 and at Harvard CMSA-Weizmann Institute of Science July 8, 2020)} (2020)
  .

\bibitem{Moore:1988ss}
G.~W. Moore and N.~Seiberg, ``{Naturality in Conformal Field Theory},''
\href{http://dx.doi.org/10.1016/0550-3213(89)90511-7}{{\em Nucl. Phys.}
  {\bfseries B313} (1989) 16--40}.
%%CITATION = NUPHA,B313,16;%%.

\bibitem{Moore:1989yh}
G.~W. Moore and N.~Seiberg, ``{Taming the Conformal Zoo},''
\href{http://dx.doi.org/10.1016/0370-2693(89)90897-6}{{\em Phys. Lett.}
  {\bfseries B220} (1989) 422--430}.
%%CITATION = PHLTA,B220,422;%%.

\bibitem{Bais:2008ni}
F.~A. Bais and J.~K. Slingerland, ``{Condensate induced transitions between
  topologically ordered phases},''
  \href{http://dx.doi.org/10.1103/PhysRevB.79.045316}{{\em Phys. Rev.}
  {\bfseries B79} (2009) 045316},
\href{http://arxiv.org/abs/0808.0627}{{\ttfamily arXiv:0808.0627
  [cond-mat.mes-hall]}}.
%%CITATION = ARXIV:0808.0627;%%.

\bibitem{KITAEV20062}
A.~Kitaev, ``Anyons in an exactly solved model and beyond,''
  \href{http://dx.doi.org/https://doi.org/10.1016/j.aop.2005.10.005}{{\em
  Annals of Physics} {\bfseries 321} no.~1, (2006) 2 -- 111}.
  \url{http://www.sciencedirect.com/science/article/pii/S0003491605002381}.
  January Special Issue.

\bibitem{Ma2019nynFeldman190402798}
K.~K.~W. Ma and D.~E. Feldman, ``{The sixteenfold way and the quantum Hall
  effect at half-integer filling factors},''
  \href{http://dx.doi.org/10.1103/PhysRevB.100.035302}{{\em Phys. Rev.}
  {\bfseries B100} no.~3, (2019) 035302},
\href{http://arxiv.org/abs/1904.02798}{{\ttfamily arXiv:1904.02798
  [cond-mat.mes-hall]}}.
%%CITATION = ARXIV:1904.02798;%%.

\bibitem{Ma2020pjs2003.05954}
R.~Ma and Y.-C. He, ``{Emergent QCD$_3$ Quantum Phase Transitions of Fractional
  Chern Insulators},'' \href{http://arxiv.org/abs/2003.05954}{{\ttfamily
  arXiv:2003.05954 [cond-mat.str-el]}}.

\bibitem{Belov2005zeMoore0505235}
D.~Belov and G.~W. Moore, ``{Classification of Abelian spin Chern-Simons
  theories},'' \href{http://arxiv.org/abs/hep-th/0505235}{{\ttfamily
  arXiv:hep-th/0505235}}.

\bibitem{Benini:2018reh}
F.~Benini, C.~C{\'o}rdova, and P.-S. Hsin, ``{On 2-Group Global Symmetries and
  their Anomalies},'' \href{http://dx.doi.org/10.1007/JHEP03(2019)118}{{\em
  JHEP} {\bfseries 03} (2019) 118},
\href{http://arxiv.org/abs/1803.09336}{{\ttfamily arXiv:1803.09336 [hep-th]}}.
%%CITATION = ARXIV:1803.09336;%%.

\bibitem{read2000Green}
N.~Read and D.~Green, ``Paired states of fermions in two dimensions with
  breaking of parity and time-reversal symmetries and the fractional quantum
  hall effect,'' \href{http://dx.doi.org/10.1103/PhysRevB.61.10267}{{\em Phys.
  Rev. B} {\bfseries 61} (Apr, 2000) 10267--10297}.

\bibitem{PhysRevB.29.6012Girvin1984}
S.~M. Girvin, ``Particle-hole symmetry in the anomalous quantum hall effect,''
  \href{http://dx.doi.org/10.1103/PhysRevB.29.6012}{{\em Phys. Rev. B}
  {\bfseries 29} (May, 1984) 6012--6014}.
  \url{https://link.aps.org/doi/10.1103/PhysRevB.29.6012}.

\bibitem{Balram2018Ajit1803.10427}
A.~C. Balram, M.~Barkeshli, and M.~S. Rudner, ``Parton construction of a wave
  function in the anti-pfaffian phase,''
  \href{http://dx.doi.org/10.1103/physrevb.98.035127}{{\em Physical Review B}
  {\bfseries 98} no.~3, (Jul, 2018) },
  \href{http://arxiv.org/abs/1803.10427}{{\ttfamily arXiv:1803.10427
  [cond-mat.str-el]}}. \url{http://dx.doi.org/10.1103/physrevb.98.035127}.

\bibitem{1804.01107PHPf}
R.~V. Mishmash, D.~F. Mross, J.~Alicea, and O.~I. Motrunich, ``{Numerical
  exploration of trial wave functions for the particle-hole-symmetric
  Pfaffian},'' \href{http://dx.doi.org/10.1103/PhysRevB.98.081107}{{\em Phys.
  Rev.} {\bfseries B98} no.~8, (2018) 081107},
\href{http://arxiv.org/abs/1804.01107}{{\ttfamily arXiv:1804.01107
  [cond-mat.str-el]}}.
%%CITATION = ARXIV:1804.01107;%%.

\bibitem{yang2017particlehole1701.03562}
J.~Yang, ``Particle-hole symmetry and the fractional quantum hall states at 5/2
  filling factor,'' 2017.

\bibitem{yang2020compressed2001.01915}
J.~Yang, ``A compressed particle-hole symmetric pfaffian state for $\nu= 5/2$
  quantum hall effect,'' 2020.

\end{thebibliography}\endgroup

\end{document}